%% file: resubmit.tex
\documentclass[aps,prb,twocolumn,showpacs,longbibliography,preprintnumbers,amsmath,amssymb,superscriptaddress]{revtex4-1}
\usepackage{graphicx}
\usepackage[T1]{fontenc}
\usepackage[colorlinks,bookmarks=false,citecolor=blue,linkcolor=red,urlcolor=blue]{hyperref}
\usepackage{color}
\usepackage{times}

\begin{document}

\title{Phase diagram of interacting spinless fermions on the honeycomb lattice:\\ a comprehensive exact diagonalization study}

\author{Sylvain Capponi}
\email{capponi@irsamc.ups-tlse.fr}
\affiliation{Laboratoire de Physique Th\'eorique, CNRS UMR 5152,
  Universit\'e Paul Sabatier, F-31062 Toulouse, France}
\author{Andreas M. L\"auchli}
\email{andreas.laeuchli@uibk.ac.at}
\affiliation{Institut f\"ur Theoretische Physik, Universit\"at Innsbruck, A-6020 Innsbruck, Austria}

\date{\today}

\begin{abstract}
We investigate the phase diagram of spinless fermions with nearest and next-nearest neighbour
 density-density interactions on the honeycomb lattice at half-filling. Using Exact
Diagonalization techniques of the full Hamiltonian and constrained subspaces,
combined with a careful choice of finite-size clusters, we determine the different charge 
orderings that occur for large interactions. In this regime we find a two-sublattice N\'eel-like
state, a charge modulated state with a tripling of the unit cell, a zig-zag phase and a novel charge ordered states
with a 12 site unit cells we call N\'eel domain wall crystal, as well as a region of phase separation 
for attractive interactions. A sizeable region of the phase diagram is classically degenerate, but it remains unclear whether
an order-by-disorder mechanism will lift the degeneracy. For intermediate repulsion we find evidence for a  Kekul\'e  or plaquette 
bond-order wave phase. 
We also investigate the possibility of a spontaneous Chern insulator phase (dubbed topological Mott insulator), as previously 
put forward by several mean-field studies. Although we are unable to detect convincing evidence for this phase based
on energy spectra and order parameters, we find an enhancement of current-current correlations
with the expected spatial structure compared to the non-interacting situation. While for the
studied $t{-}V_1{-}V_2$ model the phase transition to the putative topological Mott insulator is preempted 
by the phase transitions to the various ordered states, our findings might hint at the possibility for a topological 
Mott insulator in an enlarged Hamiltonian parameter space, where the competing
phases are suppressed.
\end{abstract}

\maketitle

\section{Introduction}

The seminal discoveries of quantum Hall phases~\cite{book_QHE} have
provided the first examples of topological phases of matter,
i.e. phases which cannot be adiabatically connected to conventional
ones.  More recently, several other topological insulators or
superconductors have been theoretically predicted and experimentally
observed~\cite{topo_review2,topo_review}, giving rise to an intense activity of the
community.  For such materials, a very thorough understanding has been
achieved thanks to the fact that a noninteracting starting point is
sufficient: indeed, topological insulators can be obtained from band
theory in the presence of strong spin-orbit coupling.

On the other hand, given the variety of electronic phases that are also generated by
strong interactions, a highly topical question is to
investigate the appearance of topological insulators due to
correlations. This new route would not require intrinsic spin-orbit
coupling, and thus could be advantageous to many applications.
Nevertheless, we face the well-known difficulty of strongly correlated
systems for which we lack unbiased tools, and often have to resort to
some hypothesis. In this context, a very promising result has been
obtained by Raghu {\it et al.}~\cite{Raghu2008} who have shown the
emergence of quantum anomalous (spin) Hall (QAH) states on the
honeycomb lattice with strong density-density repulsions between
next-nearest neighbors. These phases can be characterized by the
existence of spontaneously appearing charge (respectively spin) currents for spinless
(respectively spinful) fermions. This finding has also been reproduced 
in the spinless case using more systematic mean-field
approaches~\cite{Weeks2010,Grushin2013}. Nevertheless, in the latter
study,~\cite{Grushin2013} the QAH extension in the phase diagram has shrunk substantially due to the
stability of charge modulation with larger unit cell.

The possibility to generate topological phases with longer-range interactions has triggered a large theoretical 
activity to investigate its experimental feasibility using metallic substrate~\cite{Pereg2012}, cold-atomic gases 
on optical lattices~\cite{Dauphin2012} or RKKY interaction~\cite{Liu2014} among many others. 

In the light of several recent approximate and numerical studies~\cite{Garcia2013,Daghofer2014,Duric2014} with conflicting conclusions
regarding the presence of the putative topological Mott state and competing ordered phases,
we feel that a comprehensive state-of-the-art exact diagonalization study could shed some light on these issues. 
Based on our simulations, 
we will provide numerical evidence that quantum fluctuations at small $V_2>0$ are responsible for an enhanced response similar to the QAH state, but unfortunately our numerical data (obtained on finite clusters up to $N=42$ sites) indicates that these short-range fluctuations presumably do not turn into a true long-range ordered phase. 
In addition, we will provide a detailed analysis of the charge and bond orderings that occur at large interaction strength, allowing us to sketch a global phase diagram of the minimal model studied so far.  
In Sec.~\ref{sec:model}, we detail the model and methods we use. We then discuss the large interaction (classical) limit in Sec.~\ref{sec:Ising}. Our numerical results are given in Sec.~\ref{sec:numerics}, and we conclude in Sec.~\ref{sec:conclusion}.

\section{Model, method and phase diagram}\label{sec:model}
\subsection{Hamiltonian} 
\begin{figure}[!ht]
\includegraphics[width=0.9\linewidth]{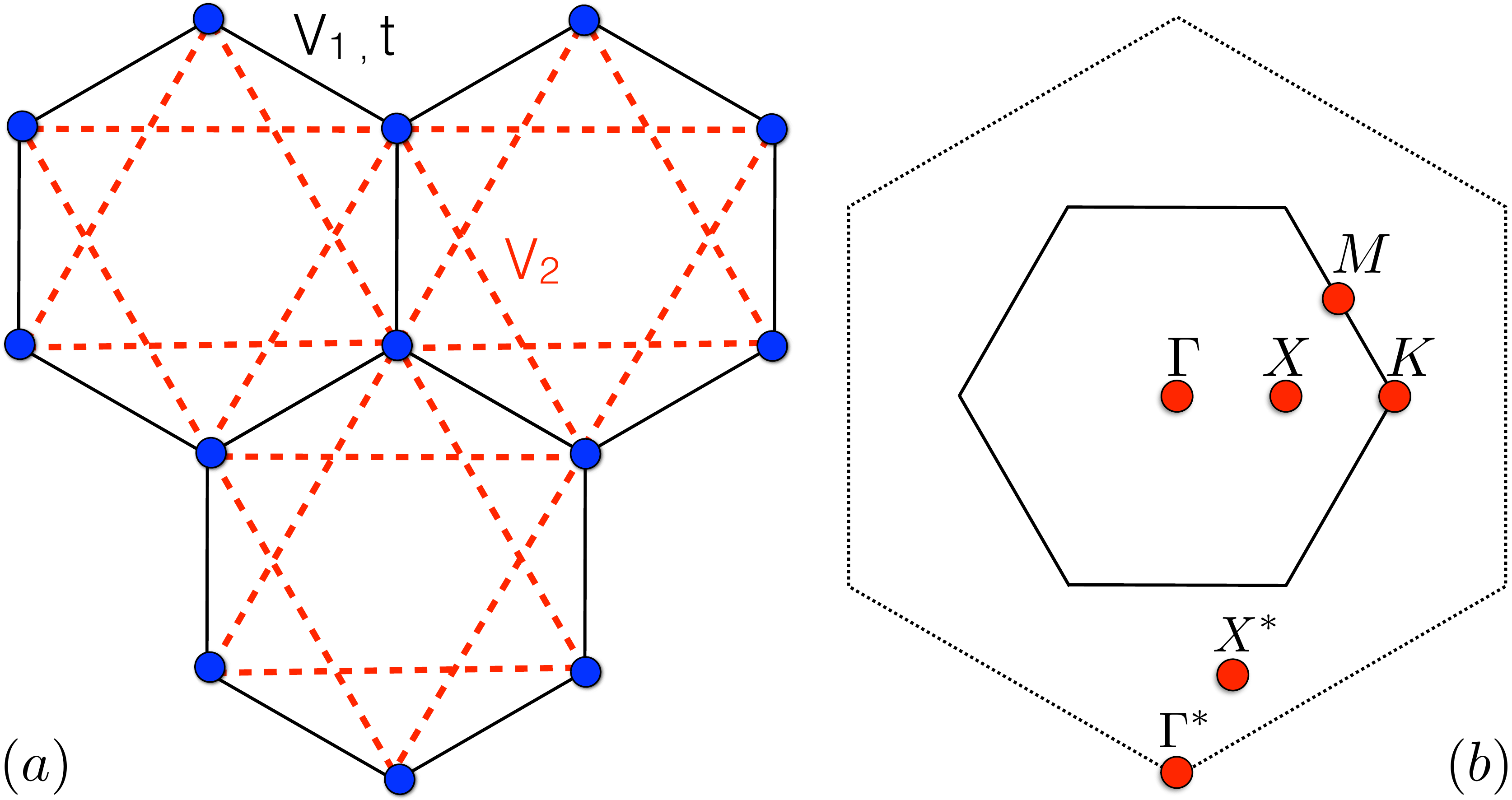}
\caption{(Color online) 
(a) Illustration of the honeycomb lattice with $V_1$, $V_2$ interactions and the hopping $t$.
(b) First (solid line) and second (dashed line) Brillouin zone of the honeycomb lattice including the 
 location of a few special points in the Brillouin zone.
}
\label{fig:latticeBZ}
\end{figure}

Following the proposal by Raghu {\it et al.}~\onlinecite{Raghu2008} of an exotic topological Mott insulator in a half-filled honeycomb lattice with $V_2$ interaction, we will consider the following Hamiltonian~:
\begin{eqnarray}\label{eq:model}
\label{eqn:Hamiltonian}
{\cal H} & = & -t\sum_{\langle i j\rangle} (c^\dagger_i c_j +h.c.) \\
&+& V_1 \sum_{\langle i j\rangle} (n_i -1/2)(n_j-1/2) \nonumber\\
&+&V_2 \sum_{\langle\langle i j\rangle\rangle} (n_i -1/2)( n_j-1/2) \nonumber 
\end{eqnarray}
depicted in Fig.~\ref{fig:latticeBZ}(a), 
where $c_i$ and $c^\dagger_i$ are the spinless fermionic operators, $t=1$ is the nearest-neighbor hopping amplitude, $V_1$ and $V_2$ are the density-density repulsion or attraction strengths respectively on nearest- and next-nearest neighbors. Note that we will focus on the half-filled case where thanks to particle-hole symmetry, the chemical potential is known to be zero exactly. 

Despite the particle-hole symmetry, it seems that quantum Monte-Carlo (QMC)
approaches exhibit a sign problem that prohibits them. 
In fact, only in the simpler case $V_2=0$ which exhibits a direct phase transition from semi-metallic (SM) phase to charge density wave (CDW) as a function of $V_1/t$, 
a very interesting recent proposal was made allowing to reformulate the problem without sign-problem~\cite{Huffman2014}, thus amenable to accurate, unbiased 
QMC simulations~\cite{Wang2014}.
Quite remarkably, the case $V_1\geq 0$ and $V_2 \leq 0$ can also be studied with QMC without a minus-sign problem using a Majorana representation~\cite{Li2015}.
In the absence of a QMC algorithm for the frustrated case, we
use Exact Diagonalization (ED) to get unbiased numerical
results. Of course, one is limited in the sizes available but, since
we are investigating in particular the topological QAH phase which is a translationally invariant ($q=0$) instability (see below), we can in principle use \emph{any}
finite-size lattice, even though some of them do not have the full
space-group symmetry. On the other hand, when looking at competing charge instabilities, it is crucial to consider only clusters compatible with such orderings 
and we will devote some discussion on this topic in the next subsection.

\subsection{Finite lattices}
Since we plan to provide a systematic study on several clusters, 
we have implemented lattices sizes
ranging from 12 to 42 sites, with various shapes and spatial symmetries. 
In particular, these clusters may or may not have some reflection or 
rotation symmetries (they all have inversion symmetry). Moreover, 
some of these lattices possess the $\pm \mathbf{K}$ points (see the Brillouin zone depicted in Fig.~\ref{fig:latticeBZ}(b)) where the
tight-binding dispersion exhibits two Dirac cones, leading to a
6-fold degeneracy of the half-filled free fermion ground-state (GS) so that
correlations require some care for these clusters. We refer to Appendix~\ref{appendix:lattice}
for further details on these finite-size clusters. 

Our large choice of clusters provides a substantial step forward with respect to other recent ED studies, where results were obtained solely on $N=18$ and $N=24$ in Ref.~\onlinecite{Garcia2013} or on $N=24$ and $N=30$ in Ref.~\onlinecite{Daghofer2014}.

\begin{figure}[!htb]
\includegraphics[width=\linewidth,clip]{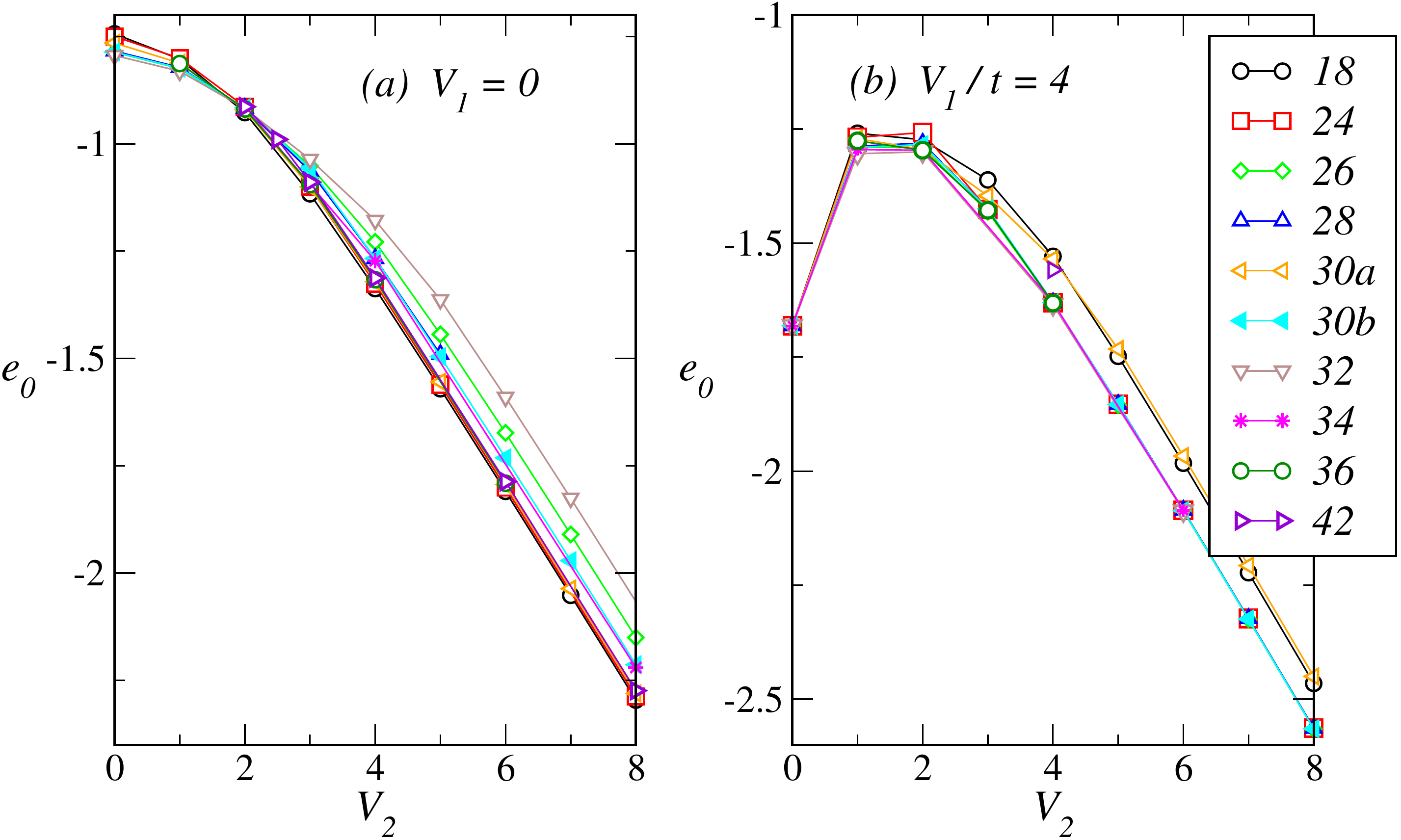}
\caption{(Color online) GS energy per site $e_0$ vs $V_2/t$ for (a) $V_1=0$ and (b) $V_1/t=4$ on various clusters. We refer to the Appendix~\ref{appendix:lattice}
for further details on these finite-size clusters. Lines are guide to the eyes.}
\label{fig:e0}
\end{figure}

In order to illustrate the need for a systematic study of cluster geometries we present some data for the ground state energy per site. 
For instance, at vanishing $V_1/t$, as shown in Fig.~\ref{fig:e0}(a), clusters with the $\mathbf{K}$ points yield the lowest energies for large $V_2/t$, i.e. are compatible with the appearing order (to be discussed later). 

At $V_1/t=4$ and large $V_2/t$, the situation becomes different as now clusters involving $\mathbf{M}$ point(s) seem to have the lowest energies, see Fig.~\ref{fig:e0}(b). 
This is already a clear indication of a phase transition between CDW (known to be stable at small $V_2/t$ and large $V_1/t$) and at least two other phases that we will characterise later. 

Clearly, the choice of clusters is important depending on the
parameters and the kind of instability. It was also pointed out
recently~\cite{Duric2014} that the choice of boundary conditions might
be crucial too. By studying the $N=18$ cluster with open boundary
conditions (OBC), as opposed to standard periodic boundary conditions (PBC), the authors of 
Ref.~\onlinecite{Duric2014} have found two
level crossings for $V_2/t \simeq 1.6$ and $V_2/t\simeq 2.9$
respectively (for $V_1=0$). This was taken as an indication for a
 stable QAH phase in this region. We have
checked that in this case, these level crossings correspond to two
low-energy states having opposite parities with respect to inversion
symmetry. However, these crossings disappear when one considers the
next largest cluster $N=32$ with OBC, so that it probably corresponds to a
short-range feature only. We generally think that using PBC is more appropriate to minimize finite-size effects, so this will be the case here.

Last, we would like to recall some features of the semi-metallic (SM) phase that exists in the absence of interactions. Since there is a vanishing density of states at the Fermi level, one needs a \emph{finite strength} of a short-ranged interaction to trigger an instability~\cite{Shankar1994,Kotov2012}, so that SM phase \emph{must have a finite extension} in the phase diagram. 

Regarding possible gap opening mechanism of the SM phase, Refs \onlinecite{Ryu2009,Herbut2009} have listed all explicit (i.e. external) weak-coupling perturbations which can open a gap. 
In the spinless case considered here, the three particle-hole related gaps are: i) the N\'eel-like charge density wave, which breaks the A-B sublattice symmetry, ii) the Kekul\'e distortion
pattern, which breaks translation symmetry by adopting a tripling of the unit-cell of modulated bond-strengths (this order parameter has a real and an imaginary part, thus corresponds to
two masses), and iii) the integer quantum Hall mass~\cite{Haldane1988}, 
induced by breaking the time-reversal invariance and parity symmetry upon adding complex Peierls phases on next-nearest neighbor hoppings, without enlarging the size of the unit-cell.
In addition to the particle-hole gaps, there is also the possibility to open gaps by the addition of superconducting order parameters~\cite{Ryu2009,Roy2010,Yao14}, we will however not address
these instabilities in this work.

The model Hamiltonian~\eqref{eqn:Hamiltonian} considered here features all the usual symmetries. If the semi-metallic phase is gapped out by interactions, then the gap opening has to happen
through the interactions by spontaneously breaking some of the symmetries. The case i) quoted before is a well known instability, since the N\'eel CDW state is an obvious strong-coupling ground
state at large $V_1/t$. The other instabilities ii) and iii) currently lack a strong coupling picture, and need to be confirmed by numerical simulations. We note however that all three particle-hole
instabilities have been reported in mean-field studies.~\cite{Raghu2008,Weeks2010,Grushin2013}
 
\subsection{Overview of the phase diagram}\label{sec:phase_diagram}

\begin{figure}[!ht]
\includegraphics[width=\linewidth]{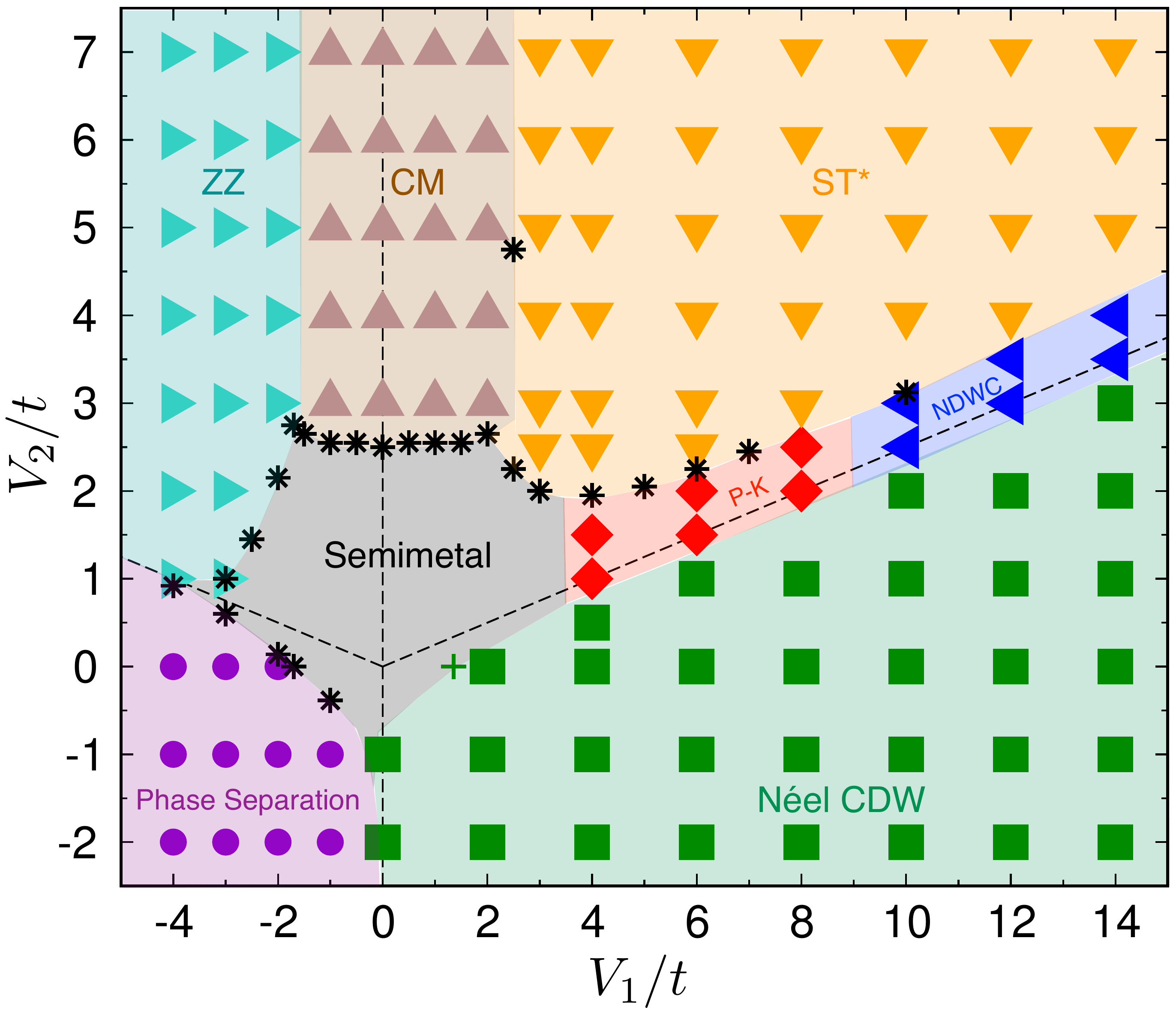}
\caption{(Color online) 
Phase diagram in the $(V_1/t,V_2/t)$ parameter space obtained from several exact diagonalization techniques (see text).  Dashed lines represent the classical transition lines, see Fig.~\ref{fig:IsingPhaseDiagram}. The semi-metal, which is the ground-state for non-interacting spinless fermions, has a finite extension in the phase diagram because of its vanishing density of states at the Fermi level. We will argue in the remainder of this article that several other phases can be stabilised for intermediate and/or large interactions: N\'eel CDW, plaquette/Kekul\'e (P-K), N\'eel domain wall crystal (NDWC), zigzag (ZZ) phase, and charge modulation (CM). The region (ST*) is degenerate at the semiclassical level, and it is presently unclear whether and how an order-by-disorder mechanism will lift the degeneracy.  Note also the large region of phase separation mostly in the attractive quadrant. 
Filled symbols correspond to numerical evidence (using level spectroscopy or measurements of correlations, see Sec.~\ref{sec:numerics}) obtained mostly on a $N=24$ cluster which contains the most important points in its Brillouin zone and features the full lattice point group symmetry of the honeycomb lattice. Star symbols denote likely first order transitions, witnessed by level crossings on the same
cluster. Our numerical results do not support any region of topological QAH phase. 
}
\label{fig:PhaseDiagram}
\end{figure}

We  start by drawing the global phase diagram that summarizes our main findings, see Fig.~\ref{fig:PhaseDiagram}. Its main features are the existence of several types of charge or bond ordering for intermediate to large $V_1$ and/or $V_2$ interactions: N\'eel CDW, charge modulation (CM), zigzag (ZZ) phase,  N\'eel domain wall crystal (NDWC), and plaquette/Kekul\'e phase (P-K) that we will clarify later. The large orange region (ST*) in the upper right part of the phase diagram features a degeneracy at the semiclassical level, and it is presently unclear whether and how 
an order-by-disorder mechanism will lift the degeneracy. While some of these phases (CM, P-K, CDW) had been predicted using mean-field studies~\cite{Grushin2013}  and confirmed numerically in some regions~\cite{Garcia2013,Daghofer2014}, the others (including NDWC and ST* phase for repulsive interactions and the ZZ phase for attractive $V_1$) had not been advocated before.
Note already that  the plaquette/Kekul\'e (P-K) phase only exists in some bounded region for intermediate ($V_1$, $V_2$) values, and does not extend to strong coupling.

There is also a large region of phase separation, mostly for strong attractive interactions, in agreement with the results of Ref.~\onlinecite{Corboz2012} for $V_2=0$. Possibly superconductivity
is present in parts of the attractive region of the phase diagram, but we did not focus on this instability here.

Last but not least, we do not have any convincing evidence for the stability of the QAH phase, as found in similar recent numerical studies~\cite{Garcia2013,Daghofer2014} but in contradiction with another numerical study using open boundary conditions and entangled-plaquette ansatz.~\cite{Duric2014}

We will now turn to the presentation of various numerical data and considerations that we have used to come up with this global phase diagram. 

\begin{figure*}[!ht]
\includegraphics[width=0.8\linewidth,clip]{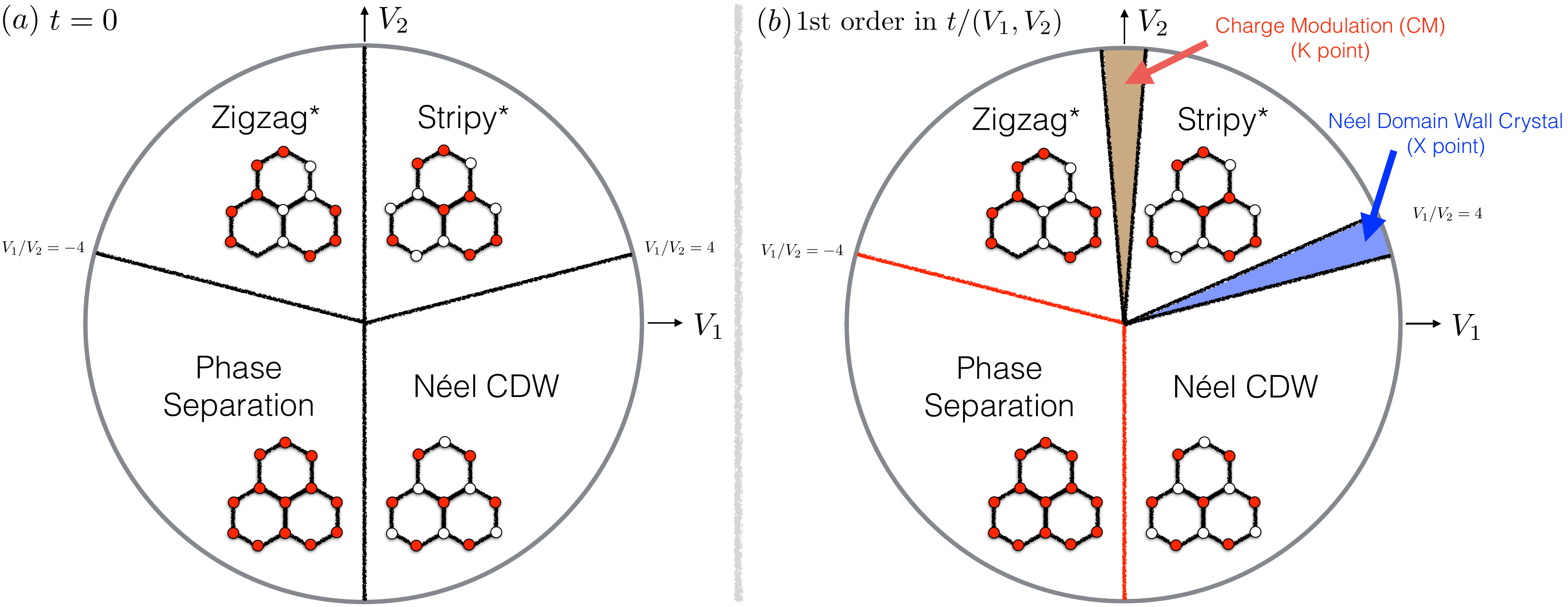}
\caption{(Color online) Classical phase diagram (a) and qualitative first order in $t/(V_1,V_2)$ phase diagram (b). The "Zigzag*" and the "Stripy*"  phases feature a
nontrivial ground state degeneracy (hence the "*" suffix in "Zigzag*" and "Stripy*"). The three points $V_2=1, V_1=\pm 4$ and $V_2=1, V_1=0$ feature an
extensive ground state degeneracy at the classical level. Upon including the first order correction due to the finite hopping $t$, two of these points spawn
new phases. The $V_2=1, V_1=0$ point develops into the charge modulation (CM) phase, while the $V_2=1, V_1=+4$ point broadens into 
a novel "N\'eel domain wall crystal" (NDWC), which is sketched in Fig.~\ref{fig:gs_eff_model_V1V2_4}. All regions and lines in (b) beyond the CM and NDWC phases
have no first order (in $t$) quantum corrections. Note that the radial direction in the two panels has no physical
relevance, i.e. it is not parametrizing $t/V_{1,2}$.}
\label{fig:IsingPhaseDiagram}
\end{figure*}
\begin{figure}[!hb]
\includegraphics[width=0.5\linewidth]{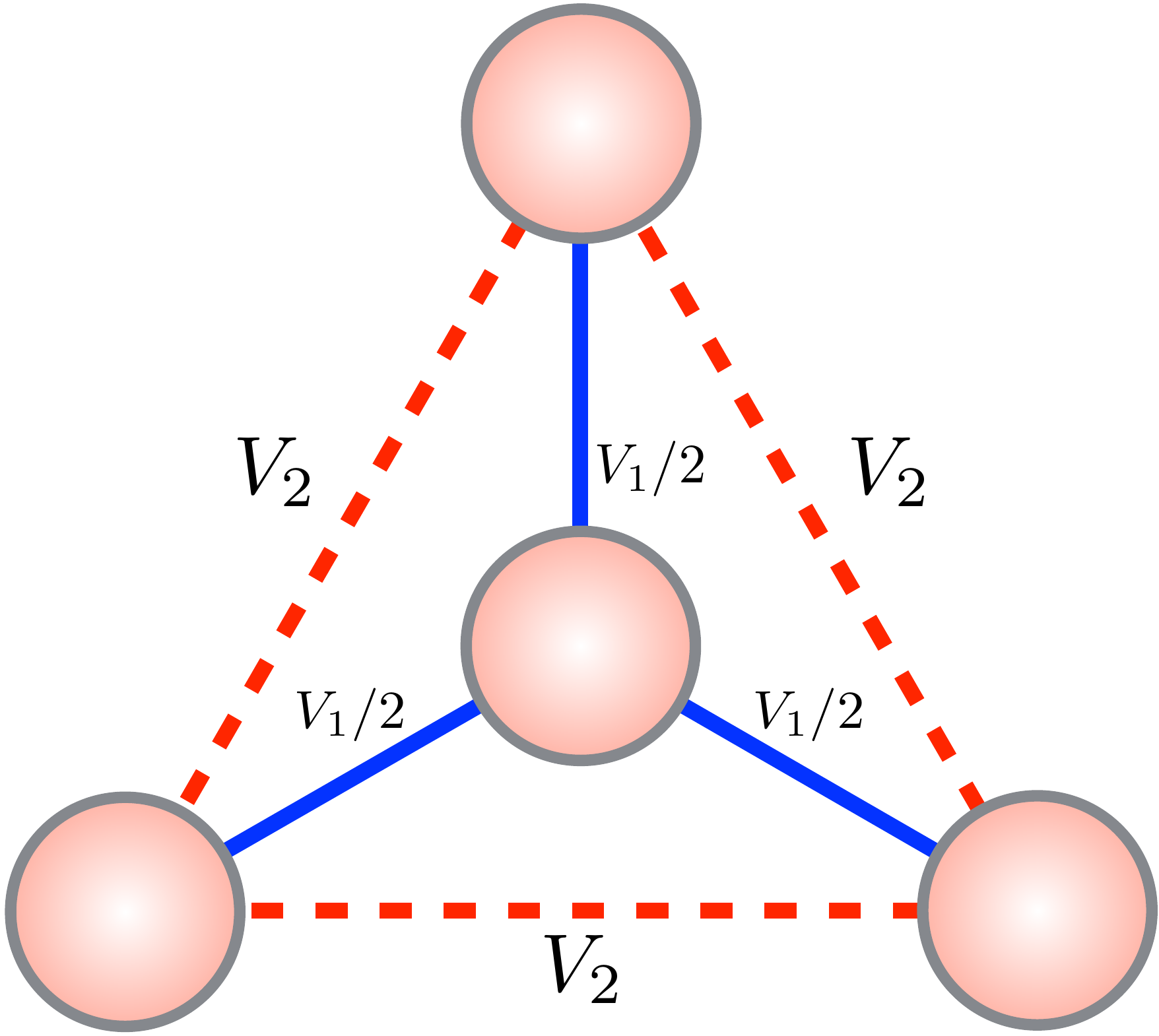}
\caption{(Color online) Sketch of a Hamiltonian unit which renders the classical $V_1{-}V_2$ problem "frustration free".
}
\label{fig:FFUnit}
\end{figure}

\section{Limit of large interactions (Ising limit)}\label{sec:Ising}

The Hamiltonian ($\ref{eqn:Hamiltonian}$) reduces to a free fermion problem in the absence of interactions, yielding a semi-metallic state with two Dirac points at half filling. 
In the opposite limit $t=0$, the model reduces to a (generally frustrated) classical Ising model with competing antiferromagnetic (for $V_{1,2}>0$) or ferromagnetic (for $V_{1,2}<0$) nearest and next-nearest neighbor interactions.
Since the phase diagram at finite but small hopping is expected to be
influenced by this (frustrated) Ising limit, we find it worthwhile to explore the classical phase diagram first.

Our results are summarized in the phase diagrams sketched in Fig.~\ref{fig:IsingPhaseDiagram} where we have considered the pure classical case $t=0$ (left panel), 
as well as the perturbative effect of a small $t$ (right panel). We will discuss these two cases and their phases in the following. We will exchangeably use the 
$V_1,V_2$ notation or the $\theta$ notation: $V_1=\cos\theta$, $V_2=\sin \theta$, with $\theta \in [0,2\pi)$. 

\subsection{Classical limit: \texorpdfstring{$\boldsymbol{t=0}$}{$t=0$}}

First we point out that this classical Hamiltonian can be written in a "frustration free" form, 
when the Hamiltonian is decomposed into parts acting on
triangles spanned by three $V_2$ bonds and their central site (this object
can alternatively be considered as a tripod, i.e.~a site plus its three nearest $V_1$ neighbors), 
whose three $V_1$ couplings are counted only with half their strength (this
is necessary because each such bond belongs to two different $V_2$
triangles, while the $V_2$ bonds belong to a unique triangle), see Fig.~\ref{fig:FFUnit}. 
In this decomposition each such tripod acts on 4 sites, and depending on $\theta$
the Hamiltonian on a tripod has a different number of local ground states. We have 
numerically verified that all ground states of the full problem are simultaneous
ground states of each tripod term, thus the characterization
"frustration free". 

By finite cluster ground state enumeration techniques (with and without exploiting the frustration free
property)
we have established the classical phase diagram displayed in the left
panel of Fig.~\ref{fig:IsingPhaseDiagram}. For $V_1=0,\ V_2=1$ ($\theta=\pi/2$), the
system decouples into two interpenetrating triangular lattices with
antiferromagnetic interactions.  The ground state manifold is thus
spanned by the product of the ground spaces of each triangular
sublattice, and is macroscopically degenerate (i.e.~featuring an
extensive ground state entropy). For $\arctan(1/4)<\theta<\pi/2$ 
the ground state manifold is much reduced, but still grows with system size, cf.~Tab.~\ref{tab:IsingCounting} in Appendix~\ref{app:isingcount}. 
On clusters featuring one or several $\mathbf{M}$ point(s) in the Brillouin zone,
some of these states can be understood as pristine stripy
ordered states (sketched in the appropriate region in Fig.~\ref{fig:IsingPhaseDiagram}), combined with line defects, as illustrated in
Fig.~\ref{fig:ChargeOrderMPoint}.  It is however not clear to us whether all
states in the ground space manifold on all lattices can be generated along these ideas.
At $\theta=\arctan(1/4)$ we find
again a single point with a rapidly growing ground state
degeneracy. For $-\pi/2<\theta < \arctan(1/4)$ we find the two
N\'eel ground states, which are obvious ground states at $\theta=0$,
corresponding to nearest neighbor interactions only.  For 
$\pi - \arctan(1/4) <\theta< 3\pi/2$ we find a region of phase separation,
where the system wants to be either empty or completely filled with fermions.
At $\theta =\pi - \arctan(1/4)\ (V_1=-4,V_2=1)$ we have another point with an
extensive ground state degeneracy, albeit with a smaller number than on the 
reflected side $\theta=\arctan(1/4)$. Finally the region $\pi/2<\theta < \pi - \arctan(1/4)$
hosts a degenerate set of states, which seem to be related to zigzag
charge configurations (sketched in the appropriate region in Fig.~\ref{fig:IsingPhaseDiagram}), 
with some defect structures in addition.

The explicit ground space counting for a number of different clusters is listed for
further reference in Tab.~\ref{tab:isingcount} in Appendix~\ref{app:isingcount}.

\subsection{Perturbed classical limit: \texorpdfstring{$\boldsymbol{t=0^+}$}{$t=0^+$}}
\label{sec:perturbed_classical}

After having established the classical phase diagram, we include the
effect of a finite hopping amplitude $t>0$ on the classical ground
states. We limit ourselves here to perturbative arguments (mostly first order in $t$), 
and check these predictions by full scale ED in the following sections.

We proceed by discussing the fate of the regions in the classical phase diagram,
and then analyze the behavior on the degenerate points.

\subsubsection{N\'eel CDW state}

We start with the least degenerate region, $-\pi/2<\theta < \arctan(1/4)$, where there
are only two N\'eel ground states. These two states are symmetry
related (and thus cannot be split by diagonal terms) and it is not
possible to connect the two states by a finite order in perturbation
theory in the thermodynamic limit. So based on perturbation theory
this phase is expected to have a finite region of stability when
$t\neq0$.

\subsubsection{Phase separation region}

The phase separation region $\pi - \arctan(1/4) <\theta< 3\pi/2$ is also 
stable at finite $t\neq 0$ because both the empty and the full system are
inert under the action of the hopping term.

\subsubsection{Zigzag state (ZZ)}\label{sec:zigzag}

\begin{figure}
\includegraphics[width=0.9\linewidth,clip]{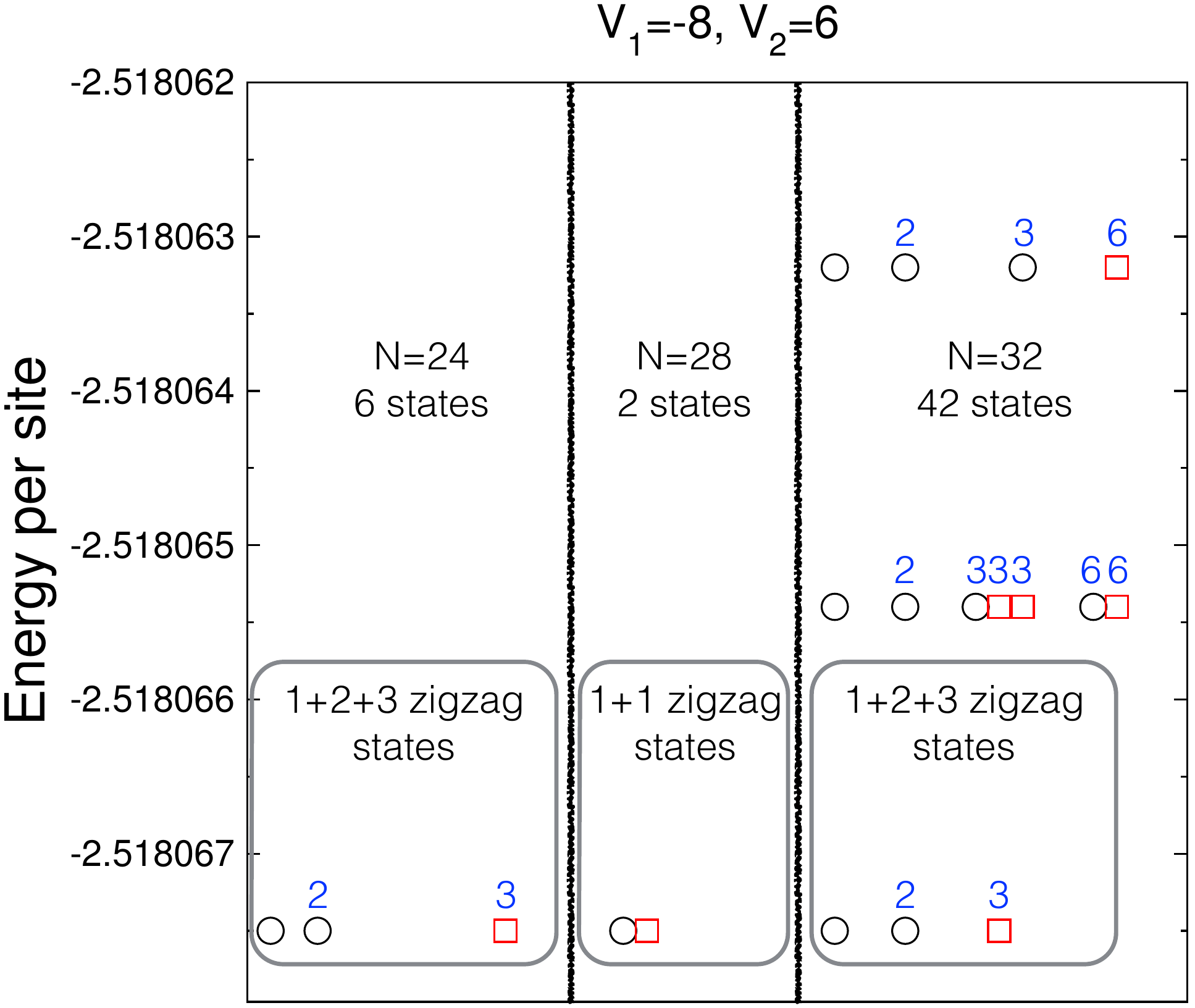}
\caption{(Color online) Splitting of the ground state degeneracy at $V_1/t=-8$, 
$V_2/t=6$, yielding the two or six zigzag states as the ground states. Black (red)
symbols denote the two different particle-hole  symmetry sectors. All levels are 
singly degenerate unless the (spatial symmetry) degeneracy is indicated by a blue number.
The perturbatively generated diagonal terms splits the ground space into symmetry
related families, whereby the two or six zigzag states become lowest in energy.
Note the very small energy difference, indicating a high-order perturbative effect.
} 
\label{fig:spectrum_V1_m8_V2_6}
\end{figure}

Next we consider the extended region $\pi/2 < \theta <\pi- \arctan(1/4)$, featuring
zigzag ground states and some types of domain walls.  To first order
in $t/V_{1,2}$ it is not possible to hop and stay in the ground
state manifold. Exact diagonalizations of the full microscopic models on various clusters (see Sec.~\ref{sec:numerics}) with true 
classical ground states reveal however an 
order-by-disorder mechanism, where a diagonal term in high-order perturbation theory 
seems to lift the degeneracy and select the pristine zigzag state, see Fig.~\ref{fig:spectrum_V1_m8_V2_6}. 
This state is six-fold degenerate. A similar state translated to spin language appears in the 
Heisenberg-Kitaev model on the honeycomb lattice.~\cite{Chaloupka2013}

\begin{figure*}[!ht]
\includegraphics[width=0.8\linewidth,clip]{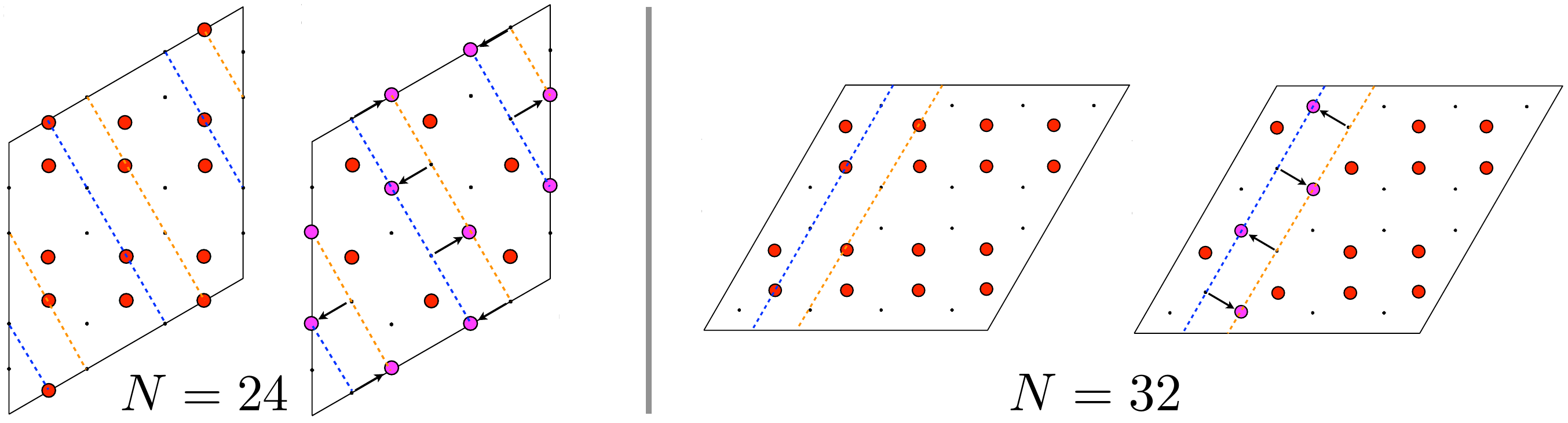}
\caption{(Color online) Illustration of the stripy charge ordering pattern and the line defects for the $N=24$ (left panel) and $N=32$ (right panel) clusters. The left
part in each panel indicates a pristine stripy configuration, while the right part shows how another ground state can be reached by a one-dimensional collective transposition of particles.
For the $N=24$ cluster the line move wraps completely around the sample and connects two ground states without defect lines, while for the $N=32$ the indicated shifts of particles
generate a ground state with a defect line, where the region along the lines resembles a rotated pristine stripy state. For the $N=24$ cluster this move is generated in $6$th order perturbation theory,
while the move on the $N=32$ cluster is effective in $4$th order.}
\label{fig:ChargeOrderMPoint}
\end{figure*}

\begin{figure}
\includegraphics[width=\linewidth,clip]{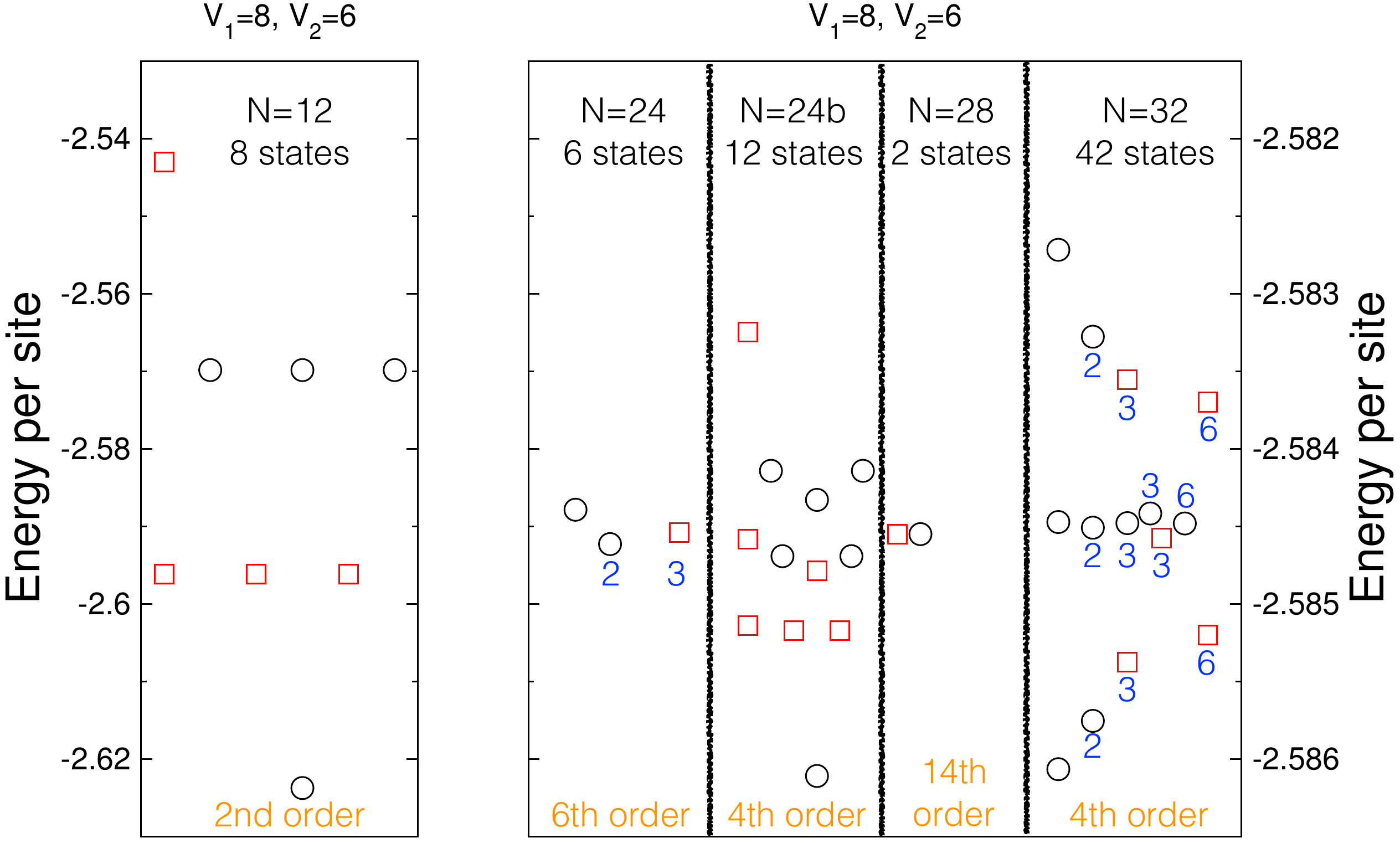}
\caption{(Color online) Splitting of the classical ground state degeneracy at $V_1/t=8$, 
$V_2/t=6$ for clusters with $N=12$ (left panel) and $N=24$, $24$b, $28$ and $32$ sites 
(right panel).  Black (red) symbols denote the two different particle-hole symmetry sectors.
All levels are singly degenerate unless the (spatial symmetry) degeneracy is indicated by a blue number.
The order at which processes of the type described in Fig.~\ref{fig:ChargeOrderMPoint}
happen on the clusters are indicated at the bottom of the frame for each system size.
In contrast to the case with attractive $V_1$ there is no obvious system-size independent 
order-by-disorder mechanism at work on the clusters considered.
}
\label{fig:spectrum_V1_8_V2_6}
\end{figure}

\subsubsection{Stripy* region}\label{sec:stripy}

Finally we consider the extended region $\arctan(1/4) <\theta < \pi/2$, featuring
stripy ground states and some degree of domain walls.  To first order
in $t/V_{1,2}$ it is not possible to hop and stay in the ground
state manifold. However on a finite cluster there is a finite order at
which the moves sketched in Fig.~\ref{fig:ChargeOrderMPoint} become
possible. As the order at which this process first occurs diverges with
the linear extent of the cluster, such tunneling events seem
irrelevant at small $t/V_{1,2}$ in the thermodynamic limit. As we currently do
not fully understand the structure of the ground space manifold, we can not rule 
out the appearance of other perturbative processes connecting ground states 
without wrapping around the torus. 

In order to shed some light on this issue, we diagonalize the full Hamiltonian at $V_1/t=8$,
$V_2/t=6$ for a number of samples. The resulting spectra are shown in Fig.~\ref{fig:spectrum_V1_8_V2_6}.
In contrast to the zigzag case studied before, here we do not observe a clearly structured
order-by-disorder selection of a new ground state. The results for these system sizes can be
rationalized as the spectrum of an abstract tight binding model, where the sites are the different 
ground states in the classical ground space, while the hopping matrix elements are of the cluster wrapping
type just discussed. As the perturbative orders of these processes strongly depend on the shape of the
cluster, the strong correlation of the bandwidth of the split ground space with the formal order
in perturbation theory seems to validate our picture. It is thus presently not clear whether the ground state
degeneracy will be lifted at a finite order in perturbation theory in the thermodynamic limit or not. The qualitative 
presence of a high-order diagonal term in the zigzag case discussed before could also carry over to the present stripy* case,
but is masked on moderate finite size samples by the leading cluster-wrapping splitting terms.

Now we turn to the points in the classical phase diagram which feature a massive ground state degeneracy. 
These are the points $\theta=\pi/2$ ($V_1=0$), $\theta=\arctan(1/4)$ $(V_1=4,\ V_2=1)$ and $\theta=\pi-\arctan(1/4)$ $(V_1=-4,\ V_2=1)$ . 

\subsubsection{Charge modulated (CM) state}
\label{sec:cm}

At $V_1=0$ we have two independent triangular sublattices with
antiferromagnetic interactions. It is known that each state which has
two or one charges on each $V_2$-triangle is a valid ground state.~\cite{Wannier}  When
working at half filling the ground space of the full
problem can be obtained by enumerating all ground states on one
triangular sublattice and then summing up the product of the two
sublattice Hilbert spaces over all particle bipartitions such that the
total number of particles amounts to half filling (i.e.~$N/2$). 

The effect of a
finite $t/V_2$ is to hop particles from one triangular sublattice to
the other one. There are indeed many states in the ground state
manifold where such first order hoppings are possible while staying in the ground
state manifold. The problem thus maps to a hopping problem on an
abstract lattice living in the constrained Hilbert space, where the sites are
the allowed configurations. This abstract lattice is not homogeneous and 
features sites with different coordination numbers. It is not known to us
how to solve this  hopping problem analytically. As a starting point 
it is instructive to identify the maximally flippable states 
(i.e.~abstract sites with the largest coordination number) of this problem. By 
explicit enumeration we have identified the 18  $uud$-$ddu$ configurations
as the maximally flippable states (the number is independent of $N$, given that
the sample geometry is compatible with $uud$-$ddu$ states). These states 
feature a threefold degenerate charge analog of the magnetic $uud$ states on one 
sublattice, while the other sublattice has three possible charge analogs of the 
magnetic $ddu$ state. Another factor of two is obtained by swapping the role of 
the two sublattices.

\begin{figure}
\includegraphics[width=\linewidth,clip]{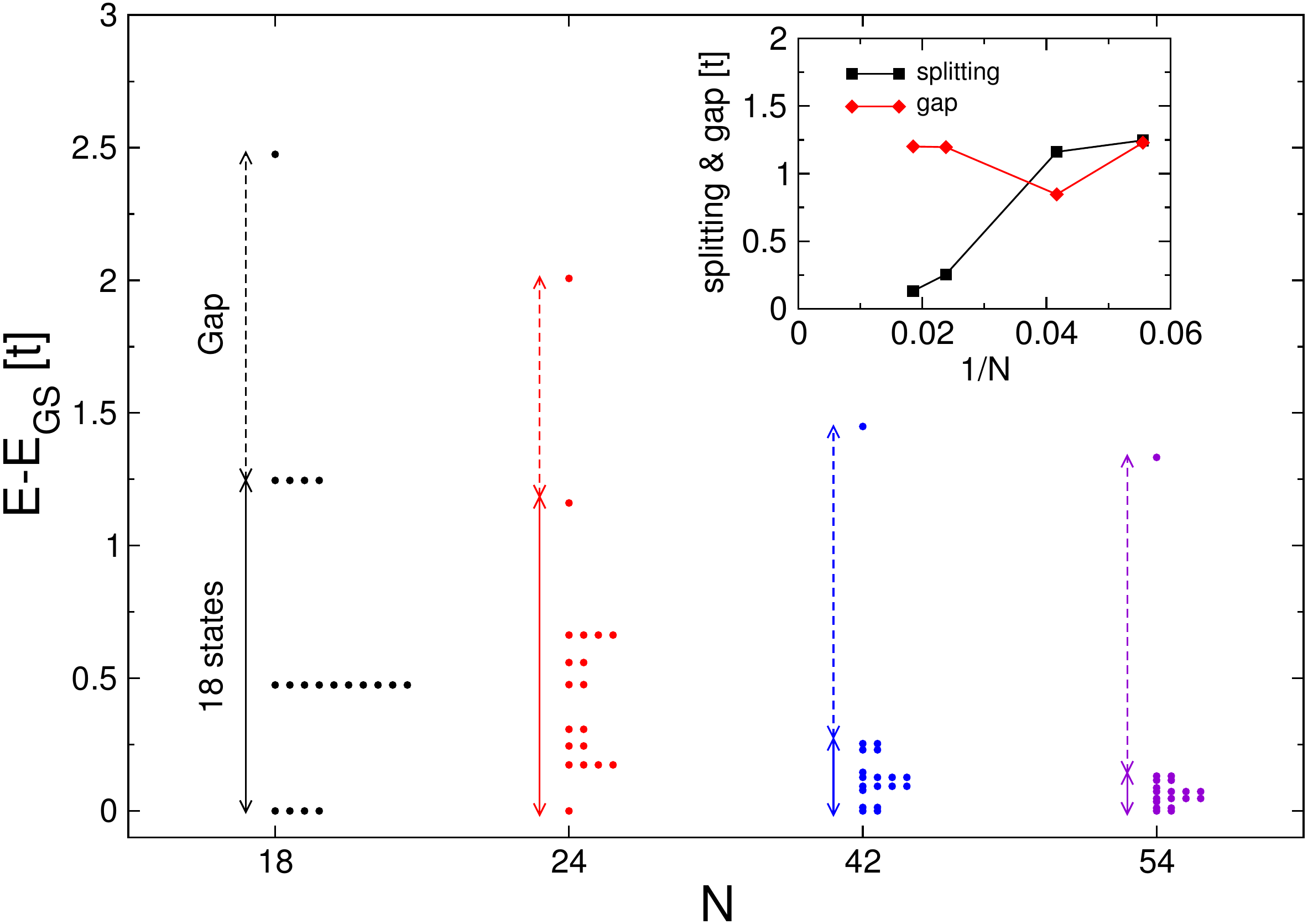}
\caption{(Color online) 
Low lying energy spectrum in units of the hopping $t$ for the CM phase at $V_1=0, V_2=1$, $t/V_2 \ll 1$.
Data for system sizes $N=18,\ 24,\ 42,\ 54$ are shown. The full arrow denotes the bandwidth of the
lowest 18 levels. Note the collapse of the bandwidth for $N=42$ and $N=54$, while the (particle-hole) gap 
above the 18 states (dashed arrows)  remains of order $t$, as illustrated in the inset.
}
\label{fig:ergs_eff_model_V_0_V2_1}
\end{figure}

\begin{figure}
\includegraphics[width=0.8\linewidth,clip]{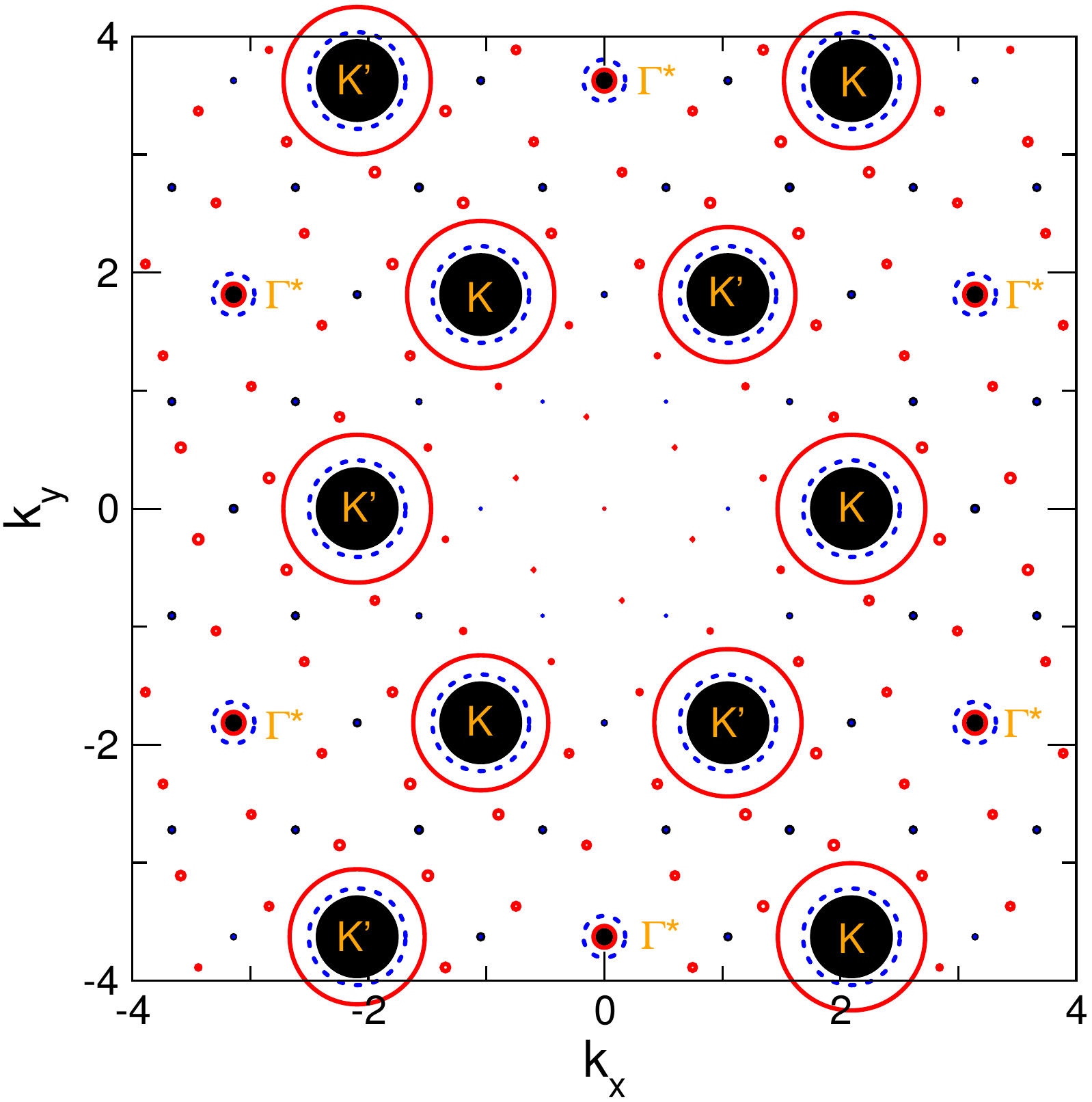}
\caption{(Color online) Charge structure factor of the effective 
model at $V_1=0, V_2=1$, $t/V_2 \ll 1$ plotted in the enlarged
Brillouin zone. The filled symbols are
data for the ground state at $N=24$ while the empty symbols are for $N=42$. The
dashed symbols denote the structure factor in an equal superposition of the
18 $uud$-$ddu$ states for $N=24$. Note the presence of small peaks at momentum
$\Gamma^*$, indicating a possibly weak sublattice imbalance.
}
\label{fig:sofq_effmodel_V1_0_V2_1}
\end{figure}

The question is now whether the ground state of the full effective model is localized
on those maximally flippable states, or whether the hopping network favors delocalized
wave functions. We have numerically solved the effective model for samples with $N=18,\ 24,\ 42$ and $54$
sites and show the low-lying energy spectra in Fig.~\ref{fig:ergs_eff_model_V_0_V2_1}. 
We note the presence of 18 low-lying states (indicated by the full arrow in Fig.~\ref{fig:ergs_eff_model_V_0_V2_1}),
separated by a gap (dashed arrow) from the higher lying states. While this picture is not yet very sharp for $N=18$ and $N=24$,
for $N=42$ and $N=54$ however the bandwidth of the 18 low-lying states is quite small compared to the gap to the higher states.
We interpret this spectrum as the localization of the low-lying eigenstates on the 18 maximally flippable states. This is
further corroborated by the fact that these eigenfunctions have their maximal amplitudes on the 18 $uud$-$ddu$ states. 
We have also obtained the static charge structure factor 
\begin{equation}
S(\mathbf{Q})=\frac{1}{N} \sum_{k,l} e^{-i\ \mathbf{Q}\cdot(\mathbf{R}_k-\mathbf{R}_l)} (n_k-1/2)(n_l-1/2)
\end{equation}
in the ground state of the effective model. In Fig.~\ref{fig:sofq_effmodel_V1_0_V2_1} we display this quantity for $N=24$ and $N=42$,
as well as for an equal superposition of the 18 $uud$-$ddu$ states on $N=24$ sites. Note that we plot the data in an extended Brillouin
zone scheme. There are obvious Bragg peaks at the momenta $\mathbf{K},\mathbf{K}'=-\mathbf{K}$ stemming from the tripling of the unit
cell in each triangular sublattice. Furthermore in the plain superposition of the 18 $uud$-$ddu$ states a sublattice imbalance is present, 
as the densities on the two triangular sublattices are different (1/3 versus 2/3), leading to another set of Bragg peaks at the $\Gamma^*$ momentum.
In the actual ground state of the effective model this peak is renormalized substantially. If our advocated picture of a localization of the ground state
wave functions on the 18 $uud$-$ddu$ states holds true, then we also expect the Bragg peaks at $\Gamma^*$ to survive, although with a sizable
reduction of the amplitude. In a recent paper Daghofer and Hohenadler~\cite{Daghofer2014} also investigated this issue and concluded an absence of the
sublattice imbalance based on exact diagonalizations of the original Hamiltonian (\ref{eqn:Hamiltonian}) on $N=18$ and $N=24$. Our results plotted in the extended
zone scheme however reveal at least the presence of a local maximum in $S(\mathbf{Q})$ at $\Gamma^*$, consistent with our claim of a weak sublattice imbalance in the 
thermodynamic limit. Note that this state would be insulating, i.e. unrelated to the presumed metallic pinball liquid phase on the triangular lattice.~\cite{Hotta2006}

 We close the discussion
of the $V_1=0$ case by noting that this point enlarges to a finite
region with respect to the addition of $V_1$ when $t>0$. The reason is
that the ground state of the hopping problem in the degenerate ground
state manifold is able to gain energy proportional to $t$ in the order-by-disorder
framework. When
switching on $V_1$, one needs a finite amount of $V_1$ to destabilize
the CM phase, because the stripy* region and the zigzag phase do not gain energy from the
hopping term at first order in $t$.

\subsubsection{N\'eel domain wall crystal (NDWC)}\label{sec:NDWC}

\begin{figure}
\includegraphics[width=\linewidth,clip]{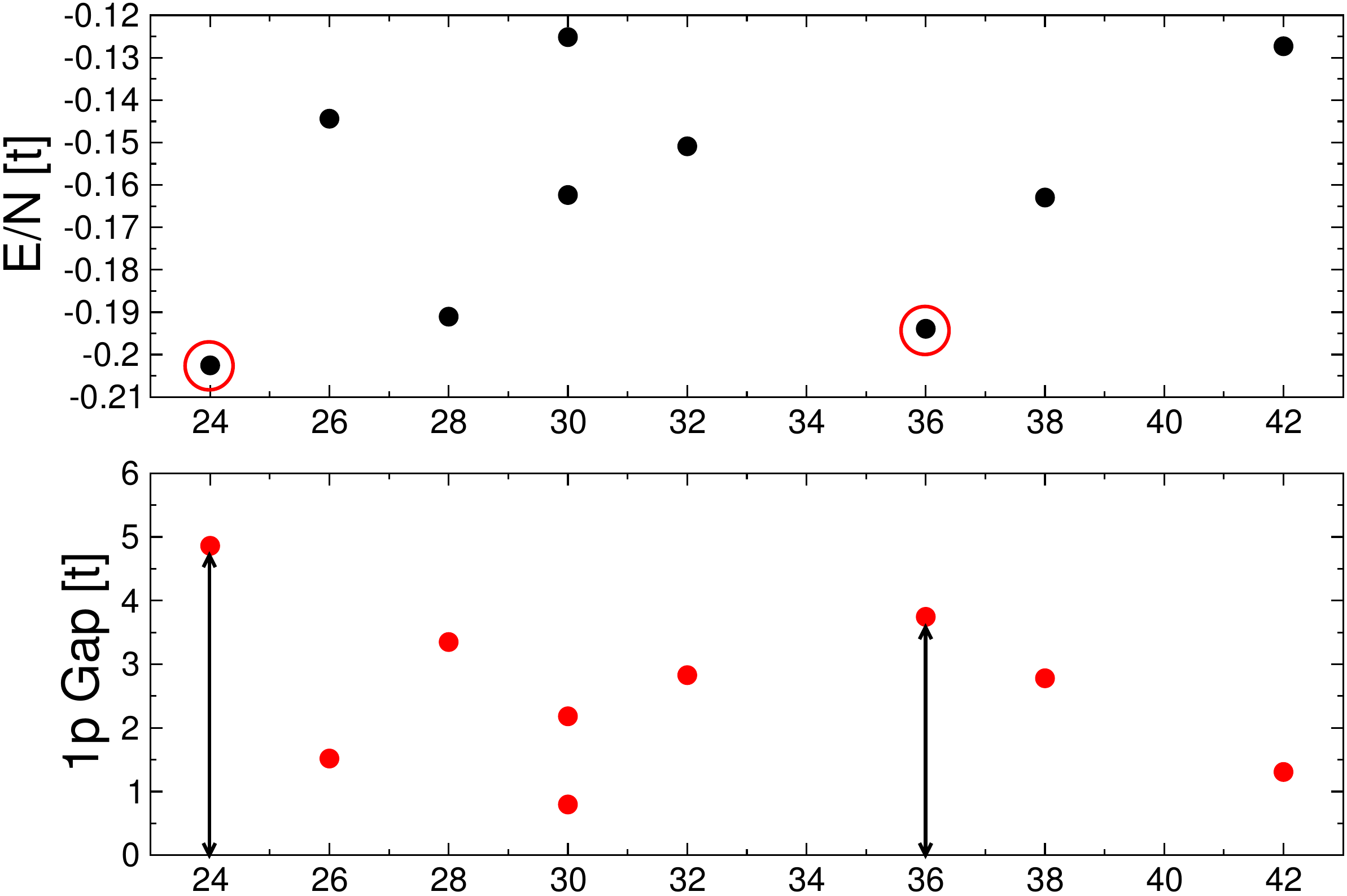}
\caption{(Color online) Energetics of the first order effective model (1st order in $t/V_{1,2}$) at 
$V_1/V_2=4$. (Top) Energy per site in units of $t$ for various cluster sizes. The clusters
with $24$ and $36$ sites have a particularly low energy per site. (Bottom) Single-particle charge
gap $\Delta_\mathrm{1p}/t$. This gap amounts to $4\sim5$ $t$ for the system sizes with a low energy per site.
}
\label{fig:ergs_eff_model_V1V2_4}
\end{figure}

\begin{figure}
\includegraphics[width=\linewidth,clip]{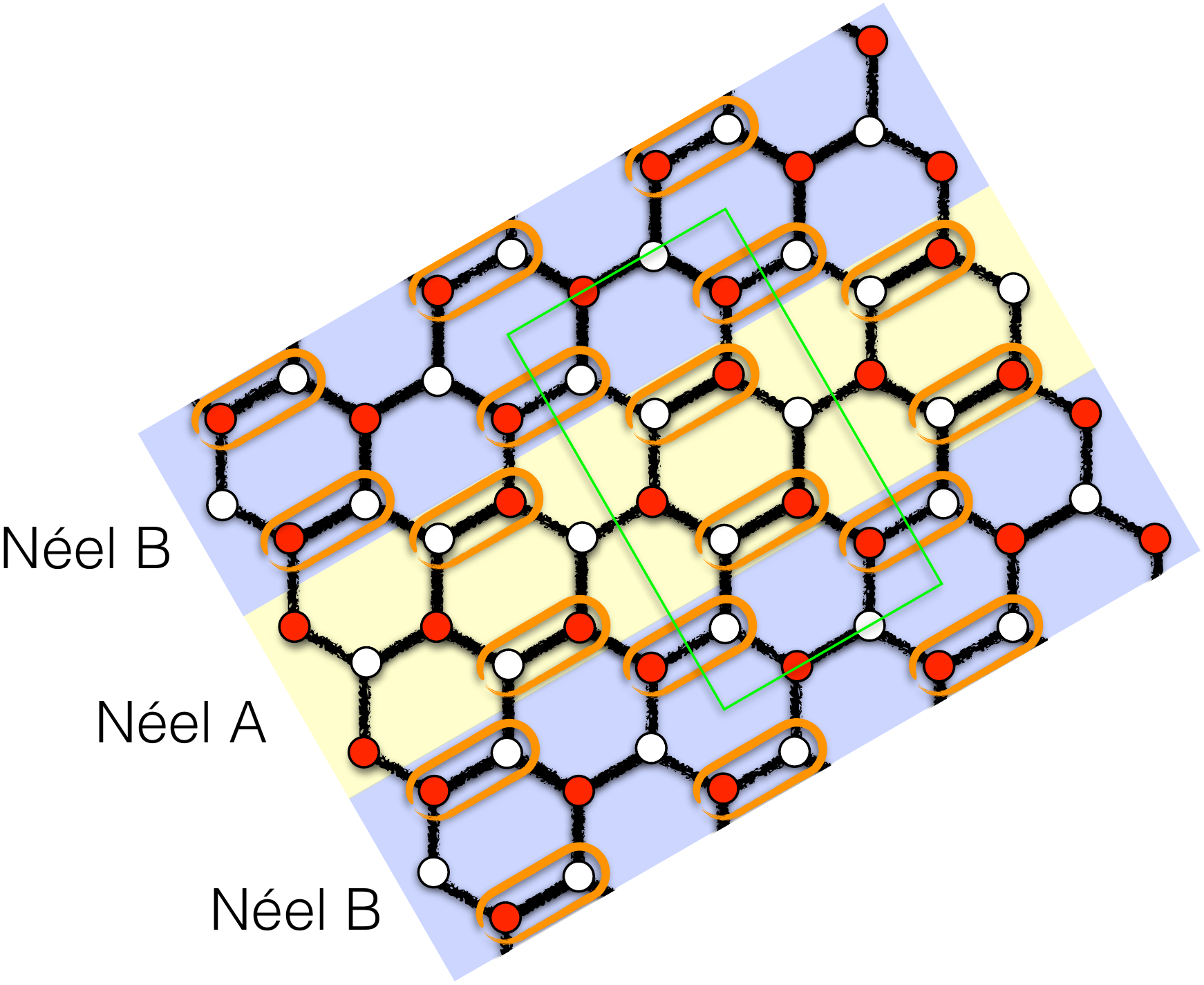}

\caption{(Color online) The N\'eel Domain Wall Crystal (NDWC): sketch of a classical ground state at $V_1=4,\ V_2=1$ on the $N=24$ sample, 
which is maximally flippable within the classical ground state manifold with respect to the hopping~$t$.
The shaded regions denote the two N\'eel domains, and the orange circled bonds along the domain walls
are flippable to first order in~$t$. The green box indicates a twelve-site unit cell.
}
\label{fig:gs_eff_model_V1V2_4}
\end{figure}

\begin{figure}
\includegraphics[width=0.9\linewidth,clip]{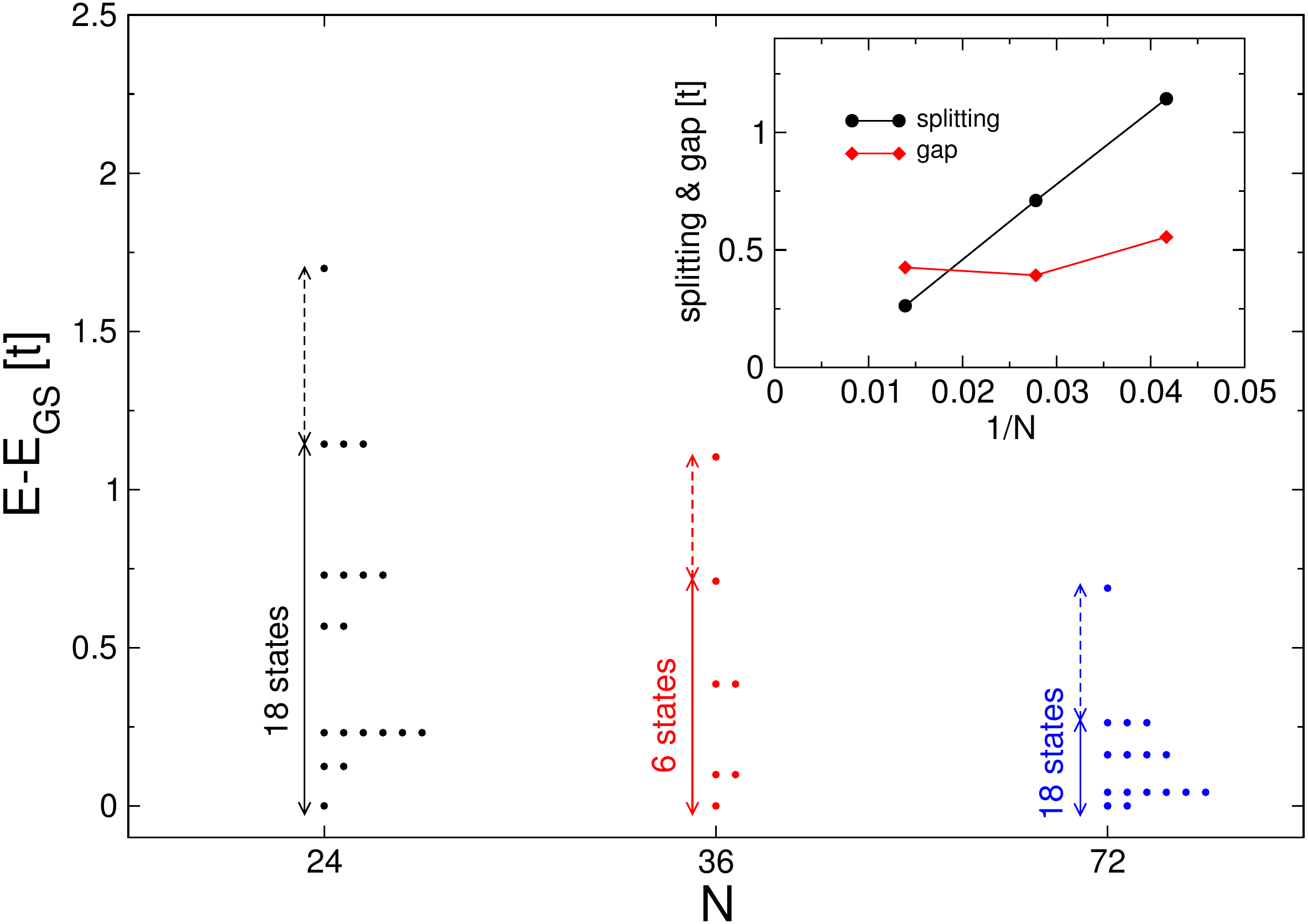}
\caption{(Color online) Low lying energy spectrum in units of the hopping $t$ for $V_1=4, V_2=1$, $t/V_2 \ll 1$.
Data for system sizes $N=24,\ 36,\ 72$ are shown. The full arrow denotes the bandwidth of the
lowest 18 levels ($N=36$ does not have rotational invariance, in that case we only expect 6 possible ground states). 
Note the collapse of the bandwidth for larger clusters, while the (particle-hole) gap 
above the 18 states (dashed arrows)  remains approximately constant, as illustrated in the inset.}
\label{fig:spec_eff_model_V1V2_4}
\end{figure}

\begin{figure}
\includegraphics[width=0.8\linewidth,clip]{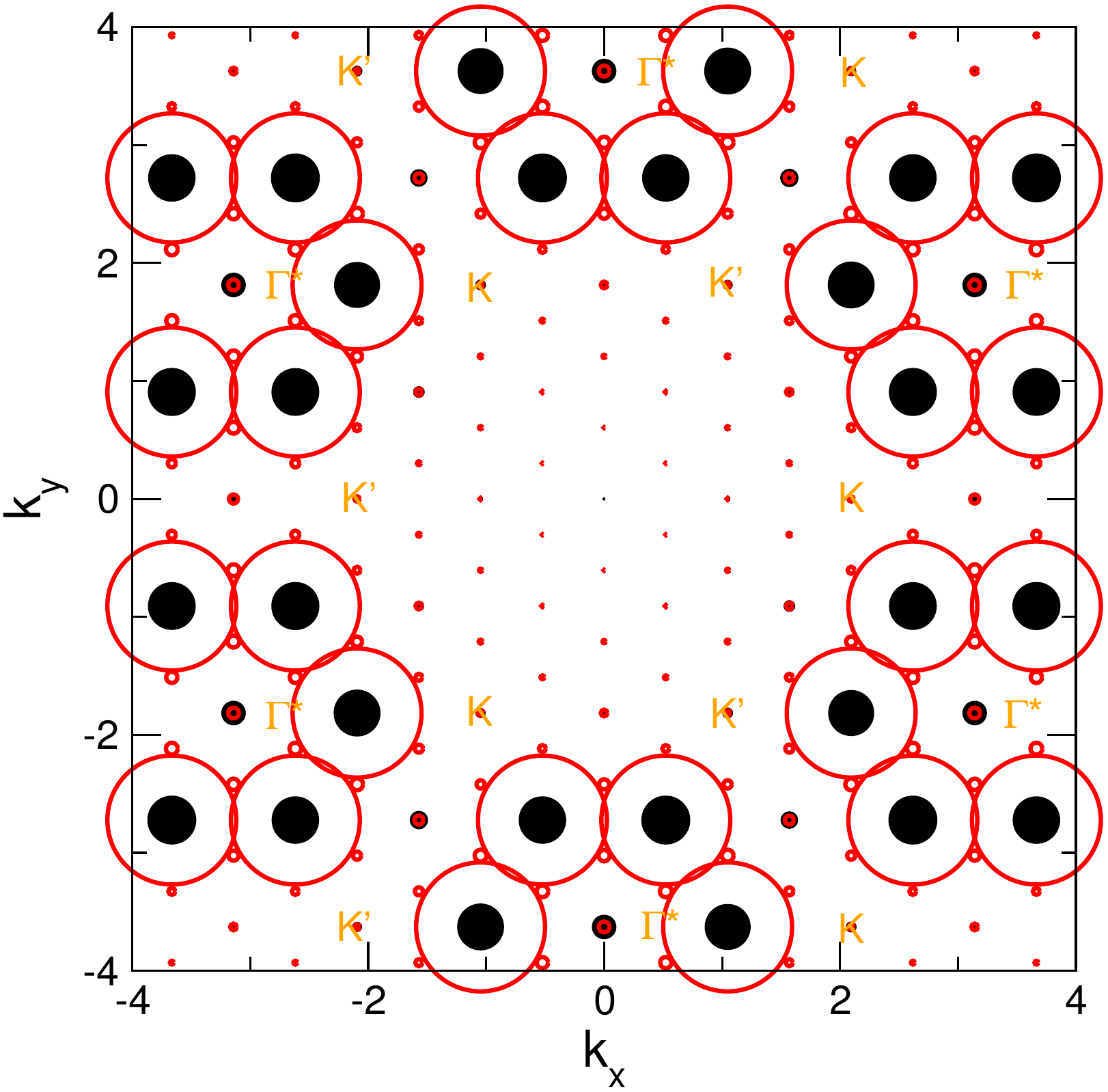}
\caption{(Color online) Charge structure factor of the effective 
model at $V_1=4, V_2=1$, $t/V_2 \ll 1$ plotted in the enlarged
Brillouin zone. The filled (empty) symbols are data for the ground state at $N=24$ ($N=72$).
The Bragg peaks appear at the $X^*$ point, c.f. Fig.~\ref{fig:latticeBZ}(b).
}
\label{fig:sofq_effmodel_V1_4_V2_1}
\end{figure}

Now let us discuss the behavior at $\theta=\arctan(1/4)$. As noted
before, this point features a sizable ground state degeneracy at the
classical level. Since - to the best of our knowledge - this manifold has
not been studied so far, we proceed by numerically diagonalizing the
first order effective Hamiltonian obtained by projecting the fermionic hopping
Hamiltonian into the classical ground state manifold.  In
Fig.~\ref{fig:ergs_eff_model_V1V2_4} we present the energy per site
for various clusters. It appears that among the studied clusters the
samples with $N=24$ and $36$ sites have a particularly low energy
per site ($N=28$ is close, but its energy per site is higher than $N=36$, so
we discard it). It seems that this low energy correlates with a high maximal
flippability of configurations on these clusters~\footnote{$N=24$: 18
  states with 8 off-diagonal matrix elements each; $N=28$: 14 states with
  10 off-diagonal matrix elements each; $N=36$: 6 states with 12
  off-diagonal matrix elements each; $N=72$: 18 states with 24
  off-diagonal matrix elements each.}, compared to the other clusters. 
 In Fig.~\ref{fig:gs_eff_model_V1V2_4} we display one such maximally
flippable state for $N=24$. Our understanding of this state is that it is formed of 
alternating strips of the two N\'eel CDW states. Along the domain walls between the
two CDW states there are many bonds which are flippable to first order in $t/V_{1,2}$. 
This state is 18 fold degenerate. There are 3 directions in which the domain walls can run, 
and for fixed orientation, there are 3 distinct bond choices  for the domain walls. The remaining
factor of two is due to the assignment of the two N\'eel CDW states. We have further checked a sample
with $N=72$ and again found 18 maximally flippable states, in agreement with our analysis. In Fig.~\ref{fig:spec_eff_model_V1V2_4}
we display the low lying eigenstates of the effective first-order Hamiltonian in the degenerate subspace
for system sizes $N=24,36$ and $72$ sites. The collapse of the lowest 18 states is clearly visible, providing strong 
evidence for symmetry breaking of the NDWC type. The results
of the effective model simulations for the charge structure factor are shown in Fig.~\ref{fig:sofq_effmodel_V1_4_V2_1}
and further corroborate this picture.

We also expect the point $V_1/V_2=4$ to enlarge to an actual phase at small but finite $t/V_{1,2}$, because the two neighboring phases in the Ising limit (the stripy* region 
and the N\'eel CDW) both do not gain energy to first order in $t/V_{1,2}$. 

\subsubsection{$V_1=-4$, $V_2=1$}

The ground state configurations at $V_1=-4$, $V_2=1$ do not seem to be flippable to first order in $t/V_1,V_2$, and we have therefore refrained from a detailed study of this line.

To conclude this section we show the resulting semiclassical qualitative phase diagram in the right part 
of Fig.~\ref{fig:IsingPhaseDiagram}.

\section{Numerical results}\label{sec:numerics}

In this section, we will present our systematic numerical study using ED and typical observables that were used to compute our global phase diagram in Fig.~\ref{fig:PhaseDiagram}. We will first present some useful tools to detect phase transitions as well as to characterize phases. Then we will provide numerical evidence for the finite extension of the phases that were discussed in the limit of large interactions in Sec.~\ref{sec:Ising}. Last but not least, we will discuss the stability of phases (with a stronger quantum character) at intermediate coupling, namely the semi-metal (SM), the plaquette/Kekul\'e phase and the absence of the conjectured topological Mott insulator phase. 

\subsection{Numerical tools}

\subsubsection{Energetics}
\begin{figure}[!htb]
\includegraphics[width=\columnwidth,clip]{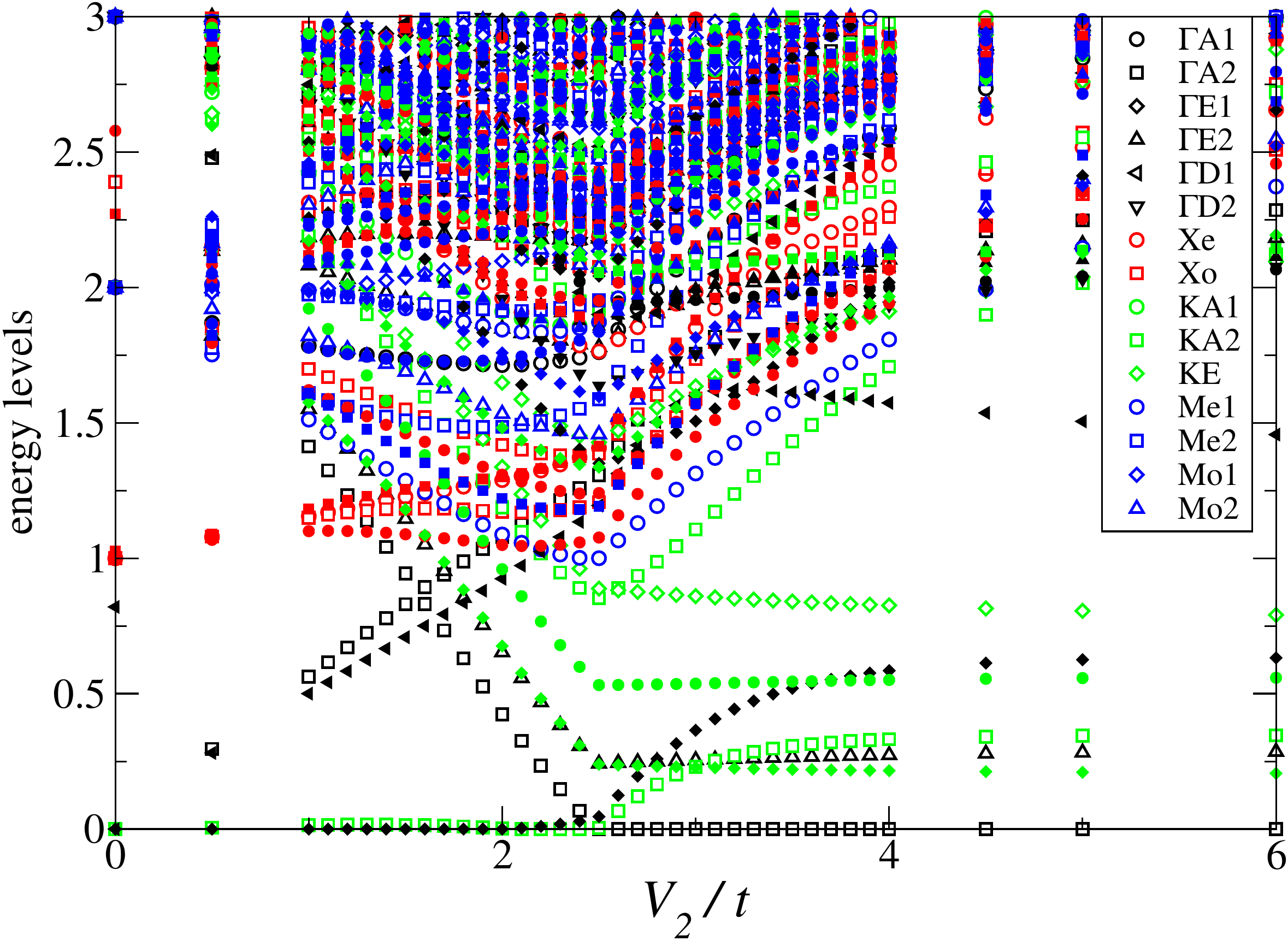}
\caption{(Color online) Full spectrum vs $V_2/t$ at fixed $V_1/t=0$ labeled with symmetry sectors on $N=24$ cluster. 
More precisely, states with momentum $\Gamma$ can be labeled with $C_{6v}$ standard point-group notations; with momentum $\mathbf{X}$ only reflection  can be used to detect even (e) vs odd (o) states; at the $\mathbf{K}$ point, we use $C_{3v}$ point-group notations; at the $\mathbf{M}$ point, we can use two reflections to label states which are even/even (e1), even/odd (e2), odd/even (o1) or odd/odd (o2). 
Open (filled) symbols denote respectively states which are even (odd) with respect to particle-hole symmetry.}
\label{fig:fullspectrum_24_V1_0}
\end{figure}

\begin{figure}[!htb]
\includegraphics[width=\columnwidth,clip]{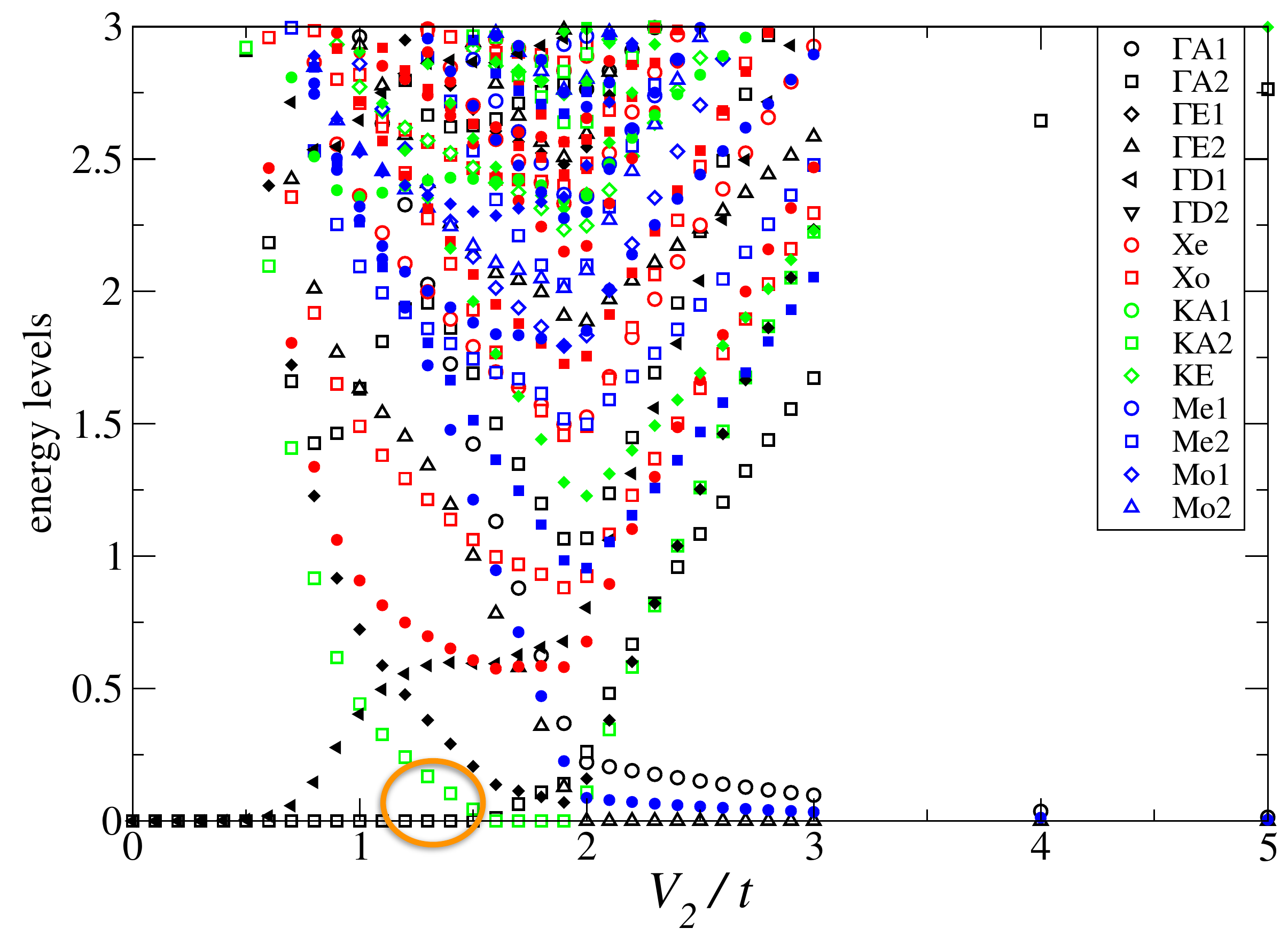}
\caption{(Color online) Same as Fig.~\ref{fig:fullspectrum_24_V1_0} at fixed $V_1/t=4$ on $N=24$ cluster. For intermediate values $V_2/t \simeq 1.5$, low-energy 
states (indicated with an ellipse) are compatible with a plaquette or Kekul\'e phase.}
\label{fig:fullspectrum_24_V1_4}
\end{figure}

Let us start by showing some full low-energy spectra, where we label each eigenstate according to its quantum numbers.~\footnote{We use the full lattice symmetry, as well as particle-hole symmetry. For space symmetries, we label sectors using momentum and irreducible representation corresponding to the point-group compatible with this momentum. For instance, on $N=24$ cluster which possesses all symmetries, we have used $C_{6v}$ group at the $\Gamma$ point, $\mathbb{Z}_2$ group at the $\mathbf{X}$ point, $C_{3v}$ group at the $\mathbf{K}$ point and $\mathbb{Z}_2 \times \mathbb{Z}_2$ group at the      $\mathbf{M}$ point.} In Fig.~\ref{fig:fullspectrum_24_V1_0}-\ref{fig:fullspectrum_24_V1_4}, we present these spectra for a fixed $V_1=0$ and $V_1/t=4$ respectively on the $N=24$ cluster which has the full point-group symmetry $C_{6v}$ and also possesses the relevant $\mathbf{K}$ and $\mathbf{M}$ points in its Brillouin zone (see Appendix~\ref{appendix:lattice}). 

When $V_1/t=0$ and $V_2/t$ is small, we expect a finite region of SM phase with low-energy excitations at the Dirac point ($\pm \mathbf{K}$). Increasing $V_2/t$, there is a sudden level crossing when $V_2/t \simeq 2.2$. Beyond this value, if we count the lowest energy levels  well separated from the others, we get 18 states, which may indicate some ordering of some kind in the thermodynamic limit (see discussion in Sec.~\ref{sec:perturbed_classical}).

When $V_1/t=4$, starting at small $V_2/t$, there are two almost-degenerate states at the $\Gamma$ point with opposite particle-hole symmetries, compatible with the N\'eel CDW state. Increasing $V_2/t$ leads to some intermediate phase $1 < V_2/t <2$ with low-energy states at momentum $\mathbf{K}$ compatible with a Kekul\'e-like (or plaquette) pattern, and finally for large $V_2/t>2$ the emergence of 6 low-energy states with momenta $\Gamma$ and $\mathbf{M}$. 

\subsection{Stability of the classical Ising-like phases}\label{subsec:Ising}

Let us now move to a more precise characterization of the possible patterns occurring for large interactions. Quite generically,  large density interactions are expected to lead to some kind of charge ordering. In the equivalent spin language, these terms are of the Ising type and the resulting Ising model would correspondingly exhibit some spin ordering. For model (\ref{eq:model}), they were discussed in Sec.~\ref{sec:Ising} and are denoted under the names: CDW, NDWC, stripy* region, CM or zigzag.

\subsubsection{CDW phase}

When $V_2=0$, since the honeycomb lattice is bipartite, a simple charge-density wave (CDW) where particles are localized on one sublattice is expected for $V_1>0$, so that degeneracy is two. Since this phase does not break translation symmetry but only inversion, symmetry analysis predicts that on a finite cluster there should be a collapse of the lowest $\Gamma A_2$ and $\Gamma D_1$ states.~
\footnote{Note that the choice of the real space Fermi normal ordering can have an impact on the absolute symmetry sectors of the finite size clusters, relative quantum numbers should however be invariant} This is indeed what is observed in Fig.~\ref{fig:fullspectrum_24_V1_4} at small $V_2/t$ for instance. When $V_1 \geq 2$, our numerical data are compatible with a  vanishing gap at zero momentum (data not shown) because of the inversion symmetry breaking in the CDW phase. Note however that the system remains an insulator. 

\begin{figure}[t]
\includegraphics[width=0.4\linewidth,clip]{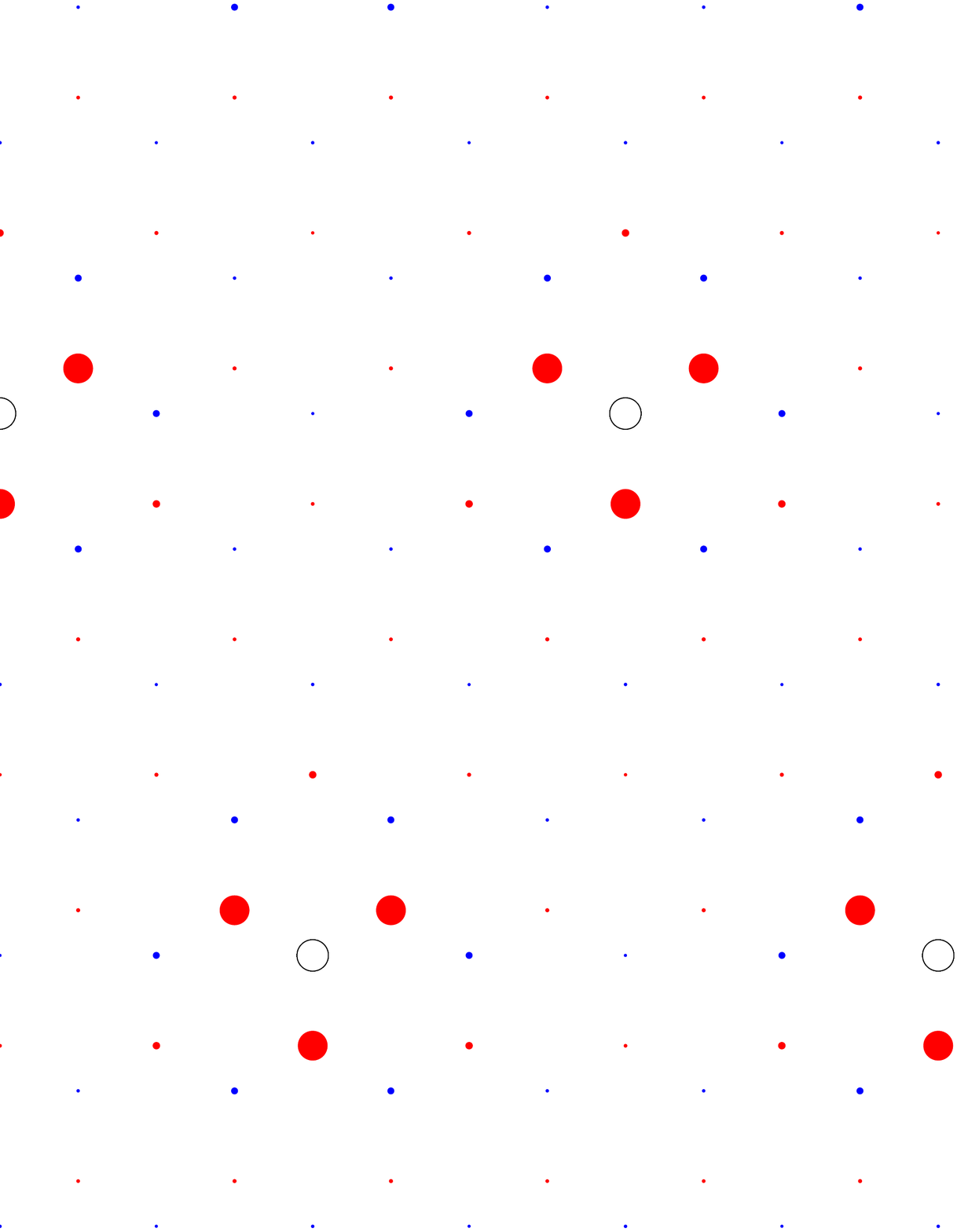}
\hspace{5mm}
\includegraphics[width=0.4\linewidth,clip]{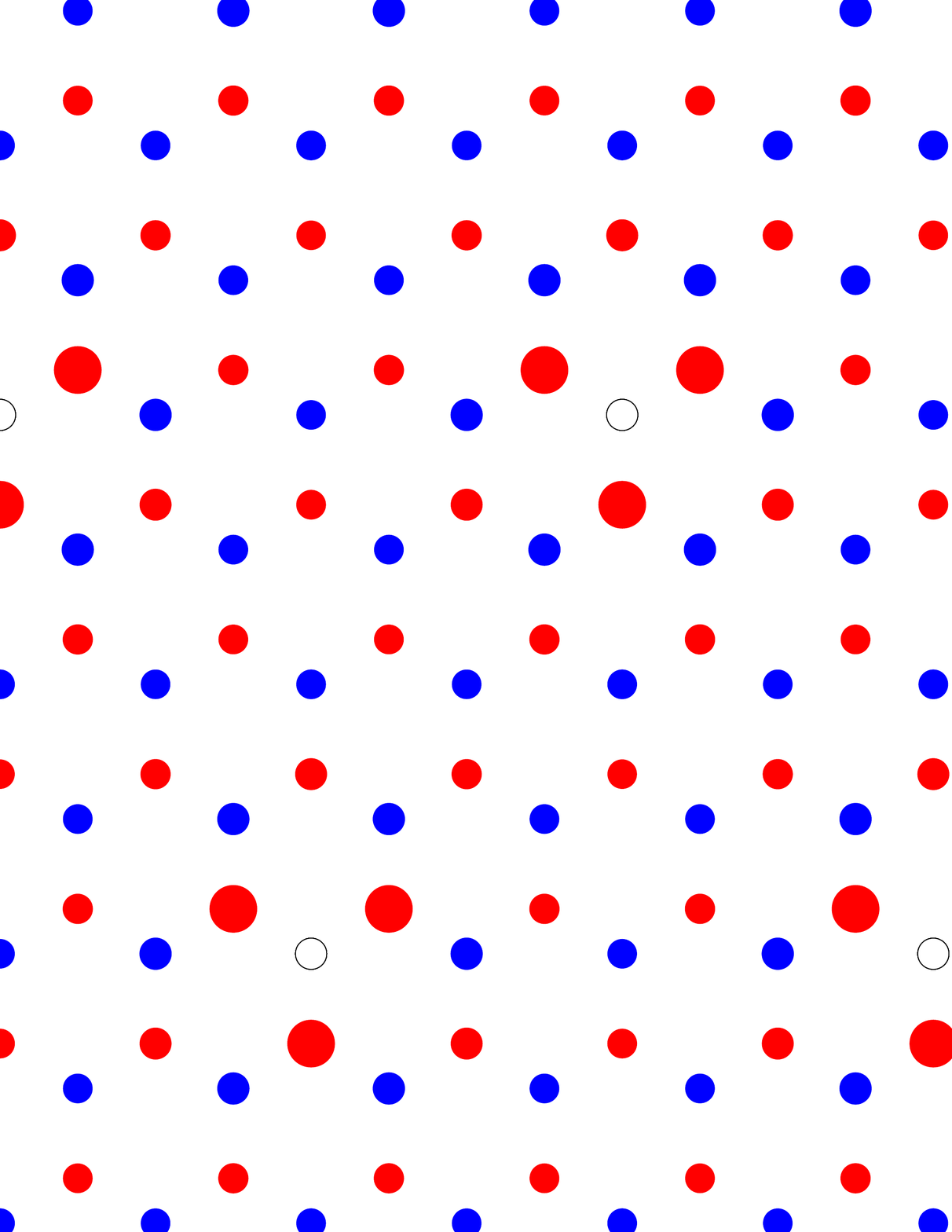}
\caption{(Color online) Connected density correlation functions between a reference site (open circle) and other sites
for $V_1=1$ and $V_1=2$ (at fixed $V_2=0$) on $N=32$ cluster. Periodic boundary conditions are
used. Blue and red correspond to positive/negative correlations.}
\label{fig:cdw}
\end{figure}

A clearer picture of this CDW state is provided by density-density correlations. On Fig.~\ref{fig:cdw}, we plot these connected correlations obtained with a $N=32$ cluster showing a clear transition from a semi-metallic phase to the N\'eel CDW for large $V_1/t\geq 2$ (at $V_2=0$). 
About the critical value, let us mention that a recent unbiased QMC approach~\cite{Wang2014} has concluded that $(V_1/t)_c=1.36$. 

Based on our previous semiclassical analysis, we expect that the CDW phase \emph{cannot} extend beyond the line $V_1=4V_2$, at least towards strong coupling.

\subsubsection{Charge modulated phase}

As we have discussed in Sec.~\ref{sec:Ising}, the large $V_2/t>0$ limit is less trivial since the classical Ising model possesses an extensive ground-state degeneracy. However, one expects on general grounds that quantum fluctuations will provide some selection mechanism within these low-energy states. Our discussion in Sec.~\ref{sec:cm} has led us to conclude in favor of
some charge-modulation (CM) state which is 18-fold degenerate and exhibits charge imbalance between the two sublattices, in agreement with Refs.~\onlinecite{Grushin2013,Garcia2013} and slightly different from the similar CM state \emph{without} charge imbalance~\cite{Daghofer2014}. 

Our ground-state energy plot in Fig.~\ref{fig:e0}(a) has indeed shown that the emergent phase is compatible with clusters having the $\mathbf{K}$ point, as is the case for the proposed tripled unit cell. Moreover, the low-energy spectra on these clusters, such as $N=24$  in Fig.~\ref{fig:fullspectrum_24_V1_0} (or similarly $N=36$, data not shown), do confirm the presence of 18 low-energy states at large $V_2$ in agreement with the expected degeneracy. The question whether all those states will collapse or if there will remain a finite gap (and thus a smaller degeneracy) is left for future studies, but our 
effective model diagonalizations in the strong coupling limit indicate a complete collapse of the 18 levels, c.f.~Fig.~\ref{fig:ergs_eff_model_V_0_V2_1}.

\subsubsection{Stripy* region}

For large $V_2 \sim V_1>0$, the lowest energies are found on clusters having the $\mathbf{M}$ point (see Fig.~\ref{fig:e0}(b)), which indicates a different instability. Moreover,  our 
symmetry resolved spectra (see Fig.~\ref{fig:fullspectrum_24_V1_4} for instance) predict a 6-fold degeneracy for large $V_2/t$ (with 3 states at momentum $\Gamma$ and 3 at equivalent momenta M) in disagreement with both the CM phase or the Kekul\'e one (see later). As discussed in Sec.~\ref{sec:stripy} based also on ED, the fate of the stripy* region at finite $t/V_1,V_2$ is not clear.
Results for $N=32$ close to the phase transitions to the neighboring phases still do not shed light on a possible selection process among the classically degenerate ground states.

\subsubsection{N\'eel domain wall crystal} 

\begin{figure}[!htb]
\includegraphics[width=\columnwidth,clip]{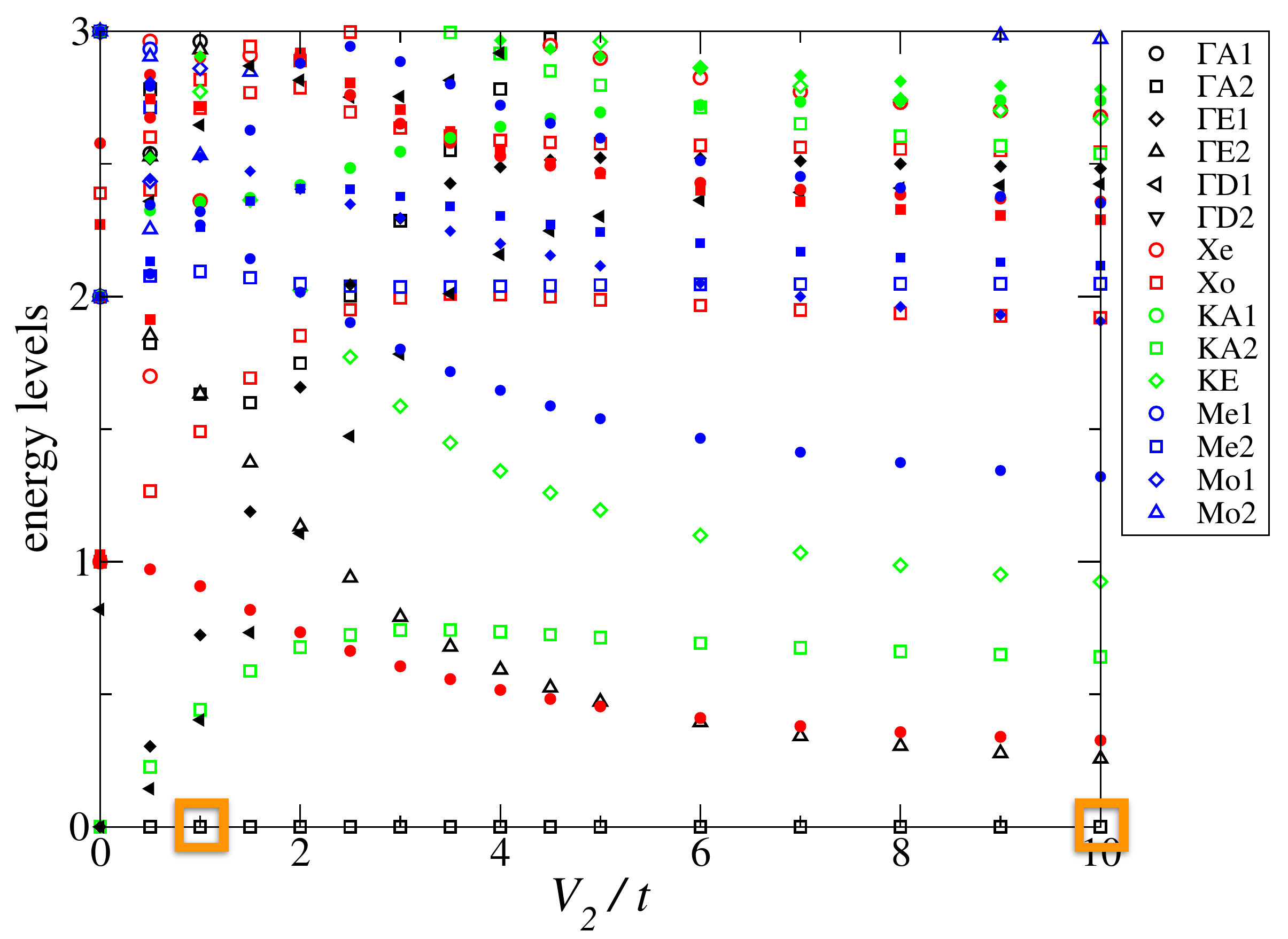}
\caption{(Color online) Same as Fig.~\ref{fig:fullspectrum_24_V1_0} as a function of $V_2/t$ at fixed $V_1/V_2=4$ on $N=24$ cluster. The orange boxes denote the two parameter
values for which the correlations shown in Fig.~\ref{fig:sketch_V1V2_4} were calculated.}
\label{fig:ergs_24_V1V2_4}
\end{figure}

\begin{figure}[!htb]
\includegraphics[width=\linewidth,clip]{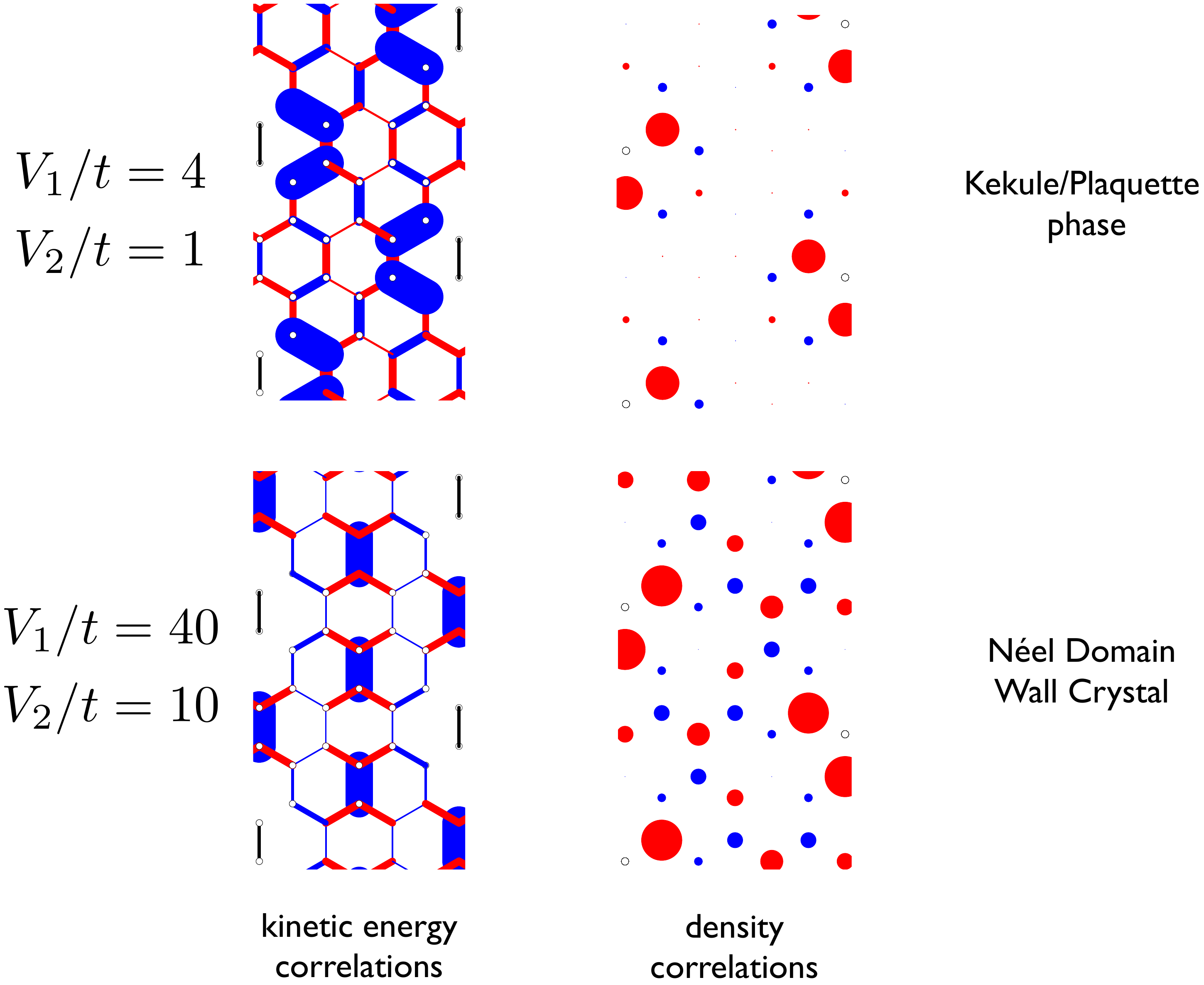}
\caption{(Color online) Kinetic energy and density correlations computed on $N=24$ cluster along the $V_1/V_2=4$ line, i.e. for $(V_1/t,V_2/t)$ respectively equal to $(4,1)$ and $(40,10)$. 
}
\label{fig:sketch_V1V2_4}
\end{figure}

Since the ratio $V_1/V_2=4$ is a special point in the classical phase
diagram we investigate here the phase diagram along the line $V_2/t$,
with the ratio $V_1/V_2=4$ fixed. In Fig.~\ref{fig:ergs_24_V1V2_4} we
display the low energy spectrum for the $N=24$ site sample along this
line. At very small $V_2/t$ we expect the stable semimetallic phase
emerging out of $V_2/t=0$.  We then detect an intermediate  
Kekul\'e or plaquette phase for values $1\lesssim V_2/t \lesssim 2$. These values 
are confirmed by directly computing kinetic correlations and checking if they 
match the expected pattern for a plaquette phase. 
However for larger values $V_2/t\gtrsim 3$ the system enters the realm of the
NDWC phase, involving the physics discussed in
subsection~\ref{sec:NDWC}.  The occurrence of two
very distinct phases beyond the semimetallic phase is also visible in the
ground state correlations shown in Fig.~\ref{fig:sketch_V1V2_4}.

\subsubsection{Zizag phase} 

\begin{figure}[!htb]
\includegraphics[width=0.4\linewidth,clip]{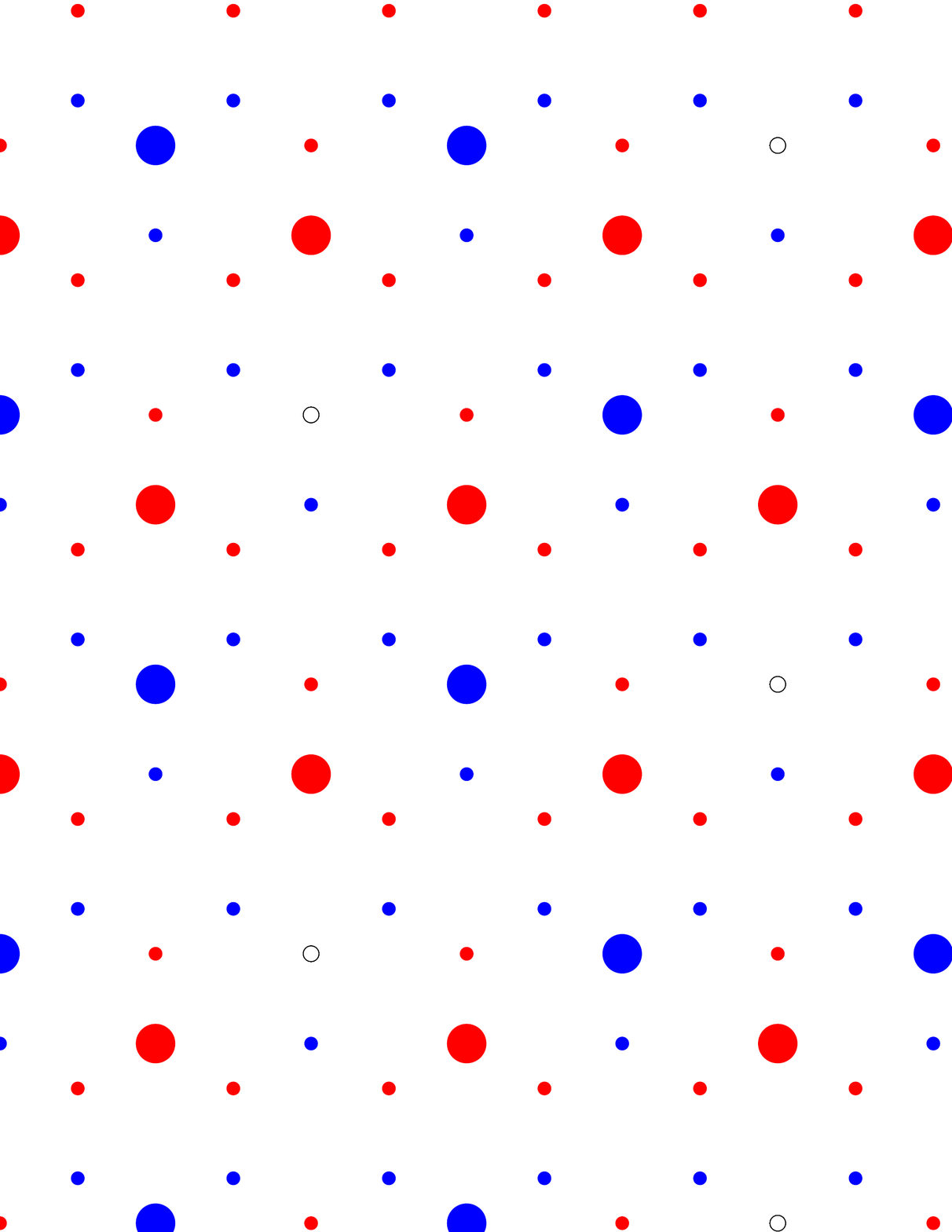}
\hspace{5mm}
\includegraphics[width=0.4\linewidth,clip]{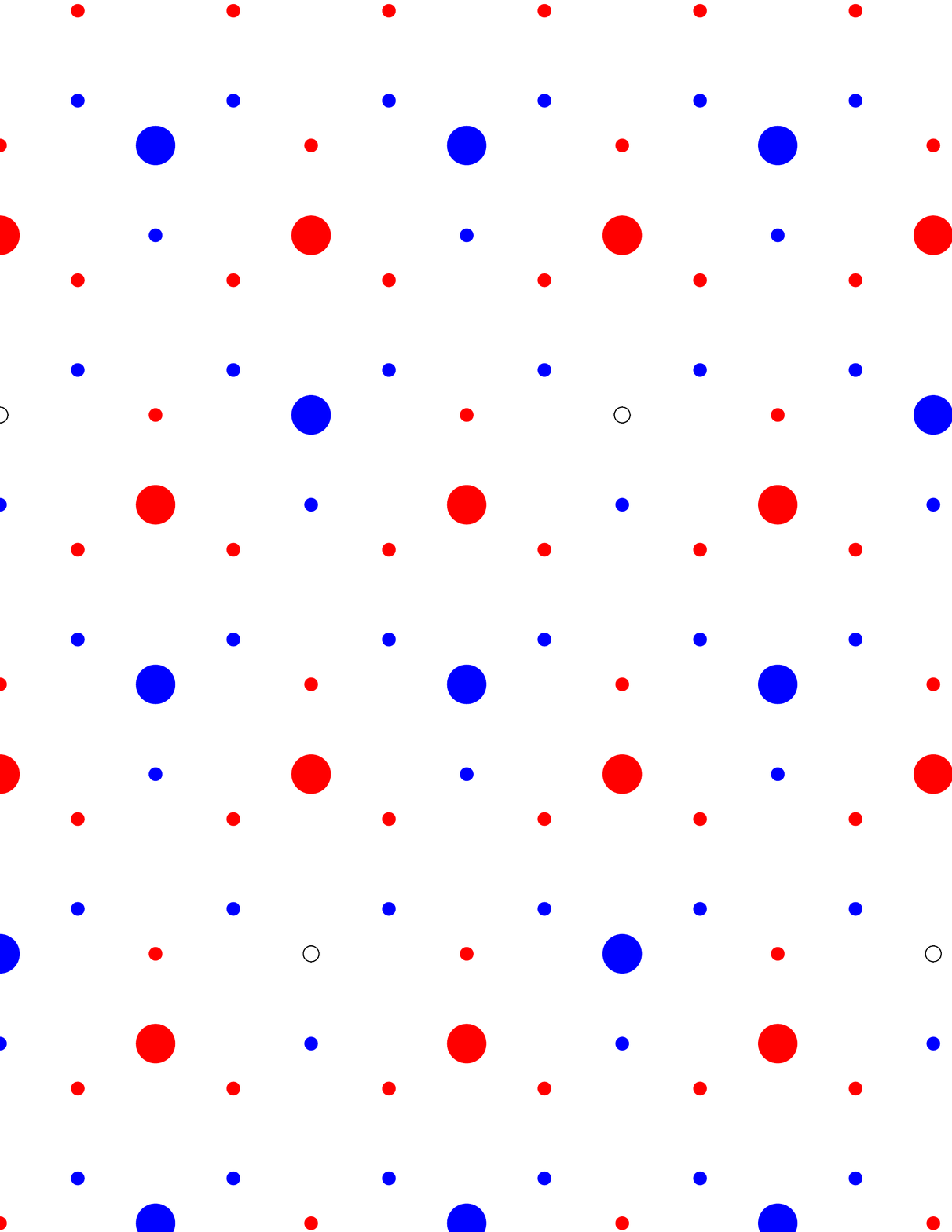}
\caption{(Color online) Connected density correlations computed on $N=24$ (left panel) and  $N=32$ (right panel) cluster for $V_1/t=-8$ and $V_2/t=6$.  Blue and red correspond to positive/negative correlations.
}
\label{fig:corr_zigzag}
\end{figure}

As already discussed in Sec.~\ref{sec:perturbed_classical}, ED with large negative $V_1/t$ and positive $V_2/t$ provide spectral evidence for a quantum order-by-disorder selection of a pristine zigzag state, as shown for instance in Fig.~\ref{fig:spectrum_V1_m8_V2_6}. For the clusters having the 3 $\mathbf{M}$ points in their Brillouin zone (see Appendix~\ref{appendix:lattice}), we do observe 6-fold quasi degenerate ground-states. By computing connected density correlations on any of these states, we obtain a pattern compatible with zigzag state. Indeed, a simple direct computation of connected density correlations on the zigzag state leads to values $\pm 1/12$ or $\pm 1/4$ depending on distances between sites, which agrees to a very high accuracy with the exact results computed on $N=24$ or $N=32$ ground-states in this region, see Fig.~\ref{fig:corr_zigzag}.

Having clarified the phases at large interaction strengths, which are connected to the classical limit, we now consider the intermediate interaction regime where new quantum phases are expected to emerge. 

\subsection{Emergence of a resonating quantum plaquette phase or Kekul\'e one}

In Refs.~\onlinecite{Grushin2013,Garcia2013}, it has been argued using mean-field or $N=18$ ED respectively that a Kekul\'e pattern can be stabilized when $V_1$ and $V_2$ are close. This would correspond to a 3-fold degeneracy~\cite{Albuquerque2011} ~\footnote{We disagree with the advocated general 4-fold degeneracy mentioned in Ref.~\onlinecite{Garcia2013}. The four-fold degeneracy is
actually an artifact of the $N=18$ sample~\cite{Corboz2013}} of the ground-state (more precisely one state at $\Gamma$ and two states at $\pm\mathbf{K}$).  

However, while the Kekul\'e phase was seen to remain stable up to  large $V_1 \sim V_2$ in Refs.~\onlinecite{Grushin2013,Garcia2013}, our numerical data on large clusters do not confirm these findings (see Sec.~\ref{subsec:Ising}) since it is replaced either by the stripy* region or the NDWC phase. However, we do confirm its existence, but for intermediate values only. 

\begin{figure}[!htb]
\includegraphics[width=0.95\columnwidth,clip]{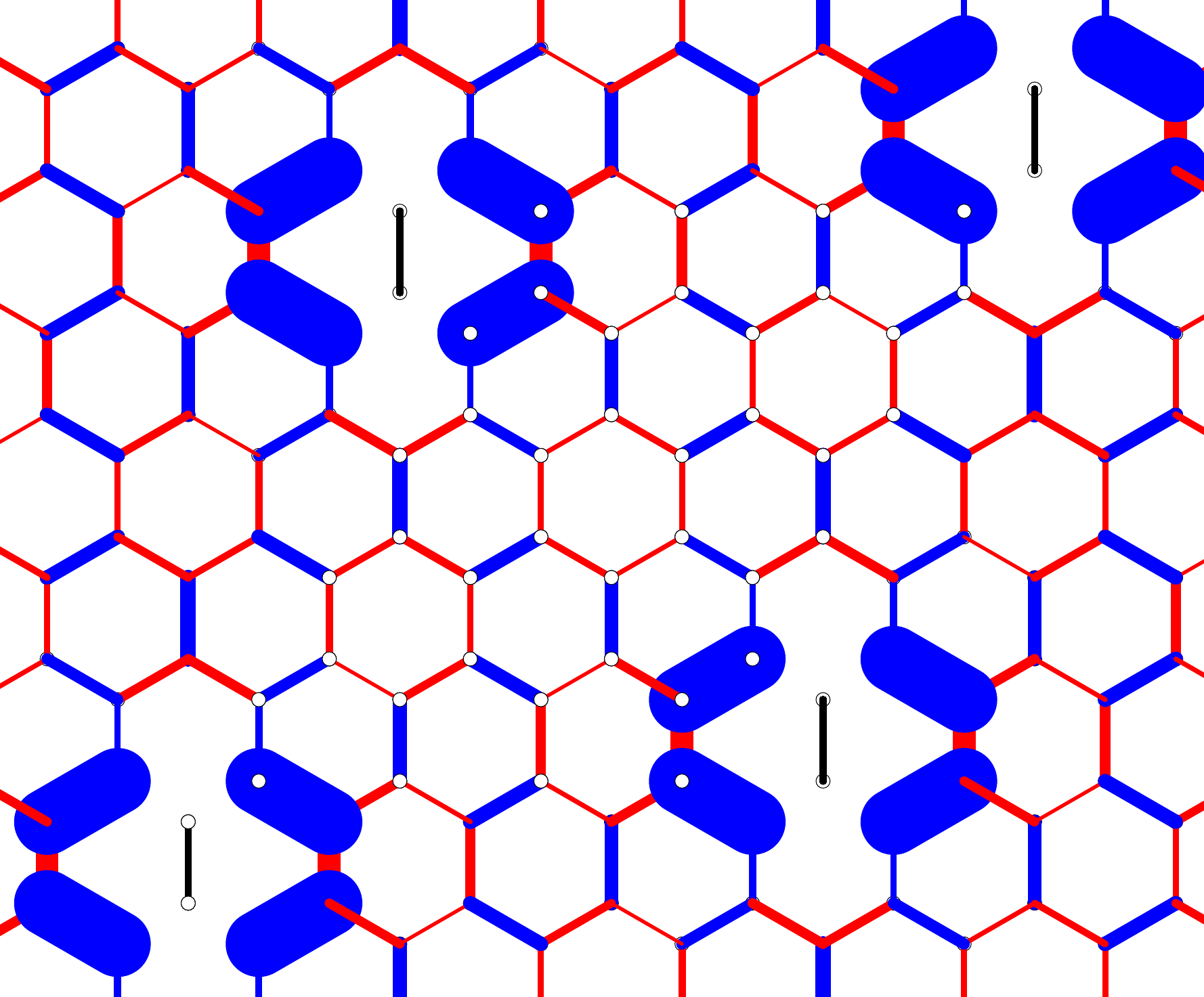}
\caption{(Color online) Kinetic energy correlations at fixed $V_1/t=8$ and $V_2/t=2$ on $N=42$ cluster. Periodic boundary conditions are used. Blue and red correspond to
positive/negative correlations. Width is proportional to the correlation.}
\label{fig:kin_42_V1_8_V2_2}
\end{figure}

A first evidence is given by the low-energy spectrum of $N=24$ at fixed $V_1/t=4$ (see Fig.~\ref{fig:fullspectrum_24_V1_4}) where in some intermediate $V_2/t$ range, the lowest energies are compatible with this plaquette/Kekul\'e phase. A second one is provided via a direct computation of the (connected) kinetic energy correlations on a larger $N=42$ cluster in Fig.~\ref{fig:kin_42_V1_8_V2_2}, which are in perfect agreement with the expected pattern.~\cite{Albuquerque2011}

There is a small caveat here, since it is rather difficult to distinguish a static Kekul\'e  pattern from a plaquette crystal, where the three strong bonds on a hexagon resonate. Indeed both phases break the same symmetries and have the same degeneracy so that a clear identification is often difficult.~\cite{Read1990,Moessner2001,Albuquerque2011} Let us mention that for $S=1/2$ Heisenberg model on the honeycomb lattice with competing $J_1$, $J_2$ (and $J_3$) interactions, there exists an intermediate region with a clear plaquette character.~\cite{Albuquerque2011,Zhu2013,Ganesh2013,Gong2013} Moreover, in this case, the plaquette phase seems connected to the ordered phase through a possibly continuous phase transition.~\cite{Albuquerque2011,Ganesh2013}
In our spinless fermionic model, we have not fully clarified the nature of this intermediate phase, but this would be an interesting project as well as the nature of the transition both to the semi-metallic and the N\'eel phases, which is unfortunately beyond what can be done using ED technique, but which might be fermionic examples of phase transitions beyond the Landau paradigm.

\subsection{Stability of the topological phase}

\begin{figure*}[!ht]
\includegraphics[width=0.3\linewidth,clip]{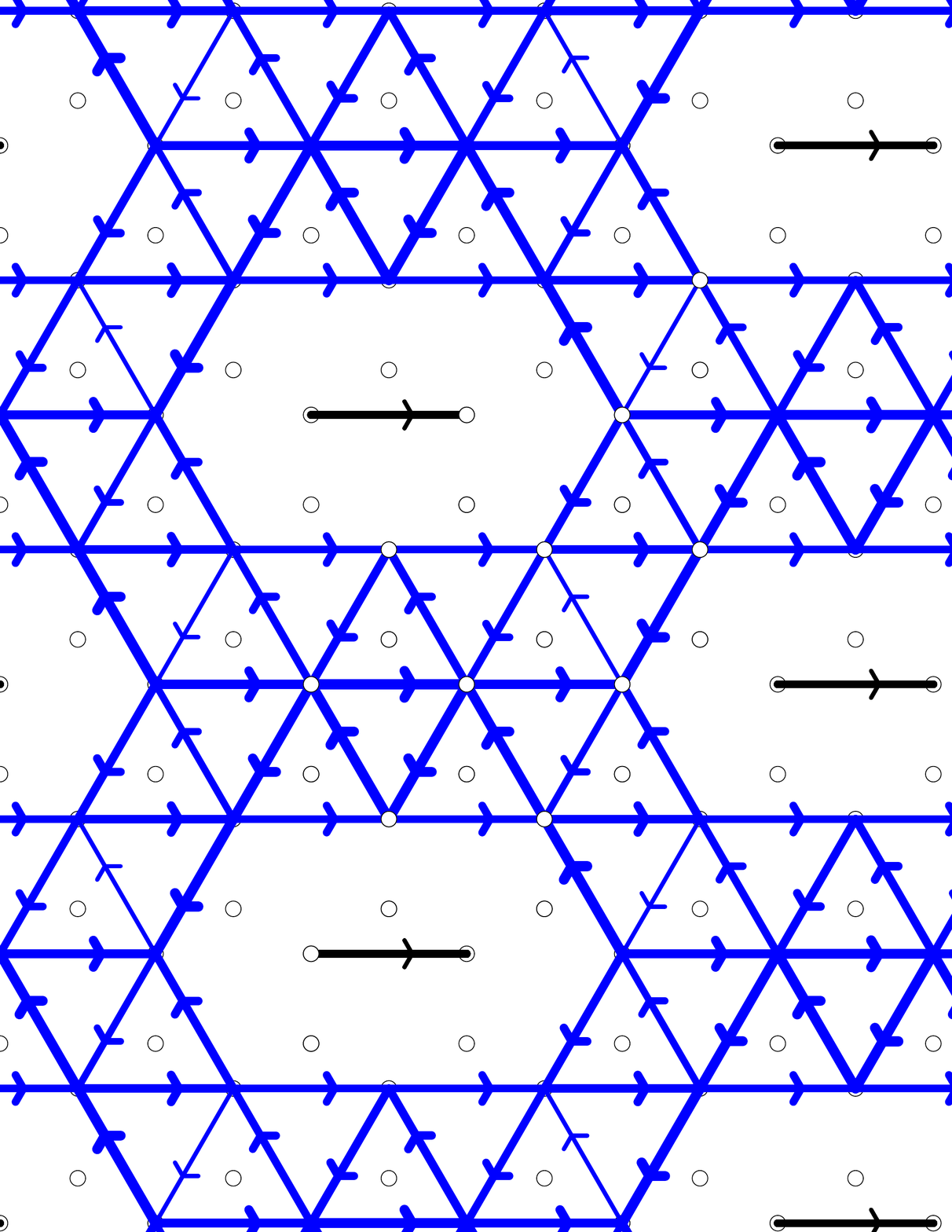}
\hspace*{0.1cm}
\includegraphics[width=0.3\linewidth,clip]{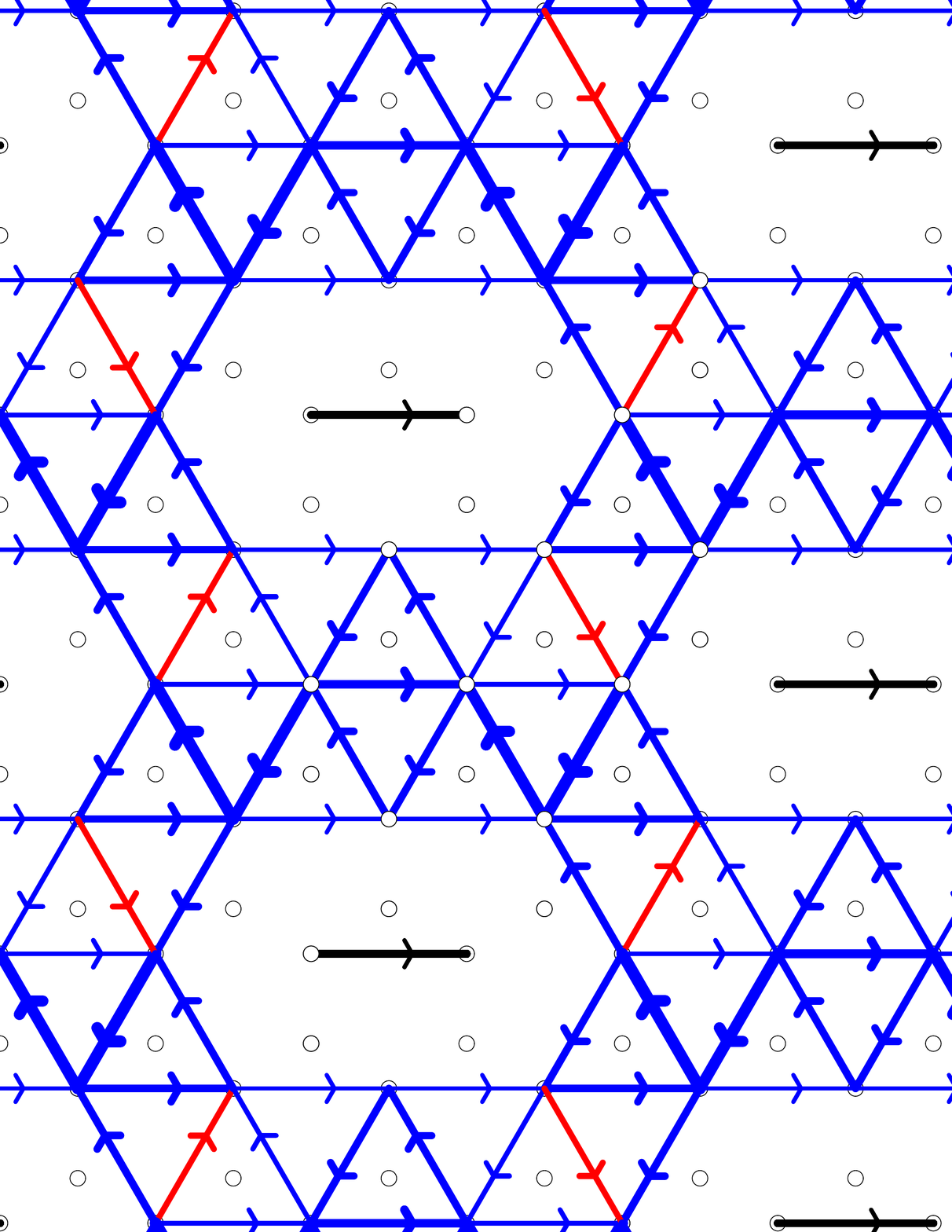}
\hspace*{0.1cm}
\includegraphics[width=0.3\linewidth,clip]{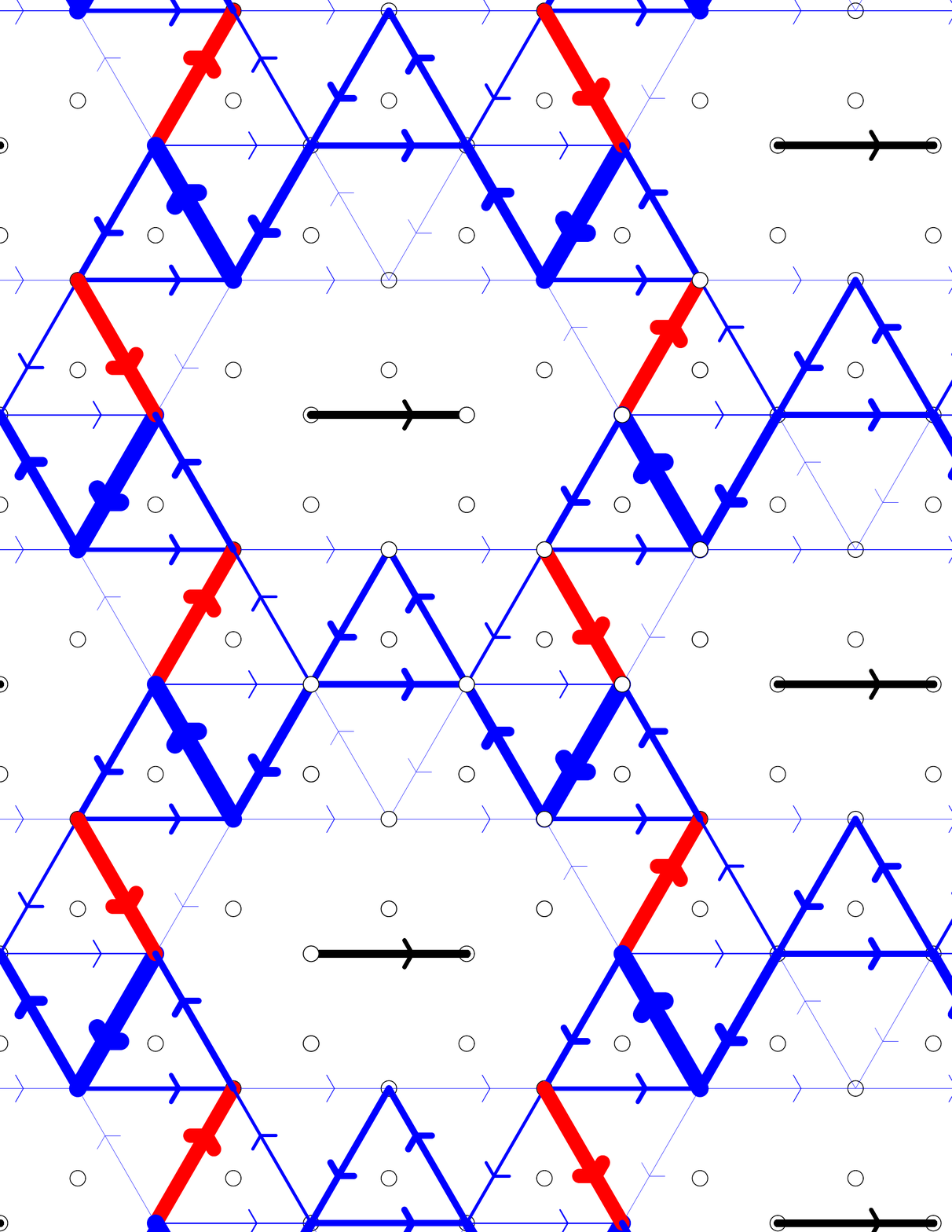}
\caption{(Color online) Current-current correlations on a given sublattice between a reference bond (black) and other bonds for $V_2/t=1$, $2$ and $3$ (from left to right) with fixed $V_1=0$ on $N=24$ sample. 
Periodic boundary conditions are used. Blue and red correspond to
positive/negative correlations according to the orientations expected in the QAH phase (taken here to be clockwise on all hexagons).
 Width is proportional to the correlation.}
\label{fig:jj24}
\end{figure*}

\begin{figure*}[!ht]
\includegraphics[width=0.2\linewidth,clip]{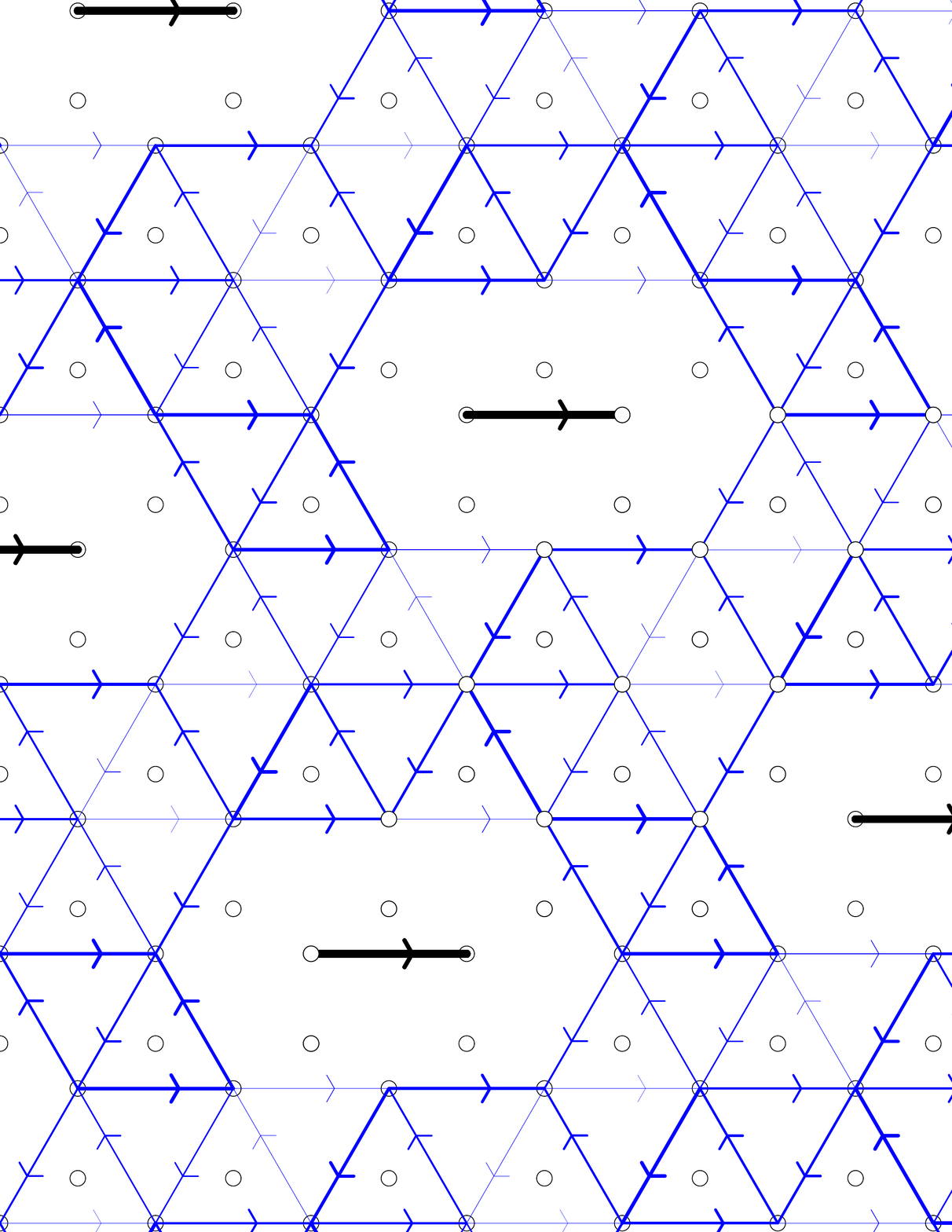}
\hspace*{0.1cm}
\includegraphics[width=0.2\linewidth,clip]{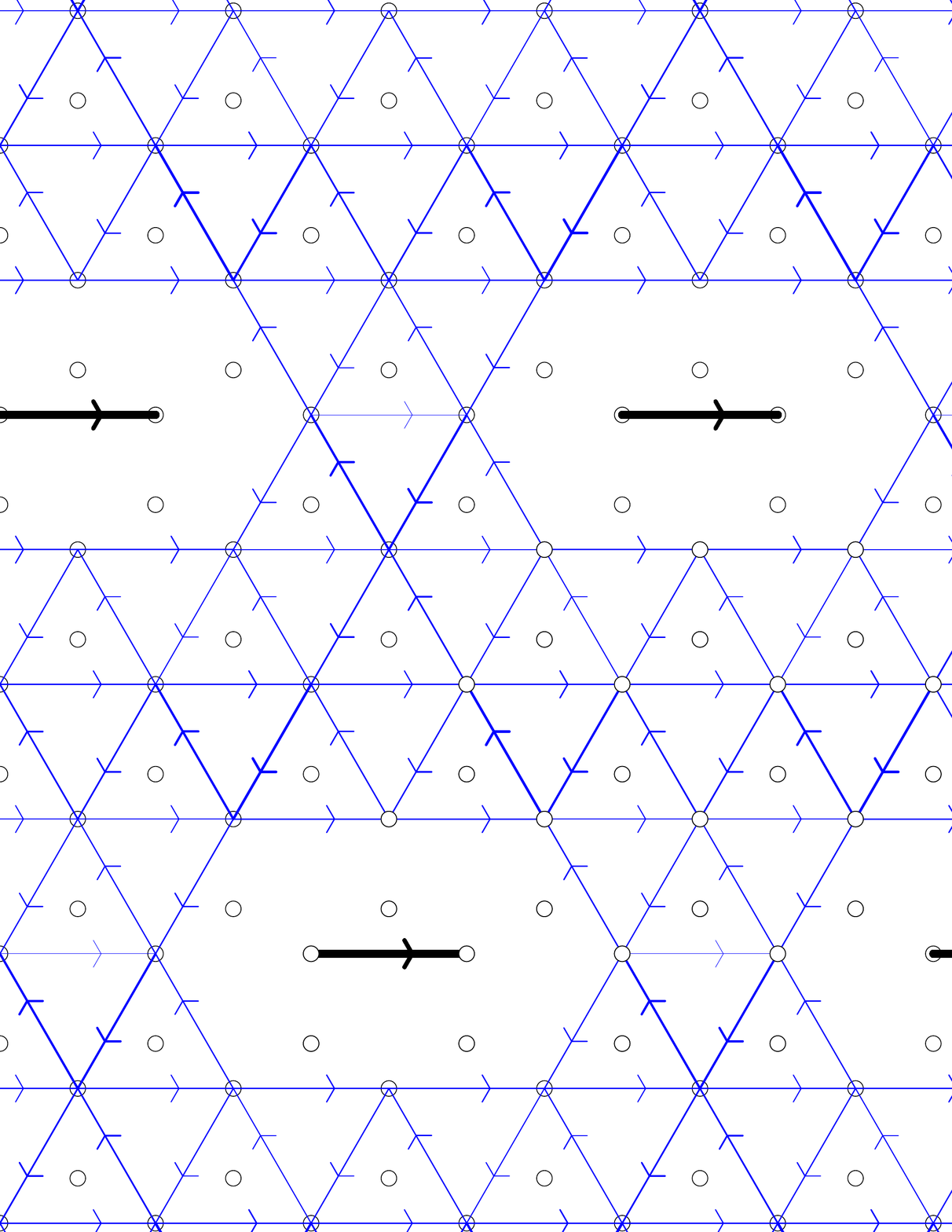}
\hspace*{0.1cm}
\includegraphics[width=0.2\linewidth,clip]{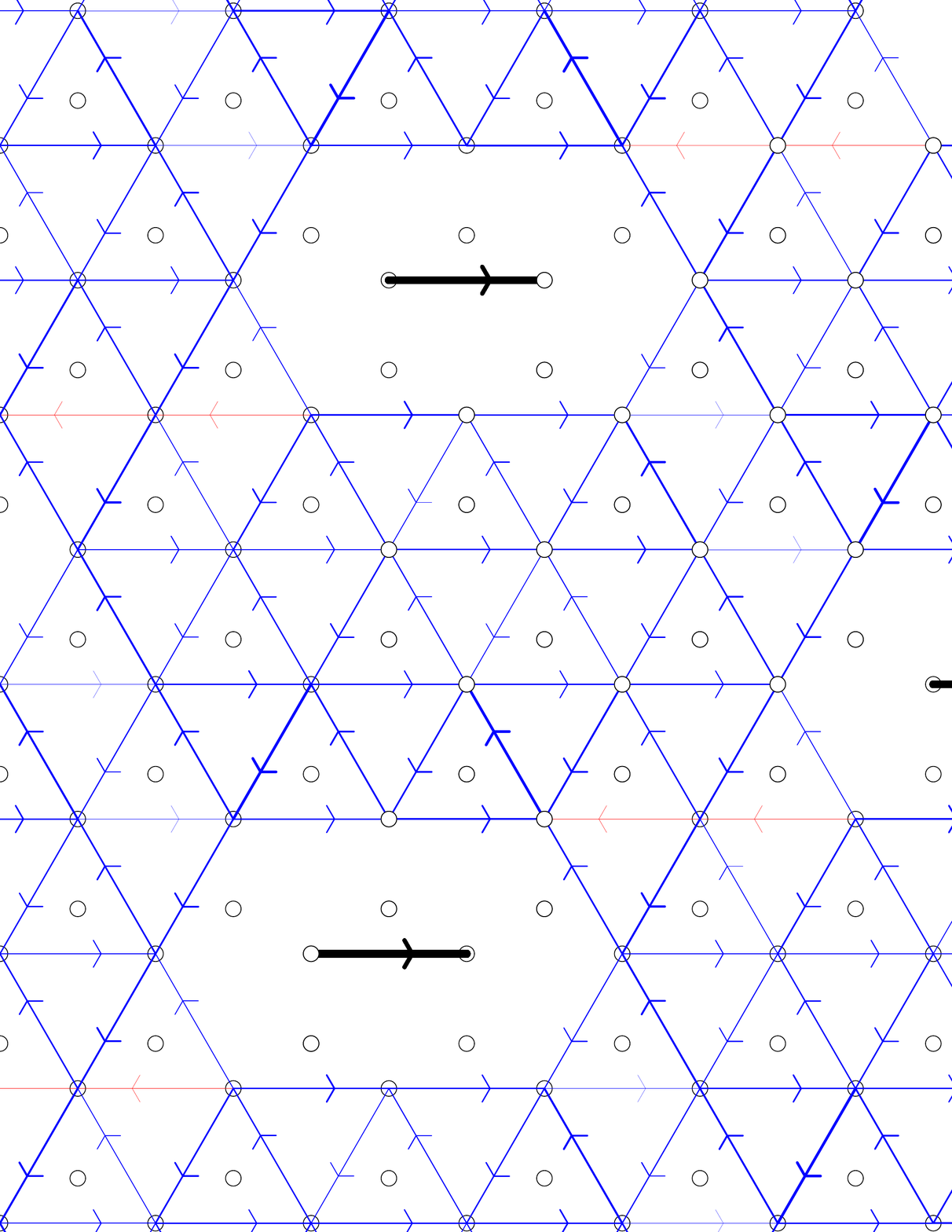}
\hspace*{0.1cm}
\includegraphics[width=0.2\linewidth,clip]{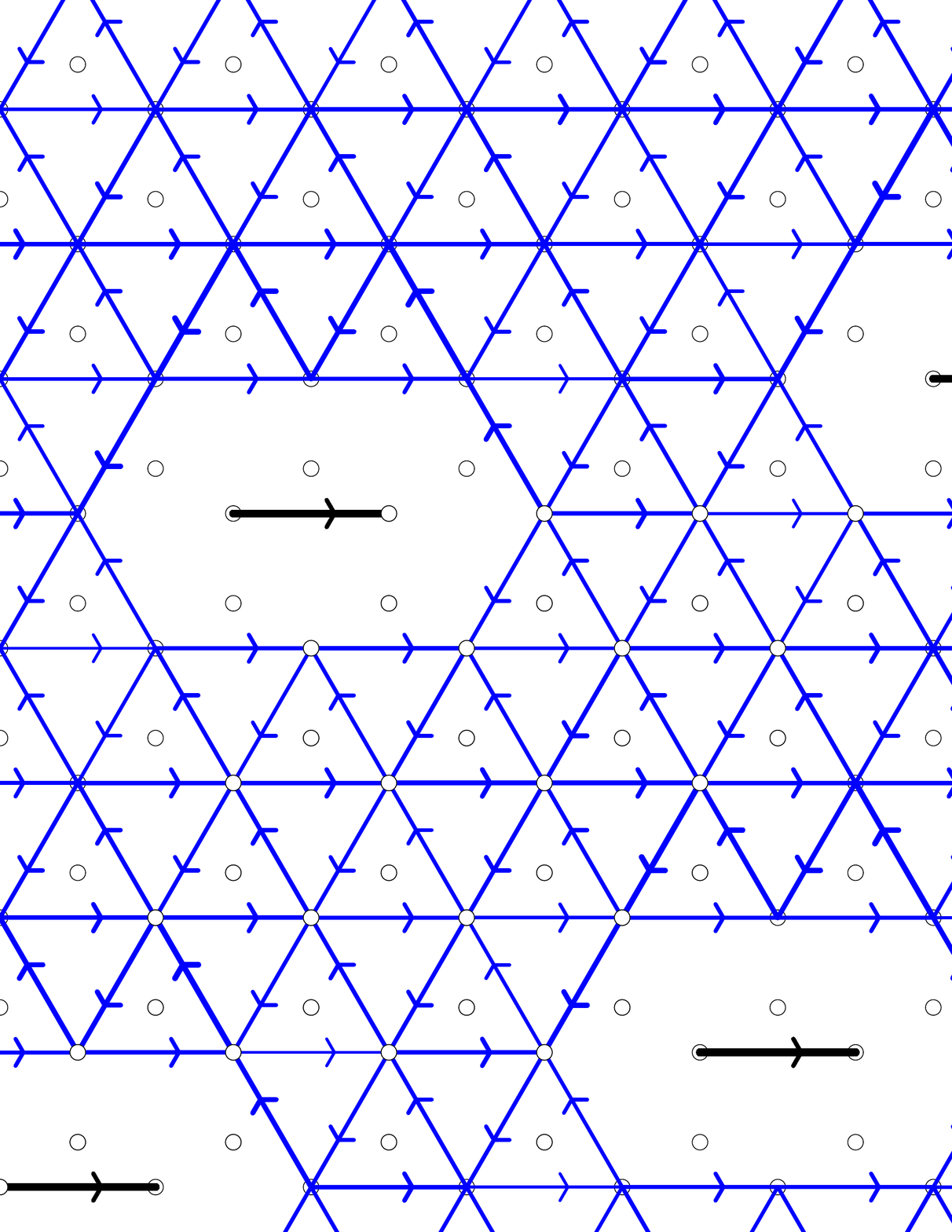}
\caption{(Color online) Same as Fig.~\ref{fig:jj24} for $V_1=0$ and $V_2/t=1$ on various clusters having $C_6$ symmetry. From left to right: $N=26$, $32$, $38$ and $42$.}
\label{fig:jj1_all}
\end{figure*}

\begin{figure*}[!ht]
\includegraphics[width=0.2\linewidth,clip]{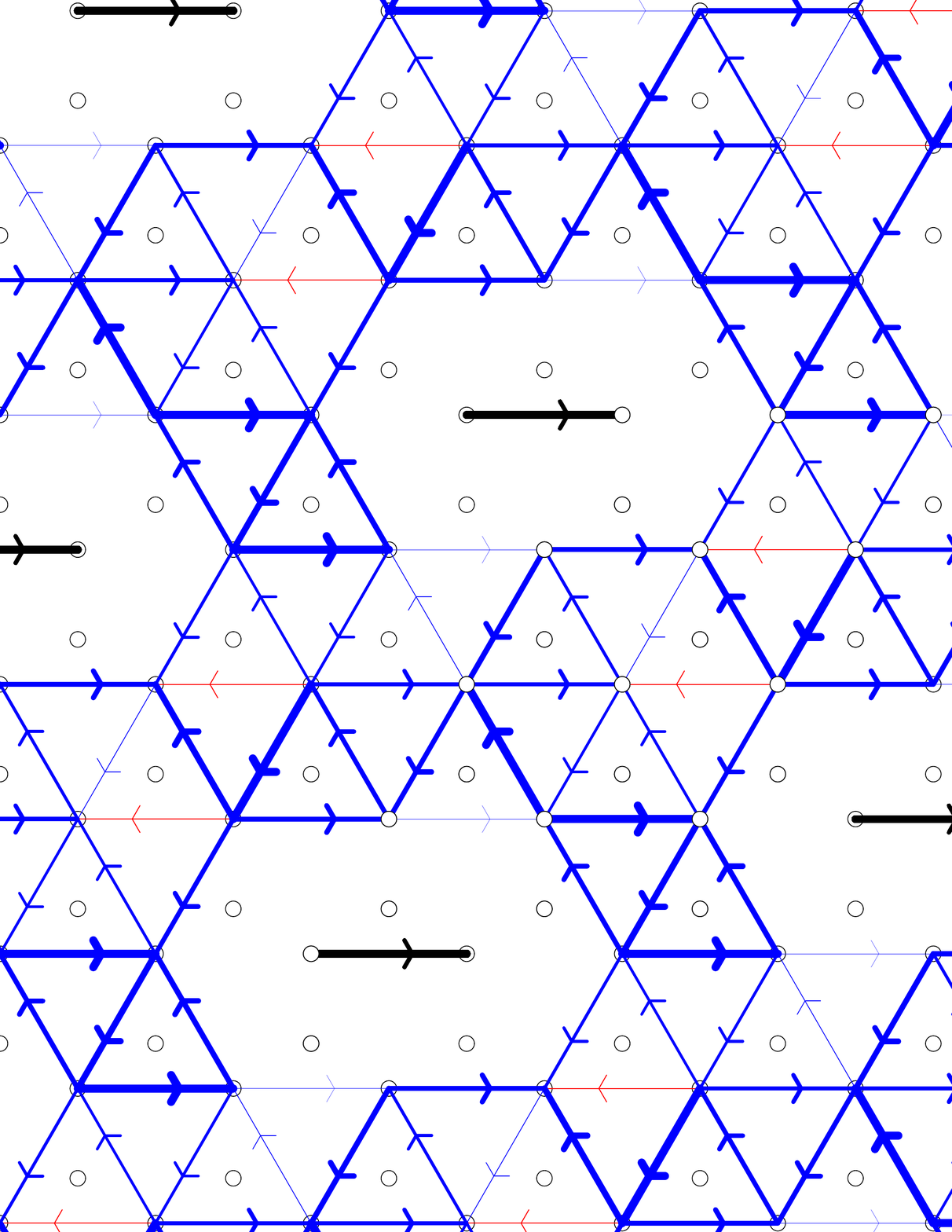}
\hspace*{0.1cm}
\includegraphics[width=0.2\linewidth,clip]{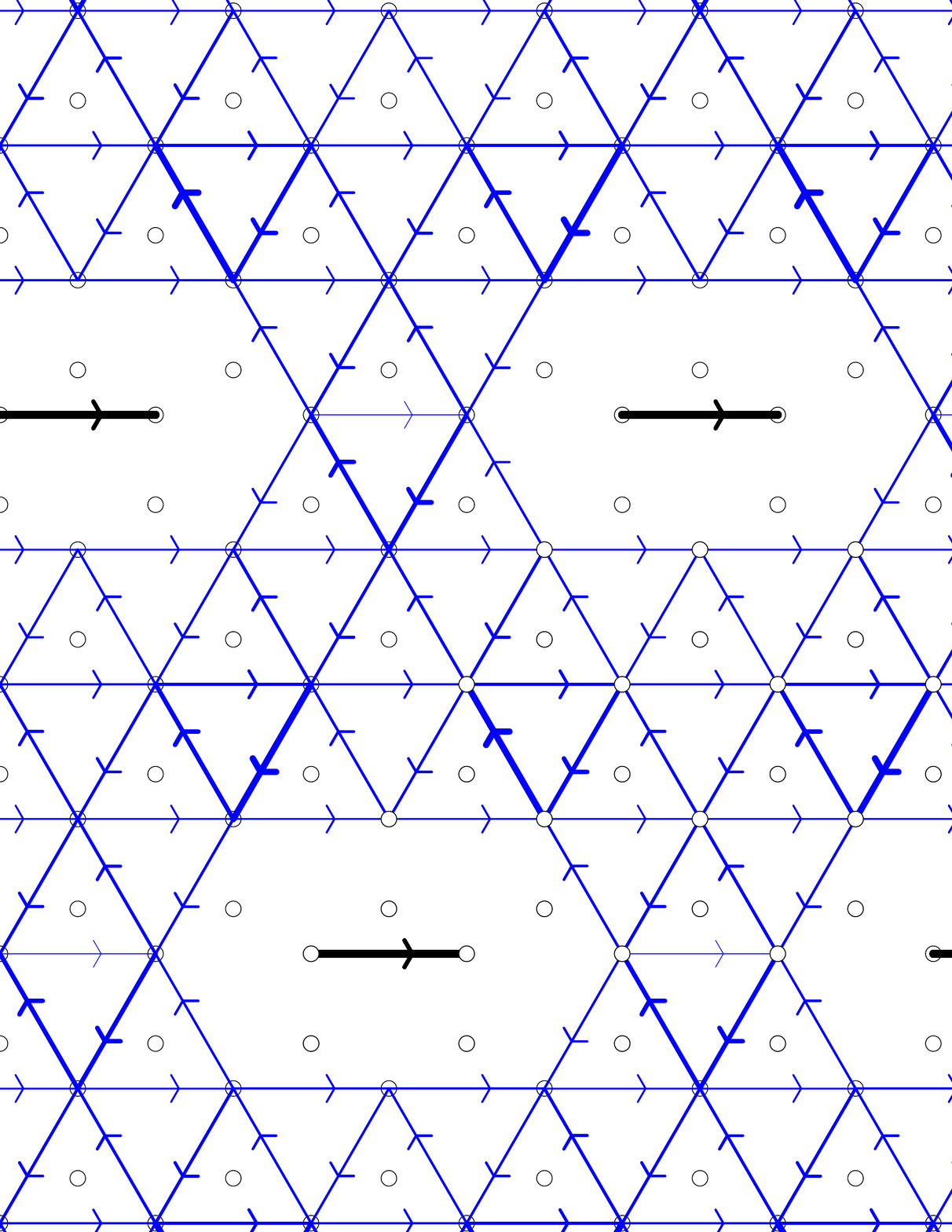}
\hspace*{0.1cm}
\includegraphics[width=0.2\linewidth,clip]{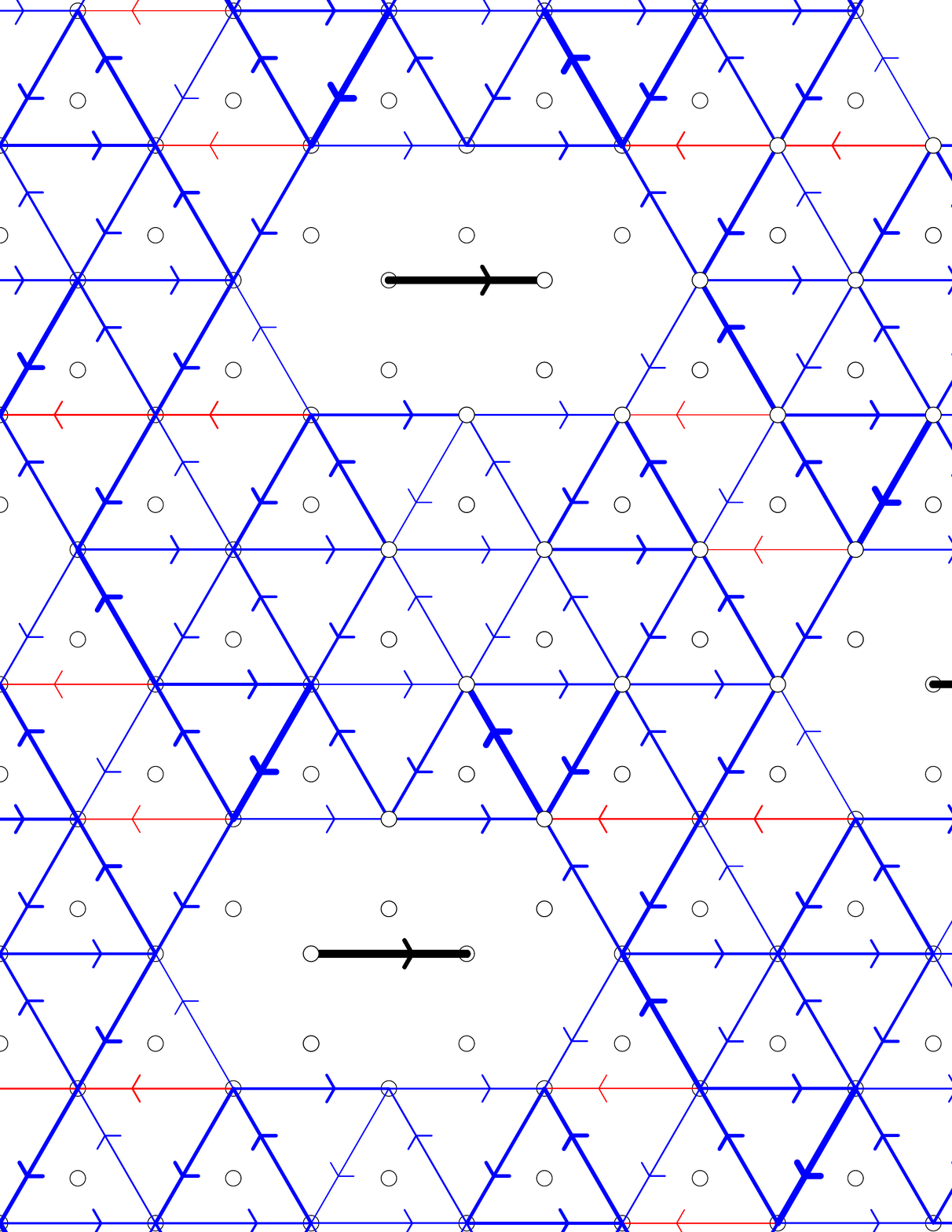}
\hspace*{0.1cm}
\includegraphics[width=0.2\linewidth,clip]{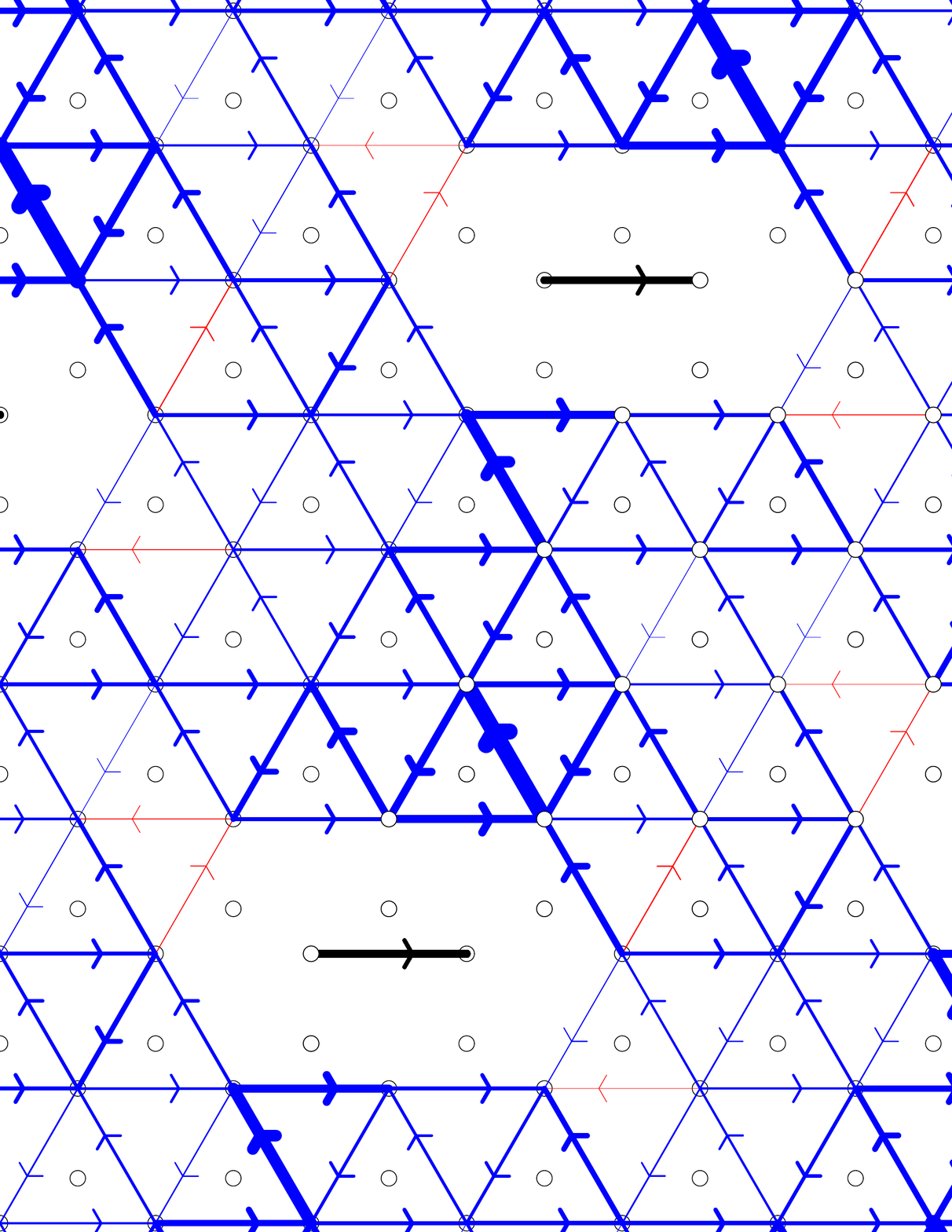}
\caption{(Color online) Same as Fig.~\ref{fig:jj1_all} for $V_1=0$ and $V_2/t=2$.}
\label{fig:jj2_all}
\end{figure*}

The proposal made by Raghu {\it et al.}\cite{Raghu2008} is that, when $V_2>0$ dominates,
the system undergoes a transition to a topologically distinct
phase that spontaneously break time-reversal symmetry. A simple
picture of this phase can be obtained by drawing circulating currents
on a given sublattice. In order to check this picture, we have
computed current-current correlations on a given sublattice. Note that
these correlations are strictly zero for all distances when
$V_1=V_2=0$. They are plotted for different $V_2$ in
Fig.~\ref{fig:jj24} in the case of $N=24$ lattice and at fixed $V_1=0$
where the instability is expected to be the strongest.\footnote{Our numerical current structure factor (see its definition below) are indeed the strongest when $V_1=0$ on clusters $N=24$ and $N=42$ for instance.} We have used
the expected $q=0$ orientation pattern of the QAH phase, i.e. where
all currents would be in the clockwise direction on each hexagon for
instance. We see on Fig.~\ref{fig:jj24} that when we increase $V_2/t$,
the signal starts by increasing in amplitude, and then some defects
(in red) start to appear on the cluster, in particular in the case
where QAH is supposed to be the strongest around $V_2/t \simeq 2$. Let
us emphasize again that QAH is perfectly compatible with all clusters
(since it is a $q=0$ instability) and as a result no
frustration is expected. Therefore, the appearance of defects reveal some
competing phases.  Still the overall pattern is almost perfect, which
is remarkable and does not occur in the $V_1$ only case for instance.  For
even larger $V_2$, the signal suddenly disappears (see Figs.~\ref{fig:jj1_all} and \ref{fig:jj2_all} for several clusters, other data not shown),
signalling the entrance in the CM phase (see above). Given that the semimetallic phase should have a finite extension for small $V_2/t$, this constrains the putative QAH phase in a small region around $V_2/t\simeq 2$.

In order to estimate the finite-size effects, we have performed
extensive simulations on a large number of samples (data not shown). 
We can clearly identify some clusters on which the patterns
are far from being perfect (i.e. have many defects), which can be
related either to lower symmetry, or to poorly two-dimensional aspect
ratio. Indeed, when looking at clusters having $C_6$ symmetry at
least, we do observe nice current correlations with no or few defects at small $V_2/t$. 
Therefore, there are definitely correlations compatible with QAH
phase, but one needs to investigate whether these are short-range
features only or not.

\begin{figure}[!htb]
\includegraphics[width=0.9\linewidth,clip]{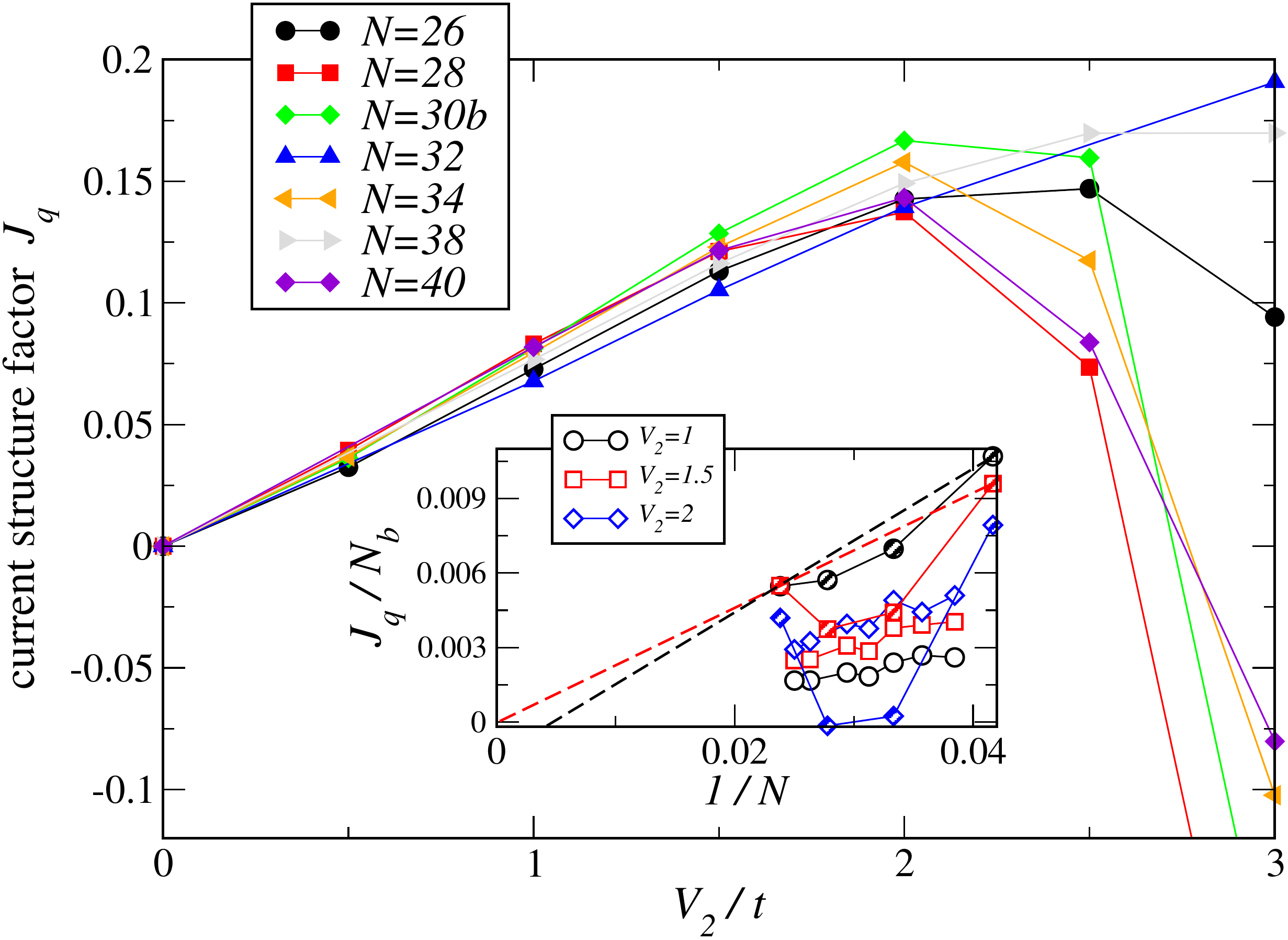}
\caption{(Color online) Current structure factor ${\cal J}_q$ as a function of $V_2/t$ for several lattices. Inset: scaling of this structure factor divided by the number of bonds $N_b=3N/2-11$ vs $1/N$ for several $V_2/t$. Open symbols are used for clusters without the $\mathbf{K}$ point. Hatched symbols are used for the clusters $N=24$ and $N=42$ which have the $\mathbf{K}$ point as well as $C_6$ symmetry at least. All data points are compatible with a vanishing signal in the thermodynamic limit.}
\label{fig:Current_Struct_Factor}
\end{figure}

In order to do so, we have to measure some order parameter. We have chosen the current structure factor, i.e. the sum of current correlations with the signs taken from the QAH pattern:
\begin{equation}\label{eq:JQ}
{\cal J}_q = \sum_{\langle\langle ij \rangle\rangle} \varepsilon_{ij} \langle {\cal J}_{ij} {\cal J}_{i_0 j_0} \rangle
\end{equation}
where the sum runs over all $N_b$ bonds on a given sublattice~\footnote{We sum over all bonds which do not share any common site with the reference bond, i.e there are $N_b=3N/2-11$ terms.} and the $\varepsilon_{ij}=\pm 1$ correspond to the expected QAH orientation (all currents clockwise or anticlockwise depending on the choice of reference bond $(i_0 j_0)$). 
 When using clusters with less symmetry, we average over the inequivalent directions of the reference bond.  
This is indeed a valid order parameter and long-range order requires that it diverges 
in the thermodynamic limit as $N_b \sim N$.  
Overall, we see in Fig.~\ref{fig:Current_Struct_Factor} a clear increase of
current correlations for moderate $V_2$, indicating that such an
interaction indeed favours current formations, in agreement with mean-field prediction. There are of course non-trivial finite-size effects due to the fact that not all clusters have the same symmetry (see Appendix~\ref{appendix:lattice}). 

In the inset of Fig.~\ref{fig:Current_Struct_Factor}, we plot this
structure factor divided by the number of terms $N_b$ for $1 \leq
V_2/t \leq 2$ where the QAH seems to be the strongest from our data,
and also from the mean-field estimate of the QAH
region~\cite{Raghu2008}: $1.5 \lessapprox V_2 \lessapprox 2.5$. 
For clusters without the $\mathbf{K}$ point, the
behaviour of this current structure factor has some non-monotonic size
dependence due to the fact that clusters have different shapes, but
its absolute value is quite small, and finite-size extrapolations are
compatible with a vanishing value.
We also plot separately the data points for clusters having the $\mathbf{K}$ point (which have a larger signal due to the degenerate ground state at half filling at $V_2=0$) but if we focus on $N=24$ and $N=42$ which are the best two-dimensional clusters of this kind (good aspect ratio and $C_6$ symmetry at least), data are also compatible with the absence of long-range order in the thermodynamic limit. Our results are in agreement with other recent numerical studies~\cite{Garcia2013,Daghofer2014} but in contrast with another work.~\cite{Duric2014} We hope that future progress in simulations of two-dimensional strongly correlated systems will allow to confirm the absence of a QAH phase for this microscopic model, but will be able to stabilise it in a larger parameter space. Indeed, it is obvious that the semi-metallic state has a strong response in the QAH channel, so that if other instabilities could be prevented (using additional interactions to frustrate them), there remains an interesting possibility to stabilise QAH phase in a truly long-range ordered fashion.

\section{Conclusion}\label{sec:conclusion}

We have investigated a model of spinless fermions on honeycomb lattice with (generally competing) density-density interactions using extensive numerical exact diagonalizations on various clusters. Based on a careful analysis, our main result is summarized in the global phase diagram in Fig.~\ref{fig:PhaseDiagram}. 
On the one hand, we have clarified the phase diagram in the strong coupling regime ($|V_{1,2}| \gg t$) using some non-trivial results on the classical model and some quantum order-by-disorder arguments. On the other hand, we have used numerical simulations of the full quantum model on finite clusters. As a result, 
for intermediate or large interactions, we have provided evidence for several crystalline phases: N\'eel CDW, CM, plaquette/Kekul\'e, N\'eel domain wall crystal, the stripy* region and the zigzag phase, some of which had not been considered yet. 

When considering the role of $V_2>0$ interaction alone, it had been suggested in the literature~\cite{Raghu2008}
that it could stabilize a topological phase, which has triggered an intense activity. We have measured the corresponding (charge) current correlations, and while there is definitely some increase due to the $V_2$ interaction, our finite-size analysis indicate that the topological phase is not present in the thermodynamic limit, and there is a direct transition between the semi-metallic phase and a charge-modulated one. However, it is very interesting to notice that $V_2$ interaction is able to create short-range current correlations, so that some additional ingredients could possibly stabilize it. Future work along these lines should also consider the spinful case where a topological spontaneous quantum spin hall phase with spin currents has been proposed.~\cite{Raghu2008}

As a last remark, let us remind that other models are known to exhibit Dirac cones and thus could potentially also host such a topological phase: for instance, on the square lattice $\pi$-flux model at half-filling, a topological insulating phase is predicted using mean-field technique~\cite{Weeks2010} but numerical study have not confirmed its stability.~\cite{Jia2013} It is also noteworthy that a generalized tight-binding model on the kagom\'e lattice always contain Dirac cones at some particular filling~\cite{Asano2011}, which leaves room to investigate the exciting possibility to stabilize a topological phase with interactions. 

{\em Note added:} While finalizing this manuscript we became aware of concurrent iDMRG work~\cite{Motruk2015} with consistent results regarding the presence and absence of the various phases
in the repulsive regime.

\acknowledgments
We thank M. Daghofer, A. Grushin, M. Hohenadler, J. Motruk and F. Pollmann for discussions. We thank F. Pollmann and collaborators for sharing their consistent iDMRG results 
prior to publication.
This work was performed using HPC resources from CALMIP and GENCI-IDRIS (Grant 2014-050225) as well as LEO2/3 and VSC2/3.
SC thanks Institut Universitaire de France (IUF) for financial support. AML was supported by the FWF (I-1310-N27/DFG FOR1807).
\appendix

\section{Lattice geometries}\label{appendix:lattice}

Since we are using several kinds of lattices, we give their definitions in Table~\ref{tab:lattice} using ${\bf a}$ and ${\bf b}$ translations to define the
torus with periodic boundary conditions. 

\begin{table}[!h]
\begin{tabular}{|c|c|c|c|c|c|c|}
\hline
Name & {\bf a} & {\bf b} & sym. group & $\mathbf{K}$ (2) & $\mathbf{M}$ (3)  & $\mathbf{X}$ (6)\\
\hline
12 & $(3,3\sqrt{3}) $ & $(3,-\sqrt{3})$ & $Z_{2}$ & no & {\bf yes}(1) & no \\
18 & $(6,0)$ & $(3,3\sqrt{3})$ & $C_{6v}$ & {\bf yes} & no & no \\
24 & $(6,2\sqrt{3})$ & $(0,4\sqrt{3})$ & $C_{6v}$ & {\bf yes} & {\bf yes} & {\bf yes} \\
24b & $(4,4\sqrt{3})$ & $(-4,2\sqrt{3})$ & $Z_{2}$ & no & {\bf yes}(1) & no \\
26 & $(7,\sqrt{3})$ & $(2,4\sqrt{3})$ & $C_{6}$ & no & no & no \\
28 & $(7,\sqrt{3})$ & $(0,4\sqrt{3})$ & $Z_2$ & no & {\bf yes} (1) & no \\
30a & $(3,5\sqrt{3})$ & $(-3,5\sqrt{3})$ & $C_2\times C_2$ & {\bf yes} & no & no \\ 
30b & $(5,3\sqrt{3})$ & $(-5,3\sqrt{3})$ & $C_2\times C_2$ & no & no & no \\
32 & $(8,0)$ & $(4,4\sqrt{3})$ & $C_{6v}$ & no & {\bf yes} & no\\
34 & $(9,\sqrt{3})$ & $(2,4\sqrt{3})$ & $Z_2$ & no & no & no\\
36 & $(6,0)$ & $(6,6\sqrt{3})$ & $C_2\times C_2$ & {\bf yes} & {\bf yes} (1) & {\bf yes} (2) \\
38 & $(8,2\sqrt{3})$ & $(1,5\sqrt{3})$ & $C_{6}$ & no & no & no\\
42 & $(9,\sqrt{3})$ & $(3,5\sqrt{3})$ & $C_{6}$ & {\bf yes} & no & no \\
54 & $(9,3\sqrt{3})$ & $(0,6\sqrt{3})$ & $C_{6v}$ & {\bf yes} & no & no \\
72 & $(8,0)$ & $(8,8\sqrt{3})$ & $C_{6v}$ & {\bf yes} &  {\bf yes} &  {\bf yes} \\
\hline
\end{tabular}
\caption{\label{tab:lattice}Finite lattices studied in this work. Listed are the number of spins
$N$; the basis vectors {\bf a}, {\bf b} in the plane; the symmetry point group; presence of the $\mathbf{K}$ point; $\mathbf{M}$ point; $\mathbf{X}$ point.}
\end{table}

\section{Counting of Ising Groundstates}
\label{app:isingcount}

In this section we provide a table with the dimensions  of the classical ground spaces
at half filling for the clusters described in Appendix~\ref{appendix:lattice}.

\begin{table}[!h]
\begin{tabular}{|c|c|c|c|c|c|}
\hline
$N_s$ &$ -4/5 $ & $-4/5$$<$$V_1$$<$$0$ & $0$ & $0$$<$$V_1$$<$$4/5$ & $4/5$  \\
\hline
$12$ &	-& 8	& - & $8$ & - \\
$18$ &	-&-	& 666 & $\emptyset$ & 38 \\
$24$ & 96	& 6 & 5'526 & 6 &  446  \\
$24$b &-	& 12 &  - & 12 & -   \\
$26$ &	-&-	& 3'042 & $\emptyset$ & 340  \\ 
$28$ &	-&-	& 8'964 & 2 &  858 \\
$30$a &-	&-	& 30'618 & $\emptyset$ &  692  \\
$30$b &-	&-	& 11'300 & $\emptyset$ & 888  \\
$32$ & 630 & 42	& 1'764 & 42 &  1'520\\
$36$ &-	&-	&  230'014 & 14 &  3'454 \\
$38$ &-	&-	& - & - & 5'208 \\
$42$ &-	&-	& 1'211'004 & - & 8'880 \\
$54$ &-	&-	& 59'711'598 & - & -  \\
$72$ &-	&186	& - & 186 & 8'705'390 \\
\hline
\end{tabular}
\caption{\label{tab:isingcount} Ground state degeneracy in the classical limit $t=0$ at {\em half filling}. $\emptyset$
denotes a ground state energy on this cluster that is higher than on other clusters. - indicates that the ground
space has not been explored for this instance. The column labels $V_1$ and we use the parametrization $V_2=1-V_1$ for $V_1>0$ and $V_2=1-|V_1|$ for $V_1<0$.}
\label{tab:IsingCounting}
\end{table}

\bibliographystyle{apsrev4-1}
\include{resubmit.bbl}
\end{document}

%% file: resubmit.bbl
%

%% file: resubmit.bbl
\begin{thebibliography}{42}%
\makeatletter
\providecommand \@ifxundefined [1]{%
 \@ifx{#1\undefined}
}%
\providecommand \@ifnum [1]{%
 \ifnum #1\expandafter \@firstoftwo
 \else \expandafter \@secondoftwo
 \fi
}%
\providecommand \@ifx [1]{%
 \ifx #1\expandafter \@firstoftwo
 \else \expandafter \@secondoftwo
 \fi
}%
\providecommand \natexlab [1]{#1}%
\providecommand \enquote  [1]{``#1''}%
\providecommand \bibnamefont  [1]{#1}%
\providecommand \bibfnamefont [1]{#1}%
\providecommand \citenamefont [1]{#1}%
\providecommand \href@noop [0]{\@secondoftwo}%
\providecommand \href [0]{\begingroup \@sanitize@url \@href}%
\providecommand \@href[1]{\@@startlink{#1}\@@href}%
\providecommand \@@href[1]{\endgroup#1\@@endlink}%
\providecommand \@sanitize@url [0]{\catcode `\\12\catcode `\$12\catcode
  `\&12\catcode `\#12\catcode `\^12\catcode `\_12\catcode `\%12\relax}%
\providecommand \@@startlink[1]{}%
\providecommand \@@endlink[0]{}%
\providecommand \url  [0]{\begingroup\@sanitize@url \@url }%
\providecommand \@url [1]{\endgroup\@href {#1}{\urlprefix }}%
\providecommand \urlprefix  [0]{URL }%
\providecommand \Eprint [0]{\href }%
\providecommand \doibase [0]{http://dx.doi.org/}%
\providecommand \selectlanguage [0]{\@gobble}%
\providecommand \bibinfo  [0]{\@secondoftwo}%
\providecommand \bibfield  [0]{\@secondoftwo}%
\providecommand \translation [1]{[#1]}%
\providecommand \BibitemOpen [0]{}%
\providecommand \bibitemStop [0]{}%
\providecommand \bibitemNoStop [0]{.\EOS\space}%
\providecommand \EOS [0]{\spacefactor3000\relax}%
\providecommand \BibitemShut  [1]{\csname bibitem#1\endcsname}%
\let\auto@bib@innerbib\@empty
\bibitem [{\citenamefont {Prange}\ and\ \citenamefont
  {Girvin}(1990)}]{book_QHE}%
  \BibitemOpen
  \bibfield  {author} {\bibinfo {author} {\bibfnamefont {R.~E.}\ \bibnamefont
  {Prange}}\ and\ \bibinfo {author} {\bibfnamefont {S.~M.}\ \bibnamefont
  {Girvin}},\ }\href@noop {} {\emph {\bibinfo {title} {The Quantum Hall Effect,
  2nd ed.}}}\ (\bibinfo  {publisher} {New York: Springer-Verlag},\ \bibinfo
  {year} {1990})\BibitemShut {NoStop}%
\bibitem [{\citenamefont {Hasan}\ and\ \citenamefont
  {Kane}(2010)}]{topo_review2}%
  \BibitemOpen
  \bibfield  {author} {\bibinfo {author} {\bibfnamefont {M.~Z.}\ \bibnamefont
  {Hasan}}\ and\ \bibinfo {author} {\bibfnamefont {C.~L.}\ \bibnamefont
  {Kane}},\ }\href {\doibase 10.1103/RevModPhys.82.3045} {\bibfield  {journal}
  {\bibinfo  {journal} {Rev. Mod. Phys.}\ }\textbf {\bibinfo {volume} {82}},\
  \bibinfo {pages} {3045} (\bibinfo {year} {2010})}\BibitemShut {NoStop}%
\bibitem [{\citenamefont {Qi}\ and\ \citenamefont {Zhang}(2011)}]{topo_review}%
  \BibitemOpen
  \bibfield  {author} {\bibinfo {author} {\bibfnamefont {X.-L.}\ \bibnamefont
  {Qi}}\ and\ \bibinfo {author} {\bibfnamefont {S.-C.}\ \bibnamefont {Zhang}},\
  }\href {\doibase 10.1103/RevModPhys.83.1057} {\bibfield  {journal} {\bibinfo
  {journal} {Rev. Mod. Phys.}\ }\textbf {\bibinfo {volume} {83}},\ \bibinfo
  {pages} {1057} (\bibinfo {year} {2011})}\BibitemShut {NoStop}%
\bibitem [{\citenamefont {Raghu}\ \emph {et~al.}(2008)\citenamefont {Raghu},
  \citenamefont {Qi}, \citenamefont {Honerkamp},\ and\ \citenamefont
  {Zhang}}]{Raghu2008}%
  \BibitemOpen
  \bibfield  {author} {\bibinfo {author} {\bibfnamefont {S.}~\bibnamefont
  {Raghu}}, \bibinfo {author} {\bibfnamefont {X.-L.}\ \bibnamefont {Qi}},
  \bibinfo {author} {\bibfnamefont {C.}~\bibnamefont {Honerkamp}}, \ and\
  \bibinfo {author} {\bibfnamefont {S.-C.}\ \bibnamefont {Zhang}},\ }\href
  {\doibase 10.1103/PhysRevLett.100.156401} {\bibfield  {journal} {\bibinfo
  {journal} {Phys. Rev. Lett.}\ }\textbf {\bibinfo {volume} {100}},\ \bibinfo
  {pages} {156401} (\bibinfo {year} {2008})}\BibitemShut {NoStop}%
\bibitem [{\citenamefont {Weeks}\ and\ \citenamefont
  {Franz}(2010)}]{Weeks2010}%
  \BibitemOpen
  \bibfield  {author} {\bibinfo {author} {\bibfnamefont {C.}~\bibnamefont
  {Weeks}}\ and\ \bibinfo {author} {\bibfnamefont {M.}~\bibnamefont {Franz}},\
  }\href {\doibase 10.1103/PhysRevB.81.085105} {\bibfield  {journal} {\bibinfo
  {journal} {Phys. Rev. B}\ }\textbf {\bibinfo {volume} {81}},\ \bibinfo
  {pages} {085105} (\bibinfo {year} {2010})}\BibitemShut {NoStop}%
\bibitem [{\citenamefont {Grushin}\ \emph {et~al.}(2013)\citenamefont
  {Grushin}, \citenamefont {Castro}, \citenamefont {Cortijo}, \citenamefont
  {de~Juan}, \citenamefont {Vozmediano},\ and\ \citenamefont
  {Valenzuela}}]{Grushin2013}%
  \BibitemOpen
  \bibfield  {author} {\bibinfo {author} {\bibfnamefont {A.~G.}\ \bibnamefont
  {Grushin}}, \bibinfo {author} {\bibfnamefont {E.~V.}\ \bibnamefont {Castro}},
  \bibinfo {author} {\bibfnamefont {A.}~\bibnamefont {Cortijo}}, \bibinfo
  {author} {\bibfnamefont {F.}~\bibnamefont {de~Juan}}, \bibinfo {author}
  {\bibfnamefont {M.~A.~H.}\ \bibnamefont {Vozmediano}}, \ and\ \bibinfo
  {author} {\bibfnamefont {B.}~\bibnamefont {Valenzuela}},\ }\href {\doibase
  10.1103/PhysRevB.87.085136} {\bibfield  {journal} {\bibinfo  {journal} {Phys.
  Rev. B}\ }\textbf {\bibinfo {volume} {87}},\ \bibinfo {pages} {085136}
  (\bibinfo {year} {2013})}\BibitemShut {NoStop}%
\bibitem [{\citenamefont {Pereg-Barnea}\ and\ \citenamefont
  {Refael}(2012)}]{Pereg2012}%
  \BibitemOpen
  \bibfield  {author} {\bibinfo {author} {\bibfnamefont {T.}~\bibnamefont
  {Pereg-Barnea}}\ and\ \bibinfo {author} {\bibfnamefont {G.}~\bibnamefont
  {Refael}},\ }\href {\doibase 10.1103/PhysRevB.85.075127} {\bibfield
  {journal} {\bibinfo  {journal} {Phys. Rev. B}\ }\textbf {\bibinfo {volume}
  {85}},\ \bibinfo {pages} {075127} (\bibinfo {year} {2012})}\BibitemShut
  {NoStop}%
\bibitem [{\citenamefont {Dauphin}\ \emph {et~al.}(2012)\citenamefont
  {Dauphin}, \citenamefont {M\"uller},\ and\ \citenamefont
  {Martin-Delgado}}]{Dauphin2012}%
  \BibitemOpen
  \bibfield  {author} {\bibinfo {author} {\bibfnamefont {A.}~\bibnamefont
  {Dauphin}}, \bibinfo {author} {\bibfnamefont {M.}~\bibnamefont {M\"uller}}, \
  and\ \bibinfo {author} {\bibfnamefont {M.~A.}\ \bibnamefont
  {Martin-Delgado}},\ }\href {\doibase 10.1103/PhysRevA.86.053618} {\bibfield
  {journal} {\bibinfo  {journal} {Phys. Rev. A}\ }\textbf {\bibinfo {volume}
  {86}},\ \bibinfo {pages} {053618} (\bibinfo {year} {2012})}\BibitemShut
  {NoStop}%
\bibitem [{\citenamefont {Liu}\ \emph {et~al.}()\citenamefont {Liu},
  \citenamefont {Dou\c{c}ot},\ and\ \citenamefont {Le~Hur}}]{Liu2014}%
  \BibitemOpen
  \bibfield  {author} {\bibinfo {author} {\bibfnamefont {T.}~\bibnamefont
  {Liu}}, \bibinfo {author} {\bibfnamefont {B.}~\bibnamefont {Dou\c{c}ot}}, \
  and\ \bibinfo {author} {\bibfnamefont {K.}~\bibnamefont {Le~Hur}},\ }\href
  {http://arxiv.org/abs/1409.6237} {\ }\bibinfo {note}
  {ArXiv:1409.6237}\BibitemShut {NoStop}%
\bibitem [{\citenamefont {Garc\'{\i}a-Mart\'{\i}nez}\ \emph
  {et~al.}(2013)\citenamefont {Garc\'{\i}a-Mart\'{\i}nez}, \citenamefont
  {Grushin}, \citenamefont {Neupert}, \citenamefont {Valenzuela},\ and\
  \citenamefont {Castro}}]{Garcia2013}%
  \BibitemOpen
  \bibfield  {author} {\bibinfo {author} {\bibfnamefont {N.~A.}\ \bibnamefont
  {Garc\'{\i}a-Mart\'{\i}nez}}, \bibinfo {author} {\bibfnamefont {A.~G.}\
  \bibnamefont {Grushin}}, \bibinfo {author} {\bibfnamefont {T.}~\bibnamefont
  {Neupert}}, \bibinfo {author} {\bibfnamefont {B.}~\bibnamefont {Valenzuela}},
  \ and\ \bibinfo {author} {\bibfnamefont {E.~V.}\ \bibnamefont {Castro}},\
  }\href {\doibase 10.1103/PhysRevB.88.245123} {\bibfield  {journal} {\bibinfo
  {journal} {Phys. Rev. B}\ }\textbf {\bibinfo {volume} {88}},\ \bibinfo
  {pages} {245123} (\bibinfo {year} {2013})}\BibitemShut {NoStop}%
\bibitem [{\citenamefont {Daghofer}\ and\ \citenamefont
  {Hohenadler}(2014)}]{Daghofer2014}%
  \BibitemOpen
  \bibfield  {author} {\bibinfo {author} {\bibfnamefont {M.}~\bibnamefont
  {Daghofer}}\ and\ \bibinfo {author} {\bibfnamefont {M.}~\bibnamefont
  {Hohenadler}},\ }\href {\doibase 10.1103/PhysRevB.89.035103} {\bibfield
  {journal} {\bibinfo  {journal} {Phys. Rev. B}\ }\textbf {\bibinfo {volume}
  {89}},\ \bibinfo {pages} {035103} (\bibinfo {year} {2014})}\BibitemShut
  {NoStop}%
\bibitem [{\citenamefont {\DJ{}uri\ifmmode~\acute{c}\else \'{c}\fi{}}\ \emph
  {et~al.}(2014)\citenamefont {\DJ{}uri\ifmmode~\acute{c}\else \'{c}\fi{}},
  \citenamefont {Chancellor},\ and\ \citenamefont {Herbut}}]{Duric2014}%
  \BibitemOpen
  \bibfield  {author} {\bibinfo {author} {\bibfnamefont {T.}~\bibnamefont
  {\DJ{}uri\ifmmode~\acute{c}\else \'{c}\fi{}}}, \bibinfo {author}
  {\bibfnamefont {N.}~\bibnamefont {Chancellor}}, \ and\ \bibinfo {author}
  {\bibfnamefont {I.~F.}\ \bibnamefont {Herbut}},\ }\href {\doibase
  10.1103/PhysRevB.89.165123} {\bibfield  {journal} {\bibinfo  {journal} {Phys.
  Rev. B}\ }\textbf {\bibinfo {volume} {89}},\ \bibinfo {pages} {165123}
  (\bibinfo {year} {2014})}\BibitemShut {NoStop}%
\bibitem [{\citenamefont {Huffman}\ and\ \citenamefont
  {Chandrasekharan}(2014)}]{Huffman2014}%
  \BibitemOpen
  \bibfield  {author} {\bibinfo {author} {\bibfnamefont {E.~F.}\ \bibnamefont
  {Huffman}}\ and\ \bibinfo {author} {\bibfnamefont {S.}~\bibnamefont
  {Chandrasekharan}},\ }\href {\doibase 10.1103/PhysRevB.89.111101} {\bibfield
  {journal} {\bibinfo  {journal} {Phys. Rev. B}\ }\textbf {\bibinfo {volume}
  {89}},\ \bibinfo {pages} {111101(R)} (\bibinfo {year} {2014})}\BibitemShut
  {NoStop}%
\bibitem [{\citenamefont {Wang}\ \emph {et~al.}(2014)\citenamefont {Wang},
  \citenamefont {Corboz},\ and\ \citenamefont {Troyer}}]{Wang2014}%
  \BibitemOpen
  \bibfield  {author} {\bibinfo {author} {\bibfnamefont {L.}~\bibnamefont
  {Wang}}, \bibinfo {author} {\bibfnamefont {P.}~\bibnamefont {Corboz}}, \ and\
  \bibinfo {author} {\bibfnamefont {M.}~\bibnamefont {Troyer}},\ }\href
  {http://stacks.iop.org/1367-2630/16/i=10/a=103008} {\bibfield  {journal}
  {\bibinfo  {journal} {New Journal of Physics}\ }\textbf {\bibinfo {volume}
  {16}},\ \bibinfo {pages} {103008} (\bibinfo {year} {2014})}\BibitemShut
  {NoStop}%
\bibitem [{\citenamefont {Li}\ \emph {et~al.}(2015)\citenamefont {Li},
  \citenamefont {Jiang},\ and\ \citenamefont {Yao}}]{Li2015}%
  \BibitemOpen
  \bibfield  {author} {\bibinfo {author} {\bibfnamefont {Z.-X.}\ \bibnamefont
  {Li}}, \bibinfo {author} {\bibfnamefont {Y.-F.}\ \bibnamefont {Jiang}}, \
  and\ \bibinfo {author} {\bibfnamefont {H.}~\bibnamefont {Yao}},\ }\href
  {\doibase 10.1103/PhysRevB.91.241117} {\bibfield  {journal} {\bibinfo
  {journal} {Phys. Rev. B}\ }\textbf {\bibinfo {volume} {91}},\ \bibinfo
  {pages} {241117(R)} (\bibinfo {year} {2015})}\BibitemShut {NoStop}%
\bibitem [{\citenamefont {Shankar}(1994)}]{Shankar1994}%
  \BibitemOpen
  \bibfield  {author} {\bibinfo {author} {\bibfnamefont {R.}~\bibnamefont
  {Shankar}},\ }\href {\doibase 10.1103/RevModPhys.66.129} {\bibfield
  {journal} {\bibinfo  {journal} {Rev. Mod. Phys.}\ }\textbf {\bibinfo {volume}
  {66}},\ \bibinfo {pages} {129} (\bibinfo {year} {1994})}\BibitemShut
  {NoStop}%
\bibitem [{\citenamefont {Kotov}\ \emph {et~al.}(2012)\citenamefont {Kotov},
  \citenamefont {Uchoa}, \citenamefont {Pereira}, \citenamefont {Guinea},\ and\
  \citenamefont {Castro~Neto}}]{Kotov2012}%
  \BibitemOpen
  \bibfield  {author} {\bibinfo {author} {\bibfnamefont {V.~N.}\ \bibnamefont
  {Kotov}}, \bibinfo {author} {\bibfnamefont {B.}~\bibnamefont {Uchoa}},
  \bibinfo {author} {\bibfnamefont {V.~M.}\ \bibnamefont {Pereira}}, \bibinfo
  {author} {\bibfnamefont {F.}~\bibnamefont {Guinea}}, \ and\ \bibinfo {author}
  {\bibfnamefont {A.~H.}\ \bibnamefont {Castro~Neto}},\ }\href {\doibase
  10.1103/RevModPhys.84.1067} {\bibfield  {journal} {\bibinfo  {journal} {Rev.
  Mod. Phys.}\ }\textbf {\bibinfo {volume} {84}},\ \bibinfo {pages} {1067}
  (\bibinfo {year} {2012})}\BibitemShut {NoStop}%
\bibitem [{\citenamefont {Ryu}\ \emph {et~al.}(2009)\citenamefont {Ryu},
  \citenamefont {Mudry}, \citenamefont {Hou},\ and\ \citenamefont
  {Chamon}}]{Ryu2009}%
  \BibitemOpen
  \bibfield  {author} {\bibinfo {author} {\bibfnamefont {S.}~\bibnamefont
  {Ryu}}, \bibinfo {author} {\bibfnamefont {C.}~\bibnamefont {Mudry}}, \bibinfo
  {author} {\bibfnamefont {C.-Y.}\ \bibnamefont {Hou}}, \ and\ \bibinfo
  {author} {\bibfnamefont {C.}~\bibnamefont {Chamon}},\ }\href {\doibase
  10.1103/PhysRevB.80.205319} {\bibfield  {journal} {\bibinfo  {journal} {Phys.
  Rev. B}\ }\textbf {\bibinfo {volume} {80}},\ \bibinfo {pages} {205319}
  (\bibinfo {year} {2009})}\BibitemShut {NoStop}%
\bibitem [{\citenamefont {Herbut}\ \emph {et~al.}(2009)\citenamefont {Herbut},
  \citenamefont {Juri\ifmmode \check{c}\else \v{c}\fi{}i\ifmmode~\acute{c}\else
  \'{c}\fi{}},\ and\ \citenamefont {Roy}}]{Herbut2009}%
  \BibitemOpen
  \bibfield  {author} {\bibinfo {author} {\bibfnamefont {I.~F.}\ \bibnamefont
  {Herbut}}, \bibinfo {author} {\bibfnamefont {V.}~\bibnamefont {Juri\ifmmode
  \check{c}\else \v{c}\fi{}i\ifmmode~\acute{c}\else \'{c}\fi{}}}, \ and\
  \bibinfo {author} {\bibfnamefont {B.}~\bibnamefont {Roy}},\ }\href {\doibase
  10.1103/PhysRevB.79.085116} {\bibfield  {journal} {\bibinfo  {journal} {Phys.
  Rev. B}\ }\textbf {\bibinfo {volume} {79}},\ \bibinfo {pages} {085116}
  (\bibinfo {year} {2009})}\BibitemShut {NoStop}%
\bibitem [{\citenamefont {Haldane}(1988)}]{Haldane1988}%
  \BibitemOpen
  \bibfield  {author} {\bibinfo {author} {\bibfnamefont {F.~D.~M.}\
  \bibnamefont {Haldane}},\ }\href {\doibase 10.1103/PhysRevLett.61.2015}
  {\bibfield  {journal} {\bibinfo  {journal} {Phys. Rev. Lett.}\ }\textbf
  {\bibinfo {volume} {61}},\ \bibinfo {pages} {2015} (\bibinfo {year}
  {1988})}\BibitemShut {NoStop}%
\bibitem [{\citenamefont {Roy}\ and\ \citenamefont {Herbut}(2010)}]{Roy2010}%
  \BibitemOpen
  \bibfield  {author} {\bibinfo {author} {\bibfnamefont {B.}~\bibnamefont
  {Roy}}\ and\ \bibinfo {author} {\bibfnamefont {I.~F.}\ \bibnamefont
  {Herbut}},\ }\href {\doibase 10.1103/PhysRevB.82.035429} {\bibfield
  {journal} {\bibinfo  {journal} {Phys. Rev. B}\ }\textbf {\bibinfo {volume}
  {82}},\ \bibinfo {pages} {035429} (\bibinfo {year} {2010})}\BibitemShut
  {NoStop}%
\bibitem [{\citenamefont {Jian}\ \emph {et~al.}(2015)\citenamefont {Jian},
  \citenamefont {Jiang},\ and\ \citenamefont {Yao}}]{Yao14}%
  \BibitemOpen
  \bibfield  {author} {\bibinfo {author} {\bibfnamefont {S.-K.}\ \bibnamefont
  {Jian}}, \bibinfo {author} {\bibfnamefont {Y.-F.}\ \bibnamefont {Jiang}}, \
  and\ \bibinfo {author} {\bibfnamefont {H.}~\bibnamefont {Yao}},\ }\href
  {\doibase 10.1103/PhysRevLett.114.237001} {\bibfield  {journal} {\bibinfo
  {journal} {Phys. Rev. Lett.}\ }\textbf {\bibinfo {volume} {114}},\ \bibinfo
  {pages} {237001} (\bibinfo {year} {2015})}\BibitemShut {NoStop}%
\bibitem [{\citenamefont {Corboz}\ \emph {et~al.}(2012)\citenamefont {Corboz},
  \citenamefont {Capponi}, \citenamefont {L\"auchli}, \citenamefont {Bauer},\
  and\ \citenamefont {Or\'us}}]{Corboz2012}%
  \BibitemOpen
  \bibfield  {author} {\bibinfo {author} {\bibfnamefont {P.}~\bibnamefont
  {Corboz}}, \bibinfo {author} {\bibfnamefont {S.}~\bibnamefont {Capponi}},
  \bibinfo {author} {\bibfnamefont {A.~M.}\ \bibnamefont {L\"auchli}}, \bibinfo
  {author} {\bibfnamefont {B.}~\bibnamefont {Bauer}}, \ and\ \bibinfo {author}
  {\bibfnamefont {R.}~\bibnamefont {Or\'us}},\ }\href
  {http://stacks.iop.org/0295-5075/98/i=2/a=27005} {\bibfield  {journal}
  {\bibinfo  {journal} {EPL}\ }\textbf {\bibinfo {volume} {98}},\ \bibinfo
  {pages} {27005} (\bibinfo {year} {2012})}\BibitemShut {NoStop}%
\bibitem [{\citenamefont {Chaloupka}\ \emph {et~al.}(2013)\citenamefont
  {Chaloupka}, \citenamefont {Jackeli},\ and\ \citenamefont
  {Khaliullin}}]{Chaloupka2013}%
  \BibitemOpen
  \bibfield  {author} {\bibinfo {author} {\bibfnamefont {J.}~\bibnamefont
  {Chaloupka}}, \bibinfo {author} {\bibfnamefont {G.}~\bibnamefont {Jackeli}},
  \ and\ \bibinfo {author} {\bibfnamefont {G.}~\bibnamefont {Khaliullin}},\
  }\href {\doibase 10.1103/PhysRevLett.110.097204} {\bibfield  {journal}
  {\bibinfo  {journal} {Phys. Rev. Lett.}\ }\textbf {\bibinfo {volume} {110}},\
  \bibinfo {pages} {097204} (\bibinfo {year} {2013})}\BibitemShut {NoStop}%
\bibitem [{\citenamefont {Wannier}(1950)}]{Wannier}%
  \BibitemOpen
  \bibfield  {author} {\bibinfo {author} {\bibfnamefont {G.~H.}\ \bibnamefont
  {Wannier}},\ }\href {\doibase 10.1103/PhysRev.79.357} {\bibfield  {journal}
  {\bibinfo  {journal} {Phys. Rev.}\ }\textbf {\bibinfo {volume} {79}},\
  \bibinfo {pages} {357} (\bibinfo {year} {1950})}\BibitemShut {NoStop}%
\bibitem [{\citenamefont {Hotta}\ and\ \citenamefont
  {Furukawa}(2006)}]{Hotta2006}%
  \BibitemOpen
  \bibfield  {author} {\bibinfo {author} {\bibfnamefont {C.}~\bibnamefont
  {Hotta}}\ and\ \bibinfo {author} {\bibfnamefont {N.}~\bibnamefont
  {Furukawa}},\ }\href {\doibase 10.1103/PhysRevB.74.193107} {\bibfield
  {journal} {\bibinfo  {journal} {Phys. Rev. B}\ }\textbf {\bibinfo {volume}
  {74}},\ \bibinfo {pages} {193107} (\bibinfo {year} {2006})}\BibitemShut
  {NoStop}%
\bibitem [{Note1()}]{Note1}%
  \BibitemOpen
  \bibinfo {note} {$N=24$: 18 states with 8 off-diagonal matrix elements each;
  $N=28$: 14 states with 10 off-diagonal matrix elements each; $N=36$: 6 states
  with 12 off-diagonal matrix elements each; $N=72$: 18 states with 24
  off-diagonal matrix elements each.}\BibitemShut {Stop}%
\bibitem [{Note2()}]{Note2}%
  \BibitemOpen
  \bibinfo {note} {We use the full lattice symmetry, as well as particle-hole
  symmetry. For space symmetries, we label sectors using momentum and
  irreducible representation corresponding to the point-group compatible with
  this momentum. For instance, on $N=24$ cluster which possesses all
  symmetries, we have used $C_{6v}$ group at the $\Gamma $ point, $\protect
  \mathbb {Z}_2$ group at the $\protect \mathbf {X}$ point, $C_{3v}$ group at
  the $\protect \mathbf {K}$ point and $\protect \mathbb {Z}_2 \times \protect
  \mathbb {Z}_2$ group at the $\protect \mathbf {M}$ point.}\BibitemShut
  {Stop}%
\bibitem [{Note3()}]{Note3}%
  \BibitemOpen
  \bibinfo {note} {Note that the choice of the real space Fermi normal ordering
  can have an impact on the absolute symmetry sectors of the finite size
  clusters, relative quantum numbers should however be invariant}\BibitemShut
  {NoStop}%
\bibitem [{\citenamefont {Albuquerque}\ \emph {et~al.}(2011)\citenamefont
  {Albuquerque}, \citenamefont {Schwandt}, \citenamefont {Het\'enyi},
  \citenamefont {Capponi}, \citenamefont {Mambrini},\ and\ \citenamefont
  {L\"auchli}}]{Albuquerque2011}%
  \BibitemOpen
  \bibfield  {author} {\bibinfo {author} {\bibfnamefont {A.~F.}\ \bibnamefont
  {Albuquerque}}, \bibinfo {author} {\bibfnamefont {D.}~\bibnamefont
  {Schwandt}}, \bibinfo {author} {\bibfnamefont {B.}~\bibnamefont {Het\'enyi}},
  \bibinfo {author} {\bibfnamefont {S.}~\bibnamefont {Capponi}}, \bibinfo
  {author} {\bibfnamefont {M.}~\bibnamefont {Mambrini}}, \ and\ \bibinfo
  {author} {\bibfnamefont {A.~M.}\ \bibnamefont {L\"auchli}},\ }\href {\doibase
  10.1103/PhysRevB.84.024406} {\bibfield  {journal} {\bibinfo  {journal} {Phys.
  Rev. B}\ }\textbf {\bibinfo {volume} {84}},\ \bibinfo {pages} {024406}
  (\bibinfo {year} {2011})}\BibitemShut {NoStop}%
\bibitem [{Note4()}]{Note4}%
  \BibitemOpen
  \bibinfo {note} {We disagree with the advocated general 4-fold degeneracy
  mentioned in Ref.~\protect \rev@citealpnum {Garcia2013}. The four-fold
  degeneracy is actually an artifact of the $N=18$ sample~\cite
  {Corboz2013}}\BibitemShut {NoStop}%
\bibitem [{\citenamefont {Read}\ and\ \citenamefont
  {Sachdev}(1990)}]{Read1990}%
  \BibitemOpen
  \bibfield  {author} {\bibinfo {author} {\bibfnamefont {N.}~\bibnamefont
  {Read}}\ and\ \bibinfo {author} {\bibfnamefont {S.}~\bibnamefont {Sachdev}},\
  }\href {\doibase 10.1103/PhysRevB.42.4568} {\bibfield  {journal} {\bibinfo
  {journal} {Phys. Rev. B}\ }\textbf {\bibinfo {volume} {42}},\ \bibinfo
  {pages} {4568} (\bibinfo {year} {1990})}\BibitemShut {NoStop}%
\bibitem [{\citenamefont {Moessner}\ \emph {et~al.}(2001)\citenamefont
  {Moessner}, \citenamefont {Sondhi},\ and\ \citenamefont
  {Chandra}}]{Moessner2001}%
  \BibitemOpen
  \bibfield  {author} {\bibinfo {author} {\bibfnamefont {R.}~\bibnamefont
  {Moessner}}, \bibinfo {author} {\bibfnamefont {S.~L.}\ \bibnamefont
  {Sondhi}}, \ and\ \bibinfo {author} {\bibfnamefont {P.}~\bibnamefont
  {Chandra}},\ }\href {\doibase 10.1103/PhysRevB.64.144416} {\bibfield
  {journal} {\bibinfo  {journal} {Physical Review B}\ }\textbf {\bibinfo
  {volume} {64}},\ \bibinfo {pages} {144416} (\bibinfo {year}
  {2001})}\BibitemShut {NoStop}%
\bibitem [{\citenamefont {Zhu}\ \emph {et~al.}(2013)\citenamefont {Zhu},
  \citenamefont {Huse},\ and\ \citenamefont {White}}]{Zhu2013}%
  \BibitemOpen
  \bibfield  {author} {\bibinfo {author} {\bibfnamefont {Z.}~\bibnamefont
  {Zhu}}, \bibinfo {author} {\bibfnamefont {D.~A.}\ \bibnamefont {Huse}}, \
  and\ \bibinfo {author} {\bibfnamefont {S.~R.}\ \bibnamefont {White}},\ }\href
  {\doibase 10.1103/PhysRevLett.110.127205} {\bibfield  {journal} {\bibinfo
  {journal} {Phys. Rev. Lett.}\ }\textbf {\bibinfo {volume} {110}},\ \bibinfo
  {pages} {127205} (\bibinfo {year} {2013})}\BibitemShut {NoStop}%
\bibitem [{\citenamefont {Ganesh}\ \emph {et~al.}(2013)\citenamefont {Ganesh},
  \citenamefont {van~den Brink},\ and\ \citenamefont {Nishimoto}}]{Ganesh2013}%
  \BibitemOpen
  \bibfield  {author} {\bibinfo {author} {\bibfnamefont {R.}~\bibnamefont
  {Ganesh}}, \bibinfo {author} {\bibfnamefont {J.}~\bibnamefont {van~den
  Brink}}, \ and\ \bibinfo {author} {\bibfnamefont {S.}~\bibnamefont
  {Nishimoto}},\ }\href {\doibase 10.1103/PhysRevLett.110.127203} {\bibfield
  {journal} {\bibinfo  {journal} {Phys. Rev. Lett.}\ }\textbf {\bibinfo
  {volume} {110}},\ \bibinfo {pages} {127203} (\bibinfo {year}
  {2013})}\BibitemShut {NoStop}%
\bibitem [{\citenamefont {Gong}\ \emph {et~al.}(2013)\citenamefont {Gong},
  \citenamefont {Sheng}, \citenamefont {Motrunich},\ and\ \citenamefont
  {Fisher}}]{Gong2013}%
  \BibitemOpen
  \bibfield  {author} {\bibinfo {author} {\bibfnamefont {S.-S.}\ \bibnamefont
  {Gong}}, \bibinfo {author} {\bibfnamefont {D.~N.}\ \bibnamefont {Sheng}},
  \bibinfo {author} {\bibfnamefont {O.~I.}\ \bibnamefont {Motrunich}}, \ and\
  \bibinfo {author} {\bibfnamefont {M.~P.~A.}\ \bibnamefont {Fisher}},\ }\href
  {\doibase 10.1103/PhysRevB.88.165138} {\bibfield  {journal} {\bibinfo
  {journal} {Phys. Rev. B}\ }\textbf {\bibinfo {volume} {88}},\ \bibinfo
  {pages} {165138} (\bibinfo {year} {2013})}\BibitemShut {NoStop}%
\bibitem [{Note5()}]{Note5}%
  \BibitemOpen
  \bibinfo {note} {Our numerical current structure factor (see its definition
  below) are indeed the strongest when $V_1=0$ on clusters $N=24$ and $N=42$
  for instance.}\BibitemShut {Stop}%
\bibitem [{Note6()}]{Note6}%
  \BibitemOpen
  \bibinfo {note} {We sum over all bonds which do not share any common site
  with the reference bond, i.e there are $N_b=3N/2-11$ terms.}\BibitemShut
  {Stop}%
\bibitem [{\citenamefont {Jia}\ \emph {et~al.}(2013)\citenamefont {Jia},
  \citenamefont {Guo}, \citenamefont {Chen}, \citenamefont {Shen},\ and\
  \citenamefont {Feng}}]{Jia2013}%
  \BibitemOpen
  \bibfield  {author} {\bibinfo {author} {\bibfnamefont {Y.}~\bibnamefont
  {Jia}}, \bibinfo {author} {\bibfnamefont {H.}~\bibnamefont {Guo}}, \bibinfo
  {author} {\bibfnamefont {Z.}~\bibnamefont {Chen}}, \bibinfo {author}
  {\bibfnamefont {S.-Q.}\ \bibnamefont {Shen}}, \ and\ \bibinfo {author}
  {\bibfnamefont {S.}~\bibnamefont {Feng}},\ }\href {\doibase
  10.1103/PhysRevB.88.075101} {\bibfield  {journal} {\bibinfo  {journal} {Phys.
  Rev. B}\ }\textbf {\bibinfo {volume} {88}},\ \bibinfo {pages} {075101}
  (\bibinfo {year} {2013})}\BibitemShut {NoStop}%
\bibitem [{\citenamefont {Asano}\ and\ \citenamefont
  {Hotta}(2011)}]{Asano2011}%
  \BibitemOpen
  \bibfield  {author} {\bibinfo {author} {\bibfnamefont {K.}~\bibnamefont
  {Asano}}\ and\ \bibinfo {author} {\bibfnamefont {C.}~\bibnamefont {Hotta}},\
  }\href {\doibase 10.1103/PhysRevB.83.245125} {\bibfield  {journal} {\bibinfo
  {journal} {Phys. Rev. B}\ }\textbf {\bibinfo {volume} {83}},\ \bibinfo
  {pages} {245125} (\bibinfo {year} {2011})}\BibitemShut {NoStop}%
\bibitem [{\citenamefont {Motruk}\ \emph {et~al.}(2015)\citenamefont {Motruk},
  \citenamefont {Grushin}, \citenamefont {de~Juan},\ and\ \citenamefont
  {Pollmann}}]{Motruk2015}%
  \BibitemOpen
  \bibfield  {author} {\bibinfo {author} {\bibfnamefont {J.}~\bibnamefont
  {Motruk}}, \bibinfo {author} {\bibfnamefont {A.~G.}\ \bibnamefont {Grushin}},
  \bibinfo {author} {\bibfnamefont {F.}~\bibnamefont {de~Juan}}, \ and\
  \bibinfo {author} {\bibfnamefont {F.}~\bibnamefont {Pollmann}},\ }\href
  {\doibase 10.1103/PhysRevB.92.085147} {\bibfield  {journal} {\bibinfo
  {journal} {Phys. Rev. B}\ }\textbf {\bibinfo {volume} {92}},\ \bibinfo
  {pages} {085147} (\bibinfo {year} {2015})}\BibitemShut {NoStop}%
\bibitem [{\citenamefont {Corboz}\ \emph {et~al.}(2013)\citenamefont {Corboz},
  \citenamefont {Lajk\'o}, \citenamefont {Penc}, \citenamefont {Mila},\ and\
  \citenamefont {L\"auchli}}]{Corboz2013}%
  \BibitemOpen
  \bibfield  {author} {\bibinfo {author} {\bibfnamefont {P.}~\bibnamefont
  {Corboz}}, \bibinfo {author} {\bibfnamefont {M.}~\bibnamefont {Lajk\'o}},
  \bibinfo {author} {\bibfnamefont {K.}~\bibnamefont {Penc}}, \bibinfo {author}
  {\bibfnamefont {F.}~\bibnamefont {Mila}}, \ and\ \bibinfo {author}
  {\bibfnamefont {A.~M.}\ \bibnamefont {L\"auchli}},\ }\href {\doibase
  10.1103/PhysRevB.87.195113} {\bibfield  {journal} {\bibinfo  {journal} {Phys.
  Rev. B}\ }\textbf {\bibinfo {volume} {87}},\ \bibinfo {pages} {195113}
  (\bibinfo {year} {2013})}\BibitemShut {NoStop}%
\end{thebibliography}
